\documentclass[conference]{IEEEtran}
\IEEEoverridecommandlockouts
\pdfoutput=1
\usepackage{cite}
\usepackage{amsmath,amssymb,amsfonts}
\usepackage{algorithmic}
\usepackage{graphicx}
\usepackage{textcomp}
\usepackage{xcolor}
\usepackage{epstopdf}
\usepackage{subfigure}
\usepackage[ruled,vlined,linesnumbered,resetcount]{algorithm2e}
\def\BibTeX{{\rm B\kern-.05em{\sc i\kern-.025em b}\kern-.08em
    T\kern-.1667em\lower.7ex\hbox{E}\kern-.125emX}}
\begin{document}

\title{Multi-attributed Community Search in Road-social Networks
}


\author{
\IEEEauthorblockN{Fangda Guo\textsuperscript{\dag}, Ye Yuan\textsuperscript{\S}, Guoren Wang\textsuperscript{\S}, Xiangguo Zhao\textsuperscript{\dag}, Hao Sun\textsuperscript{\dag}}
\IEEEauthorblockA{
\textsuperscript{\dag}\textit{School of Computer Science and Engineering, Northeastern University, Shenyang, China}\\
\textsuperscript{\S}\textit{School of Computer Science and Technology, Beijing Institute of Technology, Beijing, China}\\
\texttt{\small\{\textsuperscript{\dag}fangda@stu, \textsuperscript{\S}yuanye@, \textsuperscript{\dag}zhaoxiangguo@, \textsuperscript{\dag}andynosh@stu\}mail.neu.edu.cn,} \texttt{\small \textsuperscript{\S}wanggr@bit.edu.cn}}
}

\maketitle

\begin{abstract}
Given a location-based social network, how to find the communities that are highly relevant to query users and have top overall scores in multiple attributes according to user preferences? Typically, in the face of such a problem setting, we can model the network as a multi-attributed road-social network, in which each user is linked with location information and $d$ ($\geq\! 1$) numerical attributes. In practice, user preferences (i.e., weights) are usually inherently uncertain and can only be estimated with bounded accuracy, because a human user is not able to designate exact values with absolute precision. Inspired by this, we introduce a normative community model suitable for multi-criteria decision making, called multi-attributed community (MAC), based on the concepts of $k$-core and a novel dominance relationship specific to preferences. Given uncertain user preferences, namely, an approximate representation of weights, the MAC search reports the exact communities for each of the possible weight settings. We devise an elegant index structure to maintain the dominance relationships, based on which two algorithms are developed to efficiently compute the top-$j$ MACs. The efficiency and scalability of our algorithms and the effectiveness of MAC model are demonstrated by extensive experiments on both real-world and synthetic road-social networks.
\end{abstract}


\section{Introduction}
\newtheorem{definition}{Definition}
\newtheorem{theorem}{Theorem}
\newtheorem{example}{Example}
\newtheorem{lemma}{Lemma}
\newtheorem{corollary}{Corollary}
\newenvironment{proof}{{\it Proof}{\it :}}{\hfill $\square$\par}
\newenvironment{proof sketch}{{\it Proof Sketch}{\it :}}{\hfill $\square$\par}
Numerous real-world networks (e.g., social networks) are made up of community structures, where discovering them is an essential problem in network analysis. Recently, community search, which is a kind of query-dependent community discovery problem and is designed to find densely connected subgraphs containing query vertices, has drawn much attention among database professionals due to an ever-growing number of applications \cite{sozio2010community,cui2014local,huang2015approximate,li2015influential,fang2016effective,fang2017effective,huang2017attribute,li2018skyline}.

At the same time, with the prevalence of GPS-enabled mobile devices, location-based social networks (LBSN) are becoming more diverse and complex in recent years (e.g., Facebook Places, Foursquare). Since not only users and friendships are included, but each user is often associated with various properties, such as location information and numerical attributes. The location information is able to bridge the gap between virtual and physical worlds, while the numerical attributes obtained from user profiles or statistical information derived by various network analytics (e.g., influence, similarity, etc.) can characterize the user. For instance, in a scientific collaboration network such as Aminer, every author may have own (spatial) position/address and several numerical attributes (e.g., h-index, \#publications, activeness, diverseness, etc.). Typically, we can model such network data as a multi-attributed road-social network, in which each vertex is linked with location and $d$ ($\geq\! 1$) numerical attributes.

Given a multi-attributed road-social network, how to identify the query-user-involved communities that are not \emph{dominated} by the other communities according to user's preferences for $d$ numerical attributes? For instance, considering h-index and activeness in the Aminer network, how to find a group of collaborators who are related to and close to the query users and take the activeness in their research fields in recent years as the main criterion? Similarly, considering \#publications and diverseness, how to find a community of query users in the Aminer network such that the members provide a good tradeoff in terms of the number of publications and diverse research interests? It is noting that both the social and spatial cohesiveness of groups are incorporated.

In this regard, we want to develop a normative community model suitable for multi-criteria decision making. Considering that the traditional top-$j$ query receives a dataset of records with $d$-dimensional attributes and a weight vector $w$ of $d$ values assigning the relative importance of each dimension to the user as input, where the weighted sum of attribute values is used as the score of a record, we evolve such scoring method into a multi-attributed community model. Similarly, the weight vector $w$ denotes the user preferences and is therefore crucial in generating useful recommendations for the community. Generally, $w$ can be directly input by the user or mined from his/her past behavior or choices \cite{joachims2002optimizing,jiang2008mining}.

Driven by the fact that weight accuracy plays a vital role in the applicability and practicality of top-$j$ queries, we argue that the assumption that exact values of a weight vector are known is inherently inaccurate and almost unrealistic. To illustrate this, consider the case of manually assigning weights to the above example. By taking activeness as the main criterion, a user may specify weights 0.2 and 0.8 for h-index and activeness, respectively. At this point, leaving aside the participation of query users and spatial cohesiveness, the weighted sum of these two attribute values can be regarded as the influence (or importance) of the vertex in \cite{li2015influential}. On the other hand, based on pure intuition, she may also specify the same weight per dimension (e.g., 0.5 each). The results may be similar to the skyline community \cite{li2018skyline}, since weights are not biased towards any dimension (as verified in our experiments, the skyline community \cite{li2018skyline} is usually contained in our results). However, it is impractical and unfair for the user to require weights with absolute precision even a slight variation in the weight (e.g., from 0.2 to 0.19; see Example~\ref{example3}) may remarkably change the results. On the contrary, it would be preferable to take user inputs as generic instructions and leave room for inaccuracies in weight setting. Similarly, for the weight vector computed by preference learning techniques, it should serve only as a rough guide instead of an accurate expression of user preferences. Naturally, this issue can be dealt with by expanding the weight vector to a region\footnote{There are already preference learning techniques (e.g., \cite{qian2015learning}) to generate such a region instead of a specific weight vector.} and returning all promising results to the user, thus providing a practical and more user-centric design.

Prior work on community search problem has never incorporated the uncertainty of weight vector, thus all existing community search algorithms are unable to answer the above questions. To adequately characterize such interesting communities w.r.t. user preferences, we propose a novel community model called multi-attributed community (MAC) based on the concepts of $k$-core \cite{seidman1983network} and r-dominance (variant of traditional sense) \cite{mouratidis2018exact}. An MAC is a maximal connected $k$-core with query vertices contained and spatial cohesiveness satisfied, that is not \emph{r-dominated} by other connected $k$-cores in terms of $d$-dimensional attributes w.r.t a region of interest $R$ (see Section~\ref{section:problemStatement} for detailed definition). Importantly, since the scores of communities all depend on $w$, they are necessarily correlated and vary together as $w$ freely lies inside region $R$. As a result, a partitioning of $R$ forms the output, in which each partition is associated with the MACs when $w$ falls anywhere in that partition. The MAC model is also applicable to many interesting applications, some of which are introduced below.

\noindent
\textbf{Personalized optimum community search.} In daily life, people always want to discover the optimum community based on different needs. For example, a coach hopes to reorganize the school basketball team around certain players (as query users) to improve offense. In this application, we can limit the query scope to the school and extract three numerical attributes for each player: points, rebounds and assists. By setting the region of user preferences, we can obtain corresponding communities that are not r-dominated by the others. Similar query may also help organizations to analyze customer orientation or perform marketing/promotion activities.

\noindent
\textbf{Cohesive groups discovery in LBSNs.} In some cases, one may wish to circle the target range by finding cohesive groups to achieve identification, such as COVID-19 precaution and suspect investigation. Given several confirmed cases, possible cases are likely to be within a certain range of them (providing possibility of close contact), and the Jaccard similarity (e.g., hobbies, interests) and influence (e.g., \#neighbors) for each user can be extracted as numerical attributes; for the preliminary investigation of a case, given the victim and escape range, motive and \#criminal records of the same or different types can be used as numerical attributes. By mining the MACs, we can get the desired cohesive groups related to query users.

\noindent
\textbf{Contributions.} Efficient solutions are formulated and provided to find multi-attributed communities in road-social networks. Below, we summarize the contributions of this paper.

\noindent
\textsf{\underline{Novel community model.}} The MAC model is proposed in road-social networks, which can be used for finding communities not r-dominated by others. To the best of our knowledge, our model is the first to incorporate uncertainty of weight vectors, and our work is also the first to introduce the dominance relationship specific to preferences for community modeling.

\noindent
\textsf{\underline{New algorithms.}} An efficient global search, DFS-based algorithm, is first developed to find the top-$j$ MACs for each of the possible weight settings in user preferences. The time and space complexity of DFS-based algorithm are bounded by $O(n'^{2d})$ and $O(m'\!+\!n'\!+\!n'^2 \!\cdot\! d)$ respectively, where $n'$ and $m'$ denote the number of vertices and edges in the maximal $(k, t)$-core of a road-social network.
To further accelerate the search, we propose a more efficient local search framework, of which two striking features are that (1) its time complexity is much lower than that of global search (at least one order of magnitude faster in practice), and it can find all non-contained MACs in most cases; and (2) it enables the MACs to be output progressively during execution, which is very beneficial to applications expecting only part of the MACs.

\noindent
\textsf{\underline{Extensive experiments.}} To demonstrate the high efficiency and scalability of our proposed algorithms and to evaluate the effectiveness of the MAC model, we conduct extensive experiments and a comprehensive case study on both real-world and synthetic road-social networks, in which many significant and interesting communities are able to be discovered.

\section{Problem Statement}\label{section:problemStatement}
In this section, we formally introduce the multi-attributed community in road-social networks and its search problems. Table~\ref{tab:table1} summarizes the notations used throughout this paper.

\subsection{Preliminaries}
\noindent
\textbf{Road network.}  We model a road network as an undirected weighted graph $G_r \!=\! (V_r, E_r)$, where $V_r$ (resp. $E_r$) is the sets of vertices (resp. edges). A vertex $u \!\in\! V_r$ represents a road intersection/end. An edge $(u, v) \!\in\! E_r$ represents a road segment allowed to travel between vertices $u$ and $v$, and is associated with a non-negative weight $\omega(u, v)$ that represents its cost (e.g., distance/travel time). Let $p$ be a spatial point lying on edge $(u, v)$, and $\omega(u, p)$ be proportional to the length from $u$ to $p$. By $dist(p, p')$ we refer to the network distance between points $p$ and $p'$ in $G_r$, which is the sum of edge weights along the least costly (i.e., shortest) path from $p$ to $p'$. 

\begin{figure}[t]
\centering
\subfigure[Social network ($G_s$)] { \label{fig:social_network}
\includegraphics[width=0.45\columnwidth]{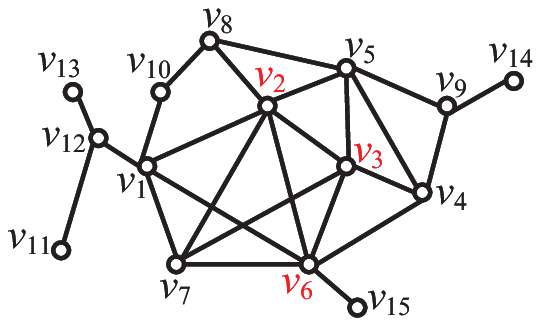}
}
\subfigure[Road network ($G_r$)] { \label{fig:road_network}
\includegraphics[width=0.45\columnwidth]{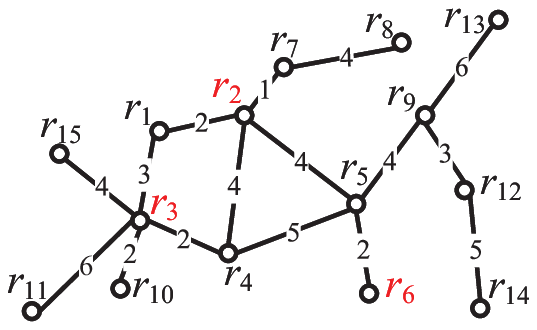}
}
\vspace{-6pt}
\caption{Example of road-social network}
\vspace{-15pt}
\label{fig:networks}
\end{figure}

\noindent
\textbf{Social network.}  We model a social network with $d$ numerical attributes as an undirected graph $G_s\!=\!(V_s, E_s, L, X)$, where $V_s (|V_s|=n)$ and $E_s (|E_s|=m)$ denote the sets of vertices (i.e., users) and edges (i.e., social relations) respectively, $L$ and $X (|X|=n)$ are the sets of mappings and $d$-dimensional vectors defined on $V_s$ respectively. Specifically, for each vertex $v \in V_s$, $L(v)$ provides a mapping of each user's location (i.e., spatial point) in the road network; and $v$ is also associated with a $d$-dimensional vector with real values, denoted by $X(v)\!=\!(x_1^v, \ldots, x_d^v)$, where $X(v) \!\in\! X$ and $x_i^v \!\in\! \mathbb{R}$.

\noindent
\textbf{Road-social network.} A road-social network is a pair of graphs, denoted by $(G_r, G_s)$, where $G_r$ is a road network and $G_s$ is a social network. Each user $u_s \!\in\! G_s$ is associated with a spatial point $p$ in $G_r$, i.e., $L(u_s) \!=\! p$, indicating the current location of $u_s$. We assume that $p$ can either be on a vertex or edge of $G_r$.

\begin{table}[t]
\caption{Summary of Notations}
\label{tab:table1}
\centering
\vspace{-6pt}
\begin{tabular}{|l|p{0.7\columnwidth}|}
  \hline
  \textbf{Notation} & \textbf{Description}\\ \hline
  $Q, j$ & query vertices and number of MACs to select among\\ \hline
  $k, t$ & coreness threshold and query distance threshold\\ \hline
  $d, R$ & dimensionality of attributes and preference domain\\ \hline
  $dg_H(v), \delta(H)$ & degree of vertex $v$ in $H$ and minimum degree of $H$\\ \hline
  $N(v, H)$ & neighbor set of vertex $v$ in subgraph $H$\\ \hline
  $L(v)$ & mapping of user $v$'s location (i.e., spatial point) in $G_r$\\ \hline
  $X(v)$ & $d$-dimensional vector with real values of user $v$\\ \hline
  $D_Q(H), S(H)$ & query distance and score of subgraph $H$ resp.\\ \hline
  $H_k^t, G_d$ & the maximal $(k, t)$-core and r-dominance graph\\ \hline
  $G_e, G_c$ & subgraphs of $G_d$ induced by $V_H$ and $V_{G_d} \!\backslash\! V_H$ resp.\\ \hline
  $l_b(G_e), l_t(G_c)$ & vertices in the bottom layer of $G_e$ and top layer of $G_c$\\
  \hline
\end{tabular}
\vspace{-15pt}
\end{table}

For example, Fig.~\ref{fig:networks} displays a road-social network, where vertices represent users and road junctions, and edges represent social relations and road segments, respectively. For simplicity, the location of user $v_i$ in Fig.~\ref{fig:social_network} is on the vertex $r_i$ of the road network in Fig.~\ref{fig:road_network}. Fig.~\ref{fig:numericalAttributes} shows the values of part of vertices in three different dimensions.

\subsection{(k, t)-Core}
We introduce a novel densely connected substructure, called $(k, t)$-core, by focusing on the following structural cohesiveness and communication cost.

\noindent
\textbf{Structural cohesiveness.} In order to model the structural cohesiveness, the generally used $k$-core model is adopted to indicate the communities \cite{seidman1983network,batagelj2003m,sozio2010community,cui2014local,li2015influential}. In particular, by $dg_{G_s}(v)$ we refer to the degree of vertex $v$ in social network $G_s$. Let $H \!=\! (V_H, E_H, L_H, X_H)$ be an induced subgraph of $G_s$, where $V_H \!\subseteq\! V_s$, $E_H \!=\! \{(u, v)|u, v \!\in\! V_H, (u, v) \!\in\! E_s\}$, $L_H \!=\! \{L(v)|v \!\in\! V_H\}$ and $X_H \!=\! \{X(v)|v \!\in\! V_H\}$.

\vskip 3pt
\begin{definition}
\text{($k$-core.)} Given a graph $G_s$ and an integer $k$, $H$ is a $k$-core of $G_s$ if each vertex $v \!\in\! V_H$ has a degree at least $k$, i.e., $dg_H(v) \!\geq\! k$.
\end{definition}
\vskip 3pt

The maximal $k$-core is the one satisfying that no super $k$-core containing it exists. We refer to the maximal $k$ in all $k$-cores containing vertex $v \!\in\! V_s$ as the core number of $v$. In order to avoid confusion, we denote a connected $k$-core as a $k$-$\widehat{core}$, since the maximal $k$-core is not necessarily connected.

\noindent
\textbf{Remarks.} Although we use $k$-core as the structural cohesiveness metric, our techniques can also be applied to other criteria such as $k$-clique \cite{cui2013online} and $k$-truss \cite{huang2015approximate}.

\noindent
\textbf{Communication cost.} For two users $u, v \in G_s$, the length of the shortest path between their locations in $G_r$ is denoted by $dist(L(u), L(v))$, which is equal to $+\infty$ if $L(u)$ and $L(v)$ are not connected. To model the communication cost in $G_r$, we utilize the notion of query distance below.

\begin{definition}\label{definition2}
\text{(Query Distance.)} Given a graph $H$ and query vertices $Q \!\subseteq\! V_H$, $\forall q \!\in\! Q$, the query distance of $v \!\in\! V_H$ is the maximum length of the shortest path from $L_H(v)$ to $L_H(q)$ in $G_r$, denoted by $D_Q(v) \!=\! \max_{q \in Q}dist(L(v), L(q))$; the query distance of $H$ is defined as $D_Q(H) \!=\! \max_{u \in V_H}D_Q(u) \!=\! \max_{u \in V_H, q \in Q}dist(L(u), L(q))$.
\end{definition}
\vskip 3pt

By query distance $D_Q(H)$, the communication cost between query vertices $Q$ and the members in $H$ can be measured. In general, a good community is considered to own a low communication cost, i.e., small $D_Q(H)$. Consider the query vertices $Q \!=\! \{v_2, v_3, v_6\}$ in Fig.~\ref{fig:networks}. The query distance of $v_7$ is $D_Q(v_7) \!=\! dist(r_7, r_6) \!=\! 7$. The query distance of the subgraph induced by $\{v_2, v_3, v_6, v_7\}$ is equal to $dist(r_3, r_6) \!=\! 9$. In the following, we propose a new notion of $(k, t)$-core, by adapting the concepts of $k$-core and query distance, to capture dense structural cohesiveness and low communication cost.

\vskip 3pt
\begin{definition}
\text{($(k, t)$-core.)} Given graphs $(G_r, G_s)$, query vertices $Q$, and numbers $k$ and $t$, $H$ is a $(k, t)$-core iff $H$ is a $k$-$\widehat{core}$ of $G_s$ containing $Q$ and $D_Q(H) \!\leq\! t$ in $G_r$.
\end{definition}
\vskip 3pt

For a $(k, t)$-core, its structural cohesiveness increases with $k$, while proximity to query vertices decreases with $t$. For instance, in Fig.~\ref{fig:networks}, the subgraph induced by $\{v_2, v_3, v_6, v_7\}$ for $Q \!=\! \{v_2, v_3, v_6\}$ is a $(k, t)$-core with $k \!=\! 3$ and $t \!=\! 9$.

\subsection{Score of Multi-Attributes}
Note that generalizing the existing community models, most of which are only for $1$-dimensional attribute such as influence \cite{li2015influential}, Euclidean distance \cite{fang2017effective} or keyword similarity \cite{fang2016effective,huang2017attribute,zhang2017engagement,zhang2019keyword}, to a comparable one with multi-dimensional attributes is nontrivial. Unlike the above, comparing two communities becomes quite tough if either can have $d$ ($d \!>\! 1$) values due to $d$ different dimensions. On the other hand, existing multi-dimensional model \cite{li2018skyline} cannot compare the pros and cons of any skyline communities and make trade-offs based on user preferences. To overcome the above issues, we introduce the \emph{r-dominance} relationship between two communities, that will be developed for defining our multi-attributed community model.

As each vertex $v \!\in\! V_s$ associated with a vector $X(v)$, the attributes define a $d$-dimensional \emph{data domain}. We assume that for each attribute a higher value is preferable, and a spatial index (e.g., R-tree \cite{guttman1984r}) is used to organize the vector set $X$.

\begin{figure}[t]
\centering
\subfigure[$3$-dimensional vectors] { \label{fig:numericalAttributes}
\includegraphics[width=0.4\columnwidth]{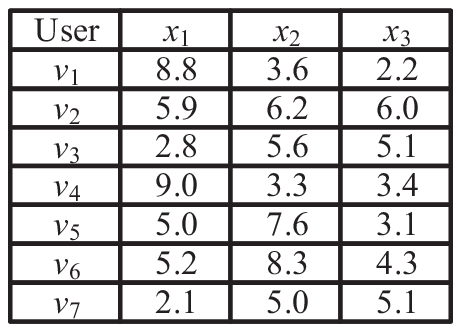}
}
\subfigure[Pref. domain and MACs] { \label{fig:MACs}
\includegraphics[width=0.4\columnwidth]{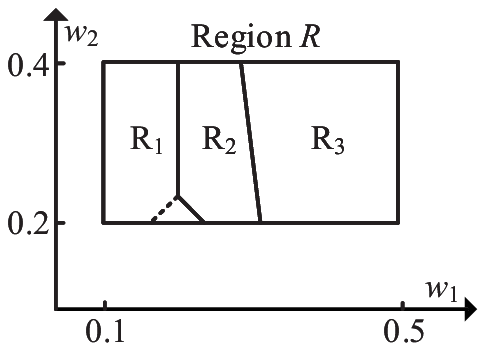}
}
\vspace{-6pt}
\caption{Numerical attributes and MACs in $R$}
\vspace{-15pt}
\label{fig:attributesAndMACs}
\end{figure}

Referring to the traditional top-$j$ queries, the score of a vertex $v$ w.r.t $X(v)$ can also be derived by inputting a weight vector $w\!=\!(w_1, w_2, \ldots, w_d)$ as
\begin{equation}\label{eq:vertexScore}
\setlength\abovedisplayskip{4pt}
\setlength\belowdisplayskip{4pt}
S(v) = \sum\nolimits_{i=1}^{d} w_i \cdot x_i^v.
\end{equation}
Thus, the top-$j$ results consist of the $j$ vertices with the highest scores. We assume w.l.o.g. that $w_i \!\in\! (0, 1)$ for $\forall i \!\in\! [1, d]$ and $\sum_{i=1}^d w_i \!=\! 1$. Such conditions cannot restrain user preferences, because score ranking depends only on the direction of $w$ instead of its magnitude \cite{chang2000onion}; but they allow $w$ to drop one weight (i.e., $w_d \!=\! 1 - \sum_{i=1}^{d-1} w_i$), thereby mapping the domain of $w$ to a ($d \!-\! 1$)-dimensional space, named the \emph{preference domain}. 
This dimensionality reduction is critical \cite{tang2017determining}, since the dimensionality directly determines the processing time of the costliest operations in our techniques (see Section~\ref{section:findSmallestScoreVertex}). In the following, $w$ refers to the ($d \!-\! 1$)-dimensional form of the weight vector. For example, given weights 0.2 and 0.3 for $x_1$ and $x_2$ respectively in Fig.~\ref{fig:numericalAttributes}, we have $w_3 \!=\! 0.5$, and $S(v_7) \!=\! 4.47$.

\noindent
\textbf{Community score.} Let $H \!=\! (V_H, E_H, L_H, X_H)$ be an induced subgraph of $G_s$. Given a weight vector $w$, we define score of $H$ as
\begin{equation}\label{eq:communityScore}
\setlength\abovedisplayskip{4pt}
\setlength\belowdisplayskip{4pt}
S(H) = \text{min}_{v \in V_H}\{S(v)\}.
\end{equation}
Here, we have a brief discussion on why the ``min" operator in Eq.~\ref{eq:communityScore} is used to define $S(H)$. The intention is the same as that in \cite{li2015influential}. By using it, all the members in $H$ can be ensured to have a score in terms of $d$ dimensions no less than $S(H)$. In other words, if $S(H)$ is large, vertices in $H$ may not have large values in each dimension but the weighted sum of their attribute values must be large. For instance, consider the previous case of taking the activeness as the main criterion (e.g., weight 0.8) in the collaboration network. Obviously, we can apply the $S(H)$ defined above to quantify the overall quality focusing on the activeness of a group of collaborators.

\vskip 3pt
\begin{definition}
\text{(r-dominance.)} Given a region $R$ in the preference domain, a community $H$ r-dominates another community $H'$ when $S(H) \!\geq\! S(H')$ for any weight vector in $R$, denoted by $H \!\succ\! H'$; and $v \!\succ\! v'$ denotes that vertex $v$ r-dominates another vertex $v'$.
\end{definition}
\vskip 3pt

Intuitively, in the traditional sense of dominance \cite{borzsony2001skyline,liu2015finding}, for any weight vector $w$ the dominator is always superior to the dominee, while r-dominance is specific to weight vectors in $R$. Although a community may not dominate another based on the skyline community model \cite{li2018skyline}, it might always have a higher score as $w$ is bounded in $R$. For ease of representation, we assume that $R$ is a hyper-rectangle parallel to axes, but our techniques is directly applicable to general convex polytopes.

For example, Fig.~\ref{fig:MACs} illustrates a preference domain (i.e., the domain of weight vector) with $d\!=\!3$, where axis $w_1$ (resp. $w_2$) corresponds to the weight for $x_1$ (resp. $x_2$) on a scale of 0 to 1, and the region $R$ is a convex polygon (i.e., an axis-parallel rectangle $[0.1, 0.5] \!\times\! [0.2, 0.4]$) representing the expanded or approximated user preferences.

\subsection{The MAC Problem}
Combining the notion of $(k, t)$-core and community score, we define the multi-attributed community (MAC) as follows.

\vskip 3pt
\begin{definition}\label{definition5}
\text{(MAC.)} Given graphs $(G_r, G_s)$, query vertices $Q \subseteq V_s$, two numbers $k$ and $t$, and a region of interest $R$, $H$ is a multi-attributed community in the road-social network, if $H$ satisfies the following conditions:
\begin{itemize}
  \item[\emph{1.}] $H$ is a $(k, t)$-core containing $Q$.
  \item[\emph{2.}] There does not exist an induced subgraph $H'$ $(V_{H'} \supset V_H)$ such that $H'$ is a $(k, t)$-core and $H' \succ H$.
\end{itemize}
\end{definition}
\vskip 3pt

In terms of structural cohesiveness and communication cost, condition (1) requires not only that the community with query vertices $Q$ contained is densely connected, but also that each vertex is spatially close to $Q$. In terms of maximality and multi-attributed community score, condition (2) ensures that the community is maximal and having the greatest score. The following example illustrates Definition~\ref{definition5}.

\vskip 3pt
\begin{example}
Consider $(G_r, G_s)$, numerical attributes and a region $R$ in Fig.~\ref{fig:networks} and~\ref{fig:attributesAndMACs}, respectively. Suppose, for instance, that $Q \!=\! \{v_2\}$, $k \!=\! 2$ and $t \!=\! 9$. By Definition~\ref{definition5}, the subgraph induced by $\{v_2, v_3, v_5, v_6, v_7\}$ is an MAC with community score equal to $S(v_7)$ if $w$ freely lies anywhere in the upper-left part of $R_1$ (divided by dotted line), as it meets both conditions. Note that the subgraph induced by $\{v_2, v_3, v_5, v_7\}$ is not an MAC. This is because it is contained in the MAC induced by $\{v_2, v_3, v_5, v_6, v_7\}$ with the same community score, thus fails to satisfy condition (2). $\hfill{} \Box$
\end{example}
\vskip 3pt

For most practical applications, we typically tend to focus on the query-user-involved communities, which score higher than (i.e., r-dominate) all other communities in the preference domain $R$. In this paper, we aim to efficiently discover such communities in road-social networks. Below, two multi-attributed community search problems are formulated.

\noindent
\textbf{Problem 1.} Given graphs $(G_r, G_s)$, query vertices $Q \subseteq V_s$, two numbers $k$ and $t$, and a region of interest $R$, the problem is to find the top-$j$ MACs with the highest community score for each possible weight vector in $R$. Although the possible weight vectors are infinite in $R$, a partitioning of $R$ can form the output, in which each partition is associated with the top-$j$ MACs when $w$ falls anywhere in that partition.

For Problem 1, an MAC may be contained in another MAC in the top-$j$ results.

\vskip 3pt
\begin{example}\label{example2}
Assume that $Q \!=\! \{v_2, v_3, v_6\}$, $k \!=\! 3$ and $t \!=\! 9$ in Fig.~\ref{fig:networks}. The top-$2$ MACs are subgraphs $H_1$ and $H_2$ induced by $\{v_2, v_3, v_6, v_7\}$ and $\{v_2, \ldots, v_7\}$ for any weight vector in $R_1$, respectively. $S(H_1) \!=\! S(v_7)$, and $S(H_2) \!=\! S(v_4)$ or $S(v_5)$ on either side of the dotted line. Clearly, $H_2$ contains $H_1$. $\hfill{} \Box$
\end{example}
\vskip 3pt

To eliminate the containment relations in the top-$j$ results, we study another problem of finding the non-contained MAC.

\vskip 3pt
\begin{definition}\label{definition6}
\text{(non-contained MAC.)} An MAC $H$ is a non-contained MAC if there does not exist an induced subgraph $H'$ $(V_{H'} \subset V_H)$ such that $H'$ is a $(k, t)$-core and $H' \succ H$.
\end{definition}
\vskip 3pt

The following example illustrates Definition~\ref{definition6}.

\begin{example}\label{example3}
Let us reconsider Example~\ref{example2}. By Definition~\ref{definition6}, subgraphs $H_1$ and $H_3$ (induced by $\{v_2, \ldots, v_6\}$) are the non-contained MACs for any weight vector in $R_1$ and in $R_2 \cup R_3$, respectively. For instance, $H_3$ (resp. $H_1$) is the top-$1$ result when $w \!=\! (0.2, 0.3)$ (resp. $w \!=\! (0.19, 0.3)$). However, $H_2$ is not a non-contained MAC as it contains $H_1$ and $H_1 \succ H_2$ for any weight vector in $R_1$. $\hfill{} \Box$
\end{example}
\vskip 3pt

\noindent
\textbf{Problem 2.} Given graphs $(G_r, G_s)$, query vertices $Q \subseteq V_s$, two numbers $k$ and $t$, and a region of interest $R$, the problem is to find the non-contained MAC for every possible weight vector in its corresponding partitioning of $R$.

\noindent
\textbf{Discussions.} Another three possible operators ``max", ``sum" and ``avg" are not appropriate for community score. The first two are monotonic w.r.t. the size of community, that is, a community scores higher than its sub-communities. Hence, the answer is always the maximal $(k, t)$-core, which is independent of numerical attributes $X$ and region $R$. The last one may cause outliers in the answer, e.g., only a few vertices score very high while the rest score low, resulting in a higher community score. Obviously, this is not an ideal community.

\noindent
\textbf{Challenges.} Solving the above two problems faces three major challenges. First, the number of $k$-$\widehat{core}$s containing $Q$ in a multi-attributed network $G_s$ can be exponentially large (even regardless of the query distance in $G_r$). Thus, enumerating all the $k$-$\widehat{core}$s to identify the MACs is intractable. Second, unlike traditional top-$j$ queries \cite{ilyas2008survey}, the score of a community may vary greatly at different parts of $R$, making it nontrivial to draw conclusions about r-dominance relationships between communities. 
Third, the MAC model enables a more flexible way to express user preferences in community search problem, which means that inherent inaccuracies in weight specification need to be taken into account. In consequence, without enumerating all the $(k, t)$-cores, devising an efficient algorithm to detect the MACs is challenging. To overcome these challenges, we will develop efficient algorithms in the following sections.

\section{Warming Up for Our Methods}\label{section:warmingUp}
According to Definition~\ref{definition5}, regardless of maximality and community score, the multi-attributed communities have to satisfy the constraints of structural cohesiveness and communication cost. Thus, we give two useful lemmas as follows.

\vskip 3pt
\begin{lemma}\label{lemma1}
For a number $t$, the vertices of $G_s$ with query distance greater than $t$ in $G_r$ cannot exist in any MAC.
\end{lemma}
\vskip 3pt

\begin{lemma}\label{lemma2}
For an integer $k$, the MACs must be contained in the maximal $k$-$\widehat{core}$ containing $Q$.
\end{lemma}
\vskip 3pt

The correctness of above lemmas can be verified by Definition~\ref{definition2} and the maximality of $k$-core, resulting in Lemma~\ref{lemma3}.

\vskip 3pt
\begin{definition}
\emph{\text{(Maximal $(k, t)$-core.)}} For graphs $(G_r, G_s)$ and query vertices $Q$, the maximal $(k, t)$-core is a $(k, t)$-core such that no super $(k, t)$-core contains it, denoted by $H_k^t$.
\end{definition}
\vskip 3pt

\begin{lemma}\label{lemma3}
For two numbers $k$ and $t$, the MACs must be contained in the maximal $(k, t)$-core.
\end{lemma}
\vskip 3pt

\begin{figure}[t]
\centering
\subfigure[$v$ r-dominates] { \label{fig:r-dominance_a}
\includegraphics[width=0.27\columnwidth]{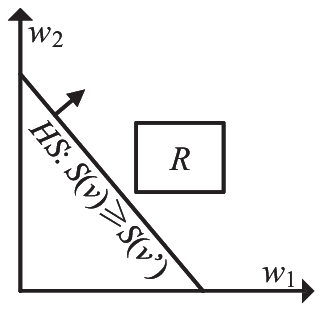}
}
\subfigure[r-incomparable] { \label{fig:r-dominance_b}
\includegraphics[width=0.27\columnwidth]{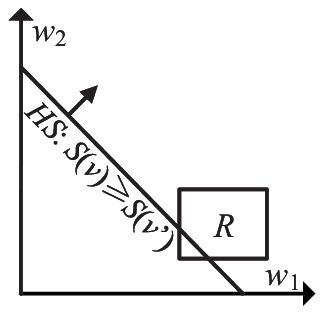}
}
\subfigure[$v$ r-dominated] { \label{fig:r-dominance_c}
\includegraphics[width=0.27\columnwidth]{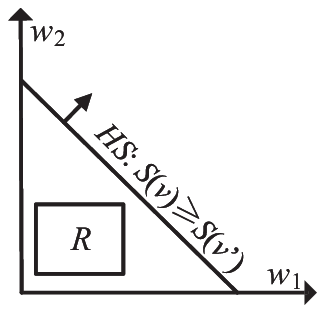}
}
\vspace{-6pt}
\caption{Cases of r-dominance for vertices $v$ and $v'$}
\vspace{-15pt}
\label{fig:r-dominance}
\end{figure}

Referring to Lemma~\ref{lemma3}, for a given $k$ and $t$, we first filter out the vertices of $G_s$ that do not satisfy the query distance threshold $t$ by range query in $G_r$, which can be accelerated by G-tree \cite{zhong2015g} or G*-tree \cite{li2019g}, to obtain the induced connected subgraph $G_s'$ by the remaining vertices. Next, we do $k$-core decomposition \cite{batagelj2003m} on the filtered social subgraph $G_s'$ to compute the maximal $k$-$\widehat{core}$ containing $Q$. It is noting that we employ the upper bound of coreness \cite{cui2014local} before core decomposition. If $k$ is larger than $\lfloor\frac{1+\sqrt{9+8(|E_s'|-|V_s'|)}}{2}\rfloor$, we immediately know there is no $k$-$\widehat{core}$ w.r.t $Q$. So far, the maximal $(k, t)$-core has been found such that the MACs can be obtained through in-depth computation. For example, the maximal $(3, 9)$-core, i.e., $H_3^9$, for $Q \!=\! \{v_2, v_3, v_6\}$ is the subgraph induced by $\{v_1, \ldots, v_7\}$, as shown in Fig.~\ref{fig:Hkt}.

After abandoning the scheme of enumerating all the $(k, t)$-cores whose number can be exponentially large even in $H_k^t$, intuitively, we may think of iteratively deleting the smallest-score vertex w.r.t. $d$-dimensional attributes until the resulting graph does not have a $k$-$\widehat{core}$ containing $Q$. However, at this time we will face another problem, that is, which vertex has the smallest score? To address this issue, we design an effective data structure and construction algorithm to preserve r-dominance relationships between vertices.

\section{R-Dominance Graph}\label{section:rdominanceGraph}
In this section, we exploit the r-dominance graph to preserve pair-wise r-dominance relationships between vertices in $H_k^t$, which will be used in our proposed search algorithms.

\subsection{R-Dominance Test}\label{section:rdominanceTest}
Consider two vertices $v$ and $v'$ where none dominates the other in terms of traditional dominance, that may not draw a reliable conclusion about which vertex ranks higher. Nevertheless, given a preference domain, the inequation $S(v) \!\geq\! S(v')$ (resp. equation $S(v) \!=\! S(v')$) corresponds to a half-space (resp. hyperplane), of which there are three different cases regarding the positioning against $R$ \cite{mouratidis2018exact}. Specifically, in Fig.~\ref{fig:r-dominance_a}, $v$ r-dominates $v'$ since half-space $HS: S(v) \!\geq\! S(v')$ completely covers $R$, which means $v$ scores higher for $\forall w \in R$; the case in Fig.~\ref{fig:r-dominance_c} is symmetric. In Fig.~\ref{fig:r-dominance_b}, $v$ scores higher in one part of $R$ but lower in another, which is called \emph{r-incomparable} as none r-dominates the other.

Clearly, the cases in Fig.~\ref{fig:r-dominance_a} and~\ref{fig:r-dominance_c} allow r-dominance conclusions to be safely drawn. In this way, we can determine r-dominance by detecting whether all \emph{polygon vertices} defining $R$ fall into the half-space $HS: S(v) \!\geq\! S(v')$. If so (resp. not), $v$ r-dominates (resp. is r-dominated by) $v'$. Otherwise, $v$ and $v'$ are r-incomparable. The inclusion detecting of each polygon vertex costs $O(d)$, so the r-dominance test requires $O(pd)$ in total, where $p$ is the number of polygon vertices defining $R$.



\subsection{Pair-Wise R-Dominance Relationship}
The computation of r-dominance relationships is somewhat similar to that of the $k$-skyband (i.e., \emph{BBS} \cite{papadias2005progressive}), but differs as follows. (1) Rather than traditional dominance, we employ r-dominance and apply its test described in Section~\ref{section:rdominanceTest} both for vertex-to-vertex and vertex-to-MBB (i.e., minimum
bounding box) dominance testing. (2) Due to the fact that $w$ is bounded in $R$, we adopt a unique sorting key for R-tree nodes (represented by MBB's upper-right corner) and vertices in the heap to accelerate search convergence by leading it to r-dominate as many members as possible first. (3) We preserve the r-dominance relationships between vertices in $H_k^t$ instead of only the top-$j$ layers. It is noting that a max-heap is utilized in our adapted \emph{BBS} and its sorting key is the score of R-tree node/vertex w.r.t. a pivot vector of $R$, whose value of each dimension is the mean of the polygon vertices of $R$ in that dimension \cite{mouratidis2018exact}. The correctness of our adaptation is guaranteed as follows. (1) The pivot vector must lie in $R$ due to $R$'s convexity \cite{mark2008computational}. (2) Vertices popped after $v$ cannot r-dominate $v$ due to pivot-based sorting (in decreasing order).



\begin{figure}[t]
\centering
\subfigure[$H_3^9$ for $Q = \{v_2, v_3, v_6\}$] { \label{fig:Hkt}
\includegraphics[width=0.352\columnwidth]{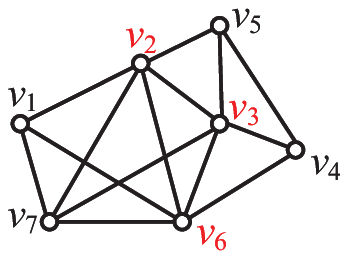}
}
\subfigure[DAG $G_d$] { \label{fig:DAG}
\includegraphics[width=0.342\columnwidth]{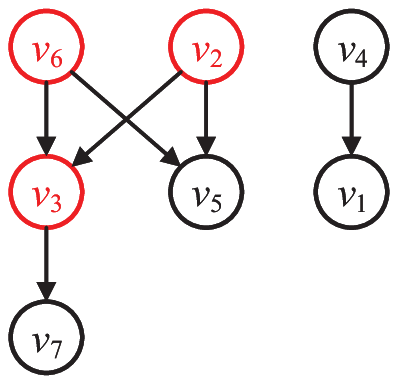}
}
\vspace{-6pt}
\caption{The maximal $(k, t)$-core and r-dominance graph}
\vspace{-15pt}
\label{fig:HktAndDAG}
\end{figure}

In addition, we adopt a directed acyclic graph (DAG) \cite{zou2011pareto,liu2015finding} to maintain all pair-wise r-dominance relationships between vertices in $H_k^t$, named \emph{r-dominance graph} (denoted by $G_d$). Fig.~\ref{fig:DAG} illustrates $G_d$ of $H_3^9$. An arc from vertex $v$ to $v'$ signifies that $v$ r-dominates $v'$. It is noting that an arc from $v_6$ or $v_2$ to $v_7$ is not needed as the transitivity of r-dominance relationship already implies this. The number of vertices that r-dominate $v$ is called $v$'s \emph{r-dominance count}.

\section{Global Search}\label{section:globalSearch}
In this section, we develop a global search algorithm for Problem~2 and discuss its generalization for Problem~1. Before proceeding further, three useful lemmas are given as follows.

\vskip 3pt
\begin{lemma} \label{lemma4}
The maximal $(k, t)$-core, i.e., $H_k^t$, is an MAC.
\end{lemma}
\vskip 3pt

\begin{lemma} \label{lemma5}
For any MAC, if we delete the smallest-score vertex w.r.t. any weight $w$ in $R$ and the resulting subgraph still has a $k$-$\widehat{core}$ $H$ containing $Q$, $H$ is an MAC.
\end{lemma}
\vskip 3pt

\begin{lemma} \label{lemma6}
For any MAC $H$, if we delete the smallest-score vertex w.r.t. any weight $w$ in $R$ but the resulting subgraph does not have a $k$-$\widehat{core}$ containing $Q$, $H$ is a non-contained MAC.
\end{lemma}
\vskip 3pt

In view of the above lemmas, we can devise an efficient algorithm based on depth-first search (DFS) for our problems.

\subsection{The DFS-based Algorithm}
The idea of the DFS-based algorithm is described in detail in Algorithm~1. First, for given $k$ and $t$, we compute the maximal $(k, t)$-core, i.e., $H_k^t$, and build the r-dominance graph $G_d$. Then, the following procedure is iteratively invoked until the resulting graph in each partition of $R$ does not have a $k$-$\widehat{core}$ containing $Q$. The procedure consists of two steps. Let $H$ and $G_d'$ be the resulting subgraph and corresponding r-dominance graph in partition $\rho$ of $R$ (Line~6). The first step is to insert sub-partitions into $\rho$ according to $G_d'$ (Lines~7-9), then find the smallest-score vertex in each sub-partition (Lines~10-11). In Line~9, note that $\rho$ is the root node of a binary tree after being passed in as a parameter, and $S$ is a set of leaf nodes of the binary tree, representing sub-partitions of $\rho$. This step is essential and will be elaborated in Section~\ref{section:findSmallestScoreVertex}. The second step is to delete all the vertices that are definitely excluded in subsequent MACs, which enables $H$ and $G_d'$ to be updated accordingly (Lines~15-20). 

\begin{figure}[t]
\centering
\includegraphics[width=0.49\textwidth]{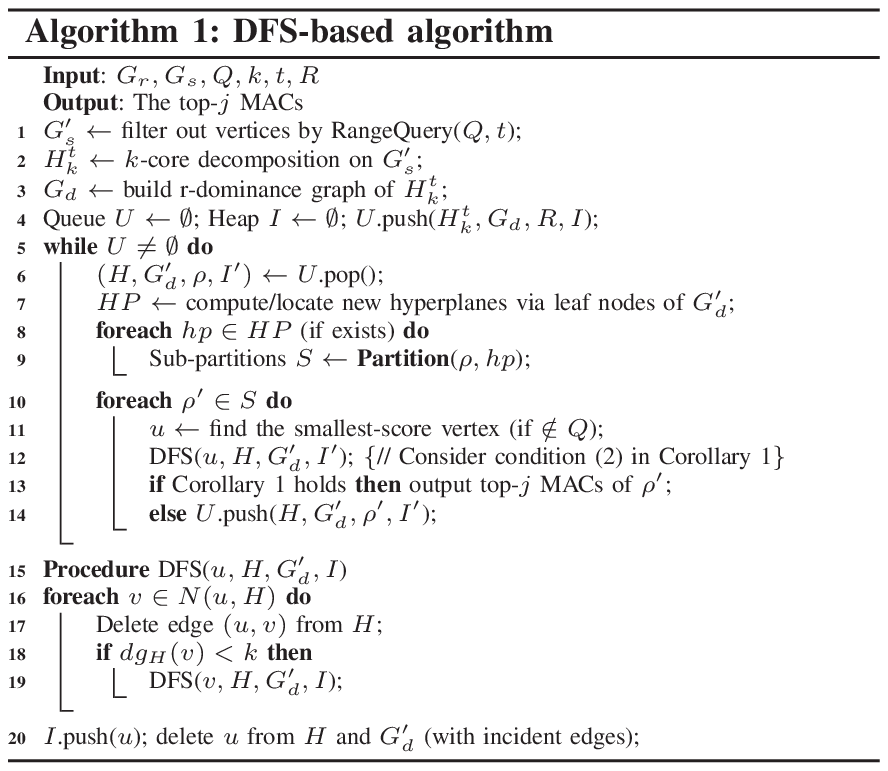}
\label{alg:1}
\vspace{-25pt}
\end{figure}

In particular, Algorithm~1 recursively deletes all the vertices violating the structural cohesiveness constraint by a DFS procedure (Lines~16-19) as long as the smallest-score vertex $u$ is found in the sub-partition. Because the degrees of the adjacent vertices of $u$ all reduce by 1 when we delete $u$. This may cause some neighbors of $u$ to violate the structural cohesiveness constraint and thus cannot be contained in subsequent MACs. Likewise, we also need to verify whether the neighbors of other hops (e.g., 2-hop, 3-hop, etc.) meet the structural cohesiveness constraint. Obviously, the DFS procedure can be used to identify and delete all these vertices.

According to Lemma~\ref{lemma6}, we have a corollary as follows.

\vskip 3pt
\begin{corollary} \label{corollary1}
Given an MAC $H$, $H$ is a non-contained MAC if the smallest-score vertex $u$ meets one of the following conditions: (1) $u \in Q$; and (2) deleting $u$ will recursively disconnect $Q$ (e.g., $\exists q \in Q$ being deleted) or make the degree of remaining vertices less than $k$.
\end{corollary}
\vskip 3pt

Note that we always consider the early termination conditions of Corollary~\ref{corollary1} in conjunction with the DFS procedure. Once either is met, it means that vertex $u$ cannot be deleted, even if $u$ is currently the one with the smallest score but $H$ is already the non-contained MAC w.r.t. partition $\rho'$ (Line~12). As a result, if Corollary~\ref{corollary1} (i.e., Lemma~\ref{lemma6}) holds, the top-$j$ MACs can be obtained by the union of top vertices in heap $I'$ (totally backtracking $j \!-\! 1$ times) and the last subgraph $H$ (Line~13). Based on Lemma~\ref{lemma4} and~\ref{lemma5}, we can easily verify that $H$ for each partition $\rho$ recursively obtained in Line~6 is an MAC. Thus, Theorem~\ref{theorem1} shows the correctness of Algorithm~1.

\vskip 3pt
\begin{theorem} \label{theorem1}
Algorithm~1 correctly finds the top-$j$ MACs.
\end{theorem}
\vskip 3pt

\begin{proof sketch}
For any partition $\rho$, as long as its current subgraph $H$ does not hold Corollary~\ref{corollary1}, $\rho$ will be divided into $|S|$ sub-partitions by $|HP|$ hyperplanes (Line 10 in Algorithm~1), e.g., $\rho_i$ for $1 \leq i \leq |S|$. Here, $\rho$ is discarded when the recursion proceeds to the promising sub-partitions of the local arrangement, i.e., $\rho_i$. Assume that $u$ is the smallest-score vertex in $H$ w.r.t. $\rho_i$, the resulting subgraph, denoted by $H_i$, obtained by invoking DFS procedure (Line 12 in Algorithm~1) can be claimed as the maximal $k$-$\widehat{core}$ of the subgraph $H \backslash u$ by contradiction, because DFS procedure recursively deletes all the vertices in $H$ whose degrees are smaller than $k$. Therefore, we have $V_{H_i} \subset V_H$ and $S(H) \leq S(H_i)$ w.r.t. $\rho_i$ for $1 \leq i \leq |S|$. In this way, the non-contained MAC corresponding to each final sub-partition can be found, as well as all the MACs. Thus, we conclude that Algorithm~1 correctly finds the top-$j$ MACs. 
\end{proof sketch}
\vskip 3pt

We analyze the complexity of Algorithm~1 in Theorem~\ref{theorem2}.

\vskip 3pt
\begin{theorem} \label{theorem2}
The time complexity and space complexity of Algorithm~1 are bounded by $O(n'^{2d})$ and $O(m'+n'+n'^2 \!\cdot\! d)$ respectively, where $n'$ and $m'$ denote the number of vertices and edges in $H_k^t$.
\end{theorem}
\vskip 3pt

\begin{proof}
The key factor determining Algorithm~1's time complexity is the construction of arrangements. In the worst case, vertices in $H_k^t$ are r-incomparable to each other, i.e., the complete arrangement of $\frac{n'(n'-1)}{2}$ half-spaces needs to be constructed, in $O(n'^{2d})$ time \cite{agarwal2000arrangements}. The algorithm only needs to store $H_k^t$, and maintains the heap $I'$ and half-space information related to $d$, which uses less than $O(m'+n'+n'^2 \!\cdot\! d)$ space complexity even in the worst case.
\end{proof}

\begin{figure}[t]
\centering
\subfigure[Partitioning $R$] { \label{fig:subregion1}
\includegraphics[width=0.4\columnwidth]{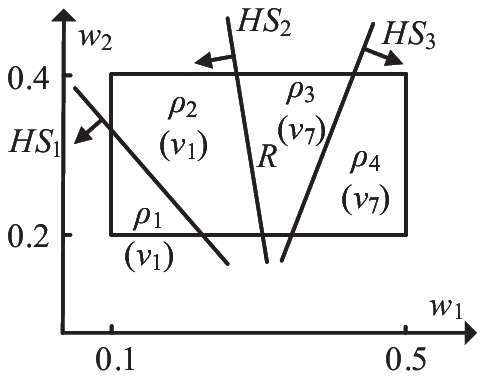}
}
\subfigure[Partitioning $\rho_1$ and $\rho_2$] { \label{fig:subregion2}
\includegraphics[width=0.4\columnwidth]{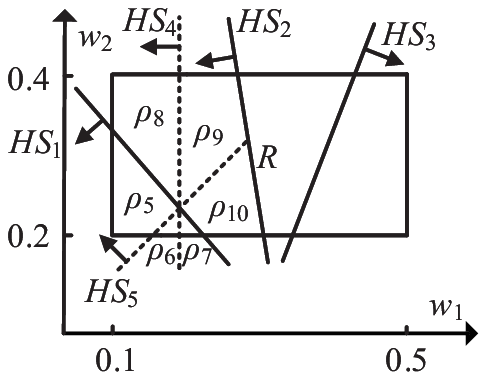}
}
\vspace{-6pt}
\caption{Arrangement and partitions in $R$}
\vspace{-15pt}
\label{fig:subregion}
\end{figure}

\subsection{Arrangement Jointing and Indexing}\label{section:findSmallestScoreVertex}
In Algorithm~1, to find the smallest-score vertex for any weight vector in partition $\rho$, we consider the vertices of $G_d'$ in a bottom-up manner. In other words, leaf vertices of the r-dominance graph will be preferred (Line~7). The reason is that if a vertex is deleted either because it does not satisfy the structural cohesiveness constraint (already considered in the DFS procedure), or because it is the smallest-score vertex, but before this, all vertices it r-dominates should be deleted first.

The verification of a leaf vertex $u$ in $G_d'$ entails partitioning $\rho$ by half-spaces $HS_i: S(u') \!\geq\! S(u)$, each corresponding to one of the remaining leaf vertex $u'$. Formally, an arrangement bounded by $R$ is defined by the supporting hyperplanes of these half-spaces, where each cell (i.e., sub-partition) is located in a set of half-spaces. The leaf vertices corresponding to these half-spaces are precisely those with scores higher than $u$ if $w$ falls in that cell, which means $u$ is the smallest-score vertex.  

Consider $G_d$ in Fig.~\ref{fig:DAG}. Initially, the leaf vertices are $v_7$, $v_5$ and $v_1$ ($G_d' \!=\! G_d, I' \!=\! \emptyset$, Line~6 in Algorithm~1). Their respective half-spaces $HS_1: S(v_7) \!\geq\! S(v_5)$, $HS_2: S(v_7) \!\geq\! S(v_1)$ and $HS_3: S(v_1) \!\geq\! S(v_5)$ are inserted into the arrangement, as shown in Fig.~\ref{fig:subregion1}. The vertex in brackets for each partition indicates the smallest-score vertex for any weight vector $w$ in that partition. For partitions $\rho_3$ and $\rho_4$ on the right, vertex $v_1$ will also be deleted by the DFS procedure due to the deletion of $v_7$, after which $v_1$ and $v_7$ are both pushed into the heap $I'$ w.r.t. the partitions (representing the vertices ignored). Thus, the resulting subgraph $H$ induced by $\{v_2, \ldots, v_6\}$ is the corresponding non-contained MAC since discarding any vertex will no longer satisfy the structural cohesiveness constraint. By backtracking the top vertices in $I'$ once (i.e., $v_1$ and $v_7$), we can easily obtain the second-ranked MAC induced by $\{v_1, \ldots, v_7\}$ in $\rho_3$ and $\rho_4$ (refer to $R_3$ in Fig.~\ref{fig:MACs}).

\begin{figure}[t]
\centering
\includegraphics[width=0.49\textwidth]{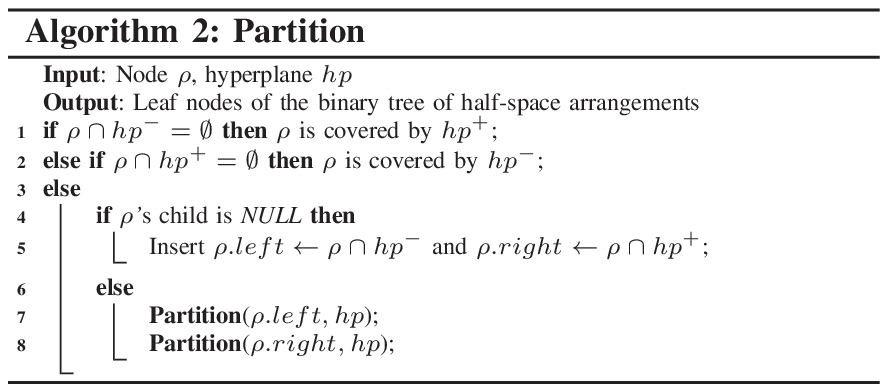}
\label{alg:2}
\vspace{-25pt}
\end{figure}

When Corollary~\ref{corollary1} does not hold, sub-partition $\rho'$ will be pushed into queue $U$ to compute the non-contained MAC in depth. At this time, vertices ignored will be discarded to update resulting subgraph $H$ and corresponding $G_d'$, so that new leaf vertex can be designated in the next round of verification and the half-spaces against other leaf vertices are inserted into the local arrangement. Consider $\rho_1$ in Fig.~\ref{fig:subregion1}, we refer to $v_4$ as the new leaf vertex after $v_1$ is deleted, i.e., $H$ and $G_d'$ are induced by $\{v_2, \ldots, v_7\}$ and $I'=\{v_1\}$. Then, new half-spaces $HS_4: v_7 \!\geq\! v_4$ and $HS_5: v_5 \!\geq\! v_4$ are inserted into a newly initialized local arrangement against partition $\rho_1$, where three sub-partitions $\rho_5$, $\rho_6$ and $\rho_7$ are produced, as shown in Fig.~\ref{fig:subregion2}. As $v_4$ and $v_5$ are pushed into $I'$ together, the non-contained MAC induced by $\{v_2, v_3, v_6, v_7\}$ is returned for each sub-partition. Likewise, same operation applies to partition $\rho_2$. Eventually, the solution in Example~\ref{example3} is obtained. Note that we can directly locate $HS_4$ and $HS_5$ for $\rho_2$ since no new half-space needs to be computed due to the same leaf vertices (in $G_d'$) as $\rho_1$. This drastically reduces repetitive computation in half-spaces, each of which is computed only once if necessary.

Specifically, for each local arrangement considered, an index is built by a recursive process \emph{Partition} (Algorithm~2). Then, the index is discarded when all relevant half-spaces are inserted, leaving only the hopeful sub-partitions of the local arrangement (if any). Note that, for any index of $\rho$ (Line~9 in Algorithm~1), the total cost of inserting the $i$-th hyperplane $hp$ is $O(i^{d-1})$ \cite{tang2017determining}. In addition, optimization of arrangement indexing and maintenance is the same as described in \cite{mouratidis2018exact}.

\section{Local Search}\label{section:localSearch}
Although the efficiency of the global search algorithm is considerable for each query, it may need to explore the entire maximal $(k, t)$-core, especially when query vertices $Q$ are located at the upper layer of the r-dominance graph $G_d$. In this section, we devise the local search algorithms for Problem~2 and investigate the generalization for Problem~1.

\begin{figure}[t]
\centering
\includegraphics[width=0.49\textwidth]{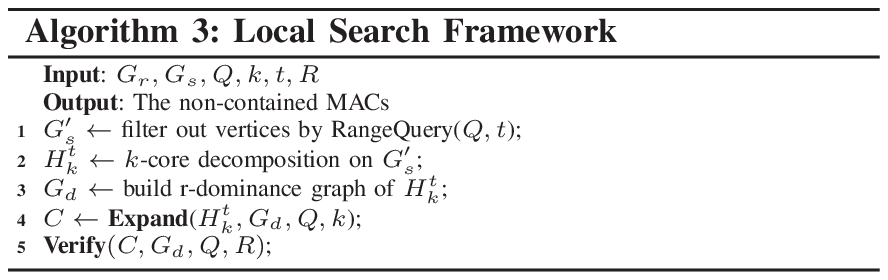}
\label{alg:3}
\vspace{-25pt}
\end{figure}

The intuition is that the non-contained MACs for $Q$ are in the vicinity of $Q$. Thus, the entire $H_k^t$ should be unnecessary to involve during the search. Nonetheless, it is intractable to enumerate all the $k$-$\widehat{core}$s containing $Q$, whose number can be exponentially large w.r.t. $H_k^t$ size. Accordingly, we only immerse in finding the communities that are most likely to be candidates for non-contained MACs. Their validity and corresponding partitions in $R$ can be quickly verified by $G_d$ alone. This inspires us to develop a framework of more efficient local search (Algorithm~3). Specifically, \emph{Expand} procedure finds candidates (i.e., $C$) by selecting the most promising vertex as we explore in the neighborhood of $Q$, and stops when each target community forms a $k$-$\widehat{core}$. \emph{Verify} procedure provides guarantee of identifying all valid non-contained MACs w.r.t. $R$ from $C$. Note that the time complexity of Algorithm~3 is bounded by $O(|C|\!\cdot\! s^d))$ (see Theorem~\ref{theorem3} and~\ref{theorem4}), which is much lower than that of Algorithm~1 ($|C|$ and $s$ are typically very small in practice). As verified in our experiments, local search is at least one order of magnitude faster than global search, and all non-contained MACs will be expanded by our candidate generation strategies in most cases. 



In this way, two problems arise: (1) how do we guarantee that the target community can be a candidate for non-contained MACs (at least form a $k$-$\widehat{core}$ containing $Q$); and (2) how do we know whether the $k$-$\widehat{core}$ is a valid non-contained MAC w.r.t. $R$? The former poses a great challenge of determining which vertex to choose and when to terminate expansion, and the latter requires an in-depth study of the characteristics of non-contained MACs. In the following, we present lemmas and algorithms for the local search strategy.

\subsection{Candidate Generation}\label{section:candidateGeneration}
To be a candidate for non-contained MACs, the current community must be at least a $k$-$\widehat{core}$ containing $Q$. It would be nice if the structural cohesiveness metric (i.e., $k$-core) is ``monotonic", which means that the larger the community, the smaller its minimum degree. So once the minimum degree drops below the given coreness threshold $k$, we can stop the search. Unfortunately, such a metric is not monotonic. Thus, the greatest challenge is to overcome non-monotonicity first, which motivates us to conduct community search only by exploring local neighborhood of $Q$. 

Now for the minimum degree of a subgraph $H$, denoted by $\delta(H)$, we make an in-depth analysis of its monotonicity. Consider the exploration starting from the query vertices, i.e., $H_0$ induced by $Q$. We add a vertex from $H_k^t$ at each step until a $k$-$\widehat{core}$ $H$ is obtained, assuming that $v_1, v_2, \ldots, v_e$ is a vertex sequence it adds. So let $H_i$ be induced by $Q \cup \{v_1, \ldots, v_i\}$. In general, $\delta(H)$ is a non-monotonic function of $H$. More formally, $\delta(H_{i+1})$ is unnecessarily greater than $\delta(H_i)$. However, the order of vertices added to $V_H$ determines the monotonicity of $\delta(H)$. Interestingly, for any $Q$ with $\delta(H_0) = 0$, we can always find a sequence of added vertices such that $\delta(H_i)$ is a non-decreasing function of $i$.

\vskip 3pt
\begin{lemma} \label{lemma7}
For any query vertices $Q \subseteq V_H$ with $\delta(H_0) = 0$ in graph $H_k^t$, there always exists an added vertex sequence $v_1, v_2, \ldots, v_e$ of $H$ starting with $Q$ such that $\forall 0 \leq i \leq e, \delta(H_i) \leq \delta(H_{i+1})$.
\end{lemma}
\vskip 3pt

\begin{proof sketch}
It is consistent with proving that vertices can be removed one by one from $V_H$ until $Q$, just ensuring that each removal of $v$ does not increase the minimal degree of the remaining vertices. Otherwise, it occurs only when $v$ is currently one of the vertices with the minimal degree. The reason is that removing a vertex with non-minimal degree will only preserve or decrease the minimal degree. 
\end{proof sketch}
\vskip 3pt


\begin{figure}[t]
\centering
\includegraphics[width=0.49\textwidth]{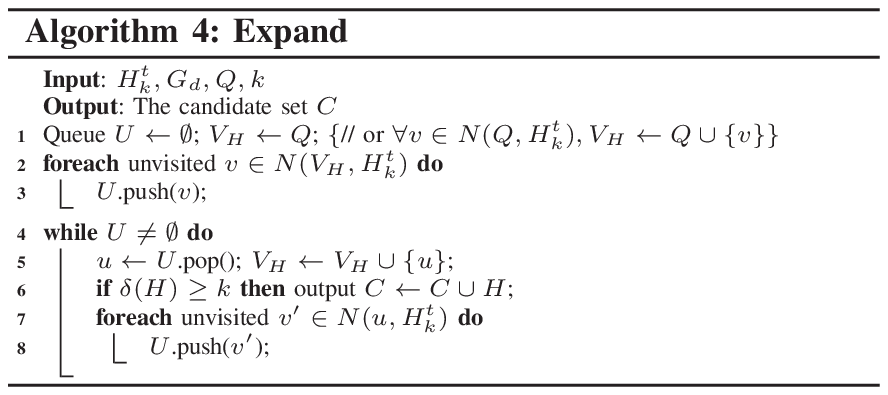}
\label{alg:4}
\vspace{-25pt}
\end{figure}

Lemma~\ref{lemma7} implies that there always exists an exploration order that leads monotonically to a community of $k$-$\widehat{core}$ containing $Q$. This can be generated by a sequence of vertices added to $V_H$ starting from $Q$ so that $\delta(H_i) \leq \delta(H_{i+1})$ for each $i$. Note that the existence of such an order is a necessary but insufficient condition for finding a valid community. To illustrate the insufficiency, consider $Q = \{v_9\}$ and $k = 2$ in Fig.~\ref{fig:social_network}. Any vertex sequence starting with $v_9, v_{14}$ cannot yield a valid solution, yet $\delta(H_1)$ is greater than $\delta(H_0)$.

In \emph{Expand} procedure (Algorithm~4), we explore from the vicinity of $Q$ by BFS and generate candidate set $C$. To converge the current community towards a candidate for non-contained MAC, we develop two intelligent candidate selection strategies according to Lemma~\ref{lemma3},~\ref{lemma6} and~\ref{lemma7}. The idea of improving candidate generation is to use priority queues such that the most promising vertex can be selected to rapidly generate a candidate. From the perspective of structural cohesiveness, the priority of a vertex $v$ can be defined as $f_1(v)=\delta(H')-\delta(H)$ or $f_2(v)=dg_{H'}(v)$, where $V_H'=V_H \cup \{v\}$. $f_1(v)$ emphasizes the improvement of minimum degree for the next step, with $f_1(v)=1$ or $0$ for any $v$; $f_2(v)$ produces the fastest increase in average degree of $H$ so that the minimum degree will increase with $H$'s density growth. From the perspective of community score, the priority of $v$ can be defined as $f_3(v)=\zeta-l(v)$, where $\zeta$ is a constant (maximum priority in $G_d$) and $l(v)$ denotes the layer of $v$ in $G_d$. $f_3(v)$ drives community score higher by adding a vertex that r-dominates as many vertices as possible. To sum up, the priority $f(v)$ is defined as 
\begin{equation}\label{eq:li}
\setlength\abovedisplayskip{4pt}
\setlength\belowdisplayskip{4pt}
f(v)= \lambda \!\cdot\! f_2(v) + f_3(v),
\end{equation}
where $\lambda$ is a trade-off against $\zeta$, or
\begin{equation}\label{eq:lc}
\setlength\abovedisplayskip{4pt}
\setlength\belowdisplayskip{4pt}
f(v) = \zeta \cdot f_1(v) + f_3(v).
\end{equation}

\vskip 3pt
\begin{theorem} \label{theorem3}
The time complexity of Algorithm~4 by Eq.~\ref{eq:li} and Eq.~\ref{eq:lc} is $O(\overline{n}+\overline{m})$ and $O(\overline{n}+\overline{m}log\overline{n})$ respectively, where $\overline{n}$ and $\overline{m}$ denote the number of vertices and edges in $C$. The space complexity is $O(\overline{n}+\overline{m})$.
\end{theorem}

\subsection{Verification}
The determinant of global search is that computing the local arrangement of all half-spaces $HS_i$ among leaf vertices is a relatively expensive process (in $O({i^*}^{d})$ time \cite{agarwal2000arrangements}, where $i^*$ is the number of half-spaces). Instead, in \emph{Verify} procedure (Algorithm~5), an empty arrangement in $R$ is initialized, into which half-spaces w.r.t. a carefully selected and therefore very small subset of vertices (i.e., \emph{competitors} below) are inserted, expecting to securely confirm or disqualify candidate $H$ without considering all other vertices. But before this, we first give a corollary to filter out unpromising candidates from $C$. Note that $G_e$ represents the r-dominance graph corresponding to $H$, which is a subgraph of $G_d$ induced by $V_H$, denoted by $G_d[V_H]$; and $G_c$ represents the rest of $G_d$, denoted by $G_d[V_{G_d} \!\backslash\! V_H]$.


\vskip 3pt
\begin{corollary} \label{corollary2}
A community $H$ can be discarded if one of the following conditions is met: \emph{(1)} $\forall v \!\in\! V_{G_d} \!\backslash\! V_H$, $v$ is a non-leaf vertex in $G_d$; and \emph{(2)} $\exists v \!\in\! V_{G_d} \!\backslash\! V_H, v' \!\in\! V_H$ and $v \!\succ\! v'$, $v$ cannot be recursively deleted by deleting vertices in $V_{G_d} \!\backslash\! V_H$.  
\end{corollary}
\vskip 3pt

\begin{proof sketch}
Vertices in $V_{G_d} \!\backslash\! V_H$ must be deleted if $H$ holds.
\end{proof sketch}
\vskip 3pt

In other words, either there exists a non-cross-layer arc in $G_c$ between a non-leaf vertex $v$ and a leaf vertex, or $v$ can be recursively deleted by the DFS procedure. We refer to the $H$ that holds Corollary~\ref{corollary2} as a \emph{promising community}. 

\vskip 3pt
\begin{lemma} \label{lemma8}
For any promising $H$, $v$ is regarded as an anchor if $H$ still forms a $k$-$\widehat{core}$ after a (non-$Q$) leaf vertex $v$ in $G_e$ is deleted.
\end{lemma}
\vskip 3pt

Now we discuss the verification process by half-space insertion, in which it may further benefit from the r-dominance relationships stored in $G_d$ as follows. 



\vskip 3pt
\begin{lemma} \label{lemma9}
Consider $u$, $u'$, and their half-space $HS_i$ inserted into the arrangement. Assume that $u''$ is a vertex r-dominated by $u'$, and $\rho$ is a partition in the arrangement not covered by $HS_i$. Thus $u$ is guaranteed to r-dominate $u''$ in partition $\rho$.
\end{lemma}
\vskip 3pt

\begin{proof}
First, from the definition of half-space $HS_i$, $S(u') \!<\! S(u)$ holds anywhere outside $HS_i$. Second, $S(u'') \!\leq\! S(u')$ holds anywhere inside $R$ by the definition of r-dominance. Thus, we derive that for each partition $\rho$ of the arrangement outside $HS_i$, $S(u'') \!<\! S(u)$, that is, $u$ r-dominates $u''$ in $\rho$.
\end{proof}
\vskip 3pt

\begin{figure}[t]
\centering
\includegraphics[width=0.49\textwidth]{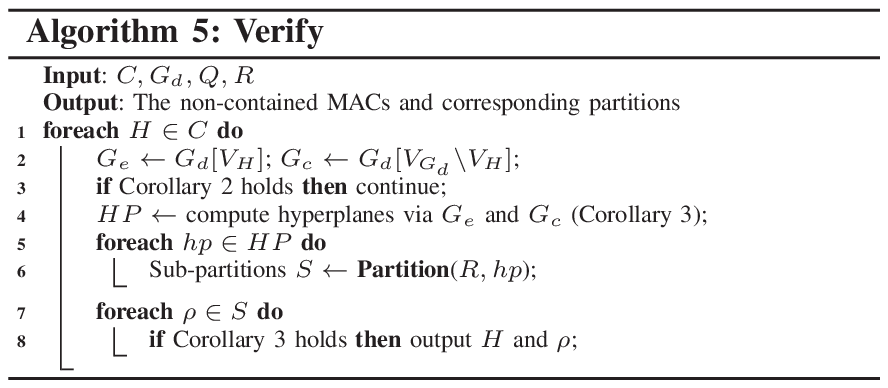}
\label{alg:5}
\vspace{-25pt}
\end{figure}

Based on Lemma~\ref{lemma9}, competitors chosen are the vertices in the bottom layer (leaf vertices) of $G_e$ and in the top layer (with r-dominance count 0) of $G_c$, denoted by $l_b(G_e)$ and $l_t(G_c)$ respectively. The fundamental is that intuitively such competitors are the strongest, which are most likely to assist in disqualifying invalid candidates w.r.t. $R$ in turn. 

\vskip 3pt
\begin{corollary} \label{corollary3}
For any promising $H$, it is a valid non-contained MAC if a partition exists in $R$ such that all vertices in $l_b(G_e)$ score higher than those in $l_t(G_c)$, and the corresponding conditions are met: 
\begin{itemize}
  \item[\emph{1.}] If Lemma~\ref{lemma8} holds, additionally, all anchors need to score higher than other leaf vertices in $G_e$.
  \item[\emph{2.}] $\exists v \!\in\! l_t(G_c)$, if $v$ can be recursively deleted by the DFS procedure starting from $l_b(G_c)$ (namely, $v$ is bound), then $l_t(G_c)$ is updated where $v$ is ignored and replaced by the vertices of its next layer in $G_c$. 
  \item[\emph{3.}] $\exists v, v' \!\in\! l_t(G_c)$, if $v$ and $v'$ are bound to each other, then vertices in $l_b(G_e)$ only need to score higher than $v$ or $v'$.
\end{itemize}
\end{corollary}
\vskip 3pt

\begin{proof}
According to Corollary~\ref{corollary2}, all vertices in $V_{G_d} \!\backslash\! V_H$ have to be deleted when $H$ is a promising community. In other words, vertices in $V_{G_d} \!\backslash\! V_H$ are either recursively deleted by deleting those in the lower layer of $G_c$, or deleted individually due to their lower scores. First, we consider the latter case. As a sufficient and necessary condition, these vertices just need to score lower than those in $l_b(G_e)$; that is, only vertices in $l_b(G_e)$ score higher than those in $l_t(G_c)$. Then, we consider the former case. That is, if condition (2) holds, it means that the restriction on half-spaces between the competitors can be relaxed by the newly updated vertices in $l_t(G_c)$, since such a vertex $v$ will be deleted anyway. Similarly, if condition (3) holds, the restriction on half-spaces between the competitors can also be relaxed through such vertices, because deletion of one will also lead to deletion of the other. On this basis, $H$ becomes a non-contained MAC when Lemma~\ref{lemma8} does not hold; otherwise, condition (1) has to be satisfied. This is because such an anchor can still be deleted as it is a non-$Q$ leaf vertex in $G_e$, i.e., possibly the current smallest-score vertex in $H$. Putting it all together, we ensure the correctness of the partition corresponding to the non-contained MAC $H$ (if any in $R$).
\end{proof}
\vskip 3pt

To illustrate Algorithm~5, let us reconsider Example~\ref{example2}. Assume that by Algorithm~4 we have three promising communities $H_1$, $H_3$ and $H_4$, where $V_{H_1} \!=\! \{v_2, v_3, v_6, v_7\}$, $V_{H_3} \!=\! \{v_2, \ldots, v_6\}$ and $V_{H_4} \!=\! \{v_1, v_2, v_3, v_6, v_7\}$. For $H_1$, $l_b(G_e) \!=\! \{v_7\}$ and $l_t(G_c) \!=\! \{v_4, v_5\}$. As $v_4$ and $v_5$ are bound to each other in $H_3^9$ (condition (3) met), we only insert $HS_1$ and $HS_4$ into $R$ in Fig.~\ref{fig:subregion2}, and choose the partitions covered by either of them. As a result, $H_1$ is a valid non-contained MAC for any weight vector of $R_1$ in Fig.~\ref{fig:MACs}. Similarly, $H_3$ is also valid w.r.t. $R_2 \cup R_3$ (condition (2) met) but $H_4$ is invalid (condition (1) met) as its partition is outside $R$.

\vskip 3pt
\begin{theorem} \label{theorem4}
The time complexity of Algorithm~5 is $O(|C|\!\cdot\!(n'\!+\!m') + c\!\cdot\! s^d)$, where $|C|$ and $c$ denote the number of candidates and the number of promising communities in $C$ respectively, and $s$ is the product of $|l_b(G_e)|$ and $|l_t(G_c)|$. The space complexity is bounded by $O(c\!\cdot\! s \!\cdot\! d + \overline{n}+\overline{m} + n'+m')$.
\end{theorem}
\vskip 3pt

Finally, we can simply generalize local search for Problem~1. As the non-contained MAC $H$ with corresponding partition $\rho$ is known, we insert sub-partitions into $\rho$ according to $G_c$ (in a up-bottom manner) and add the highest-score vertex to $H$. As long as the current $H$ contains an MAC, it will be output. The process terminates until all top-$j$ MACs in $\rho$ are acquired. Consider $H_1$ and its $G_c$, half-spaces among vertices in $l_t(G_c)$ (i.e., $HS_5$) are inserted first into $R_1$. Then $v_5$ is added to $V_{H_1}$ for $\rho_5$ and $\rho_8$ shown in Fig.~\ref{fig:subregion2}. As $v_4$ r-dominates $v_1$, we have the second MAC induced by $\{v_2, \ldots, v_7\}$ in $\rho_5$ and $\rho_8$. The same applies to $\rho_6$ and $\rho_7$.

\section{Experiments}\label{section:experiments}

\begin{table}[t]
\caption{Datasets (K=$10^3$ and M=$10^6$)}
\label{tab:table2}
\centering
\vspace{-6pt}
\begin{tabular}{|c|c|c|c|c|c|}
  \hline
  \textbf{Dataset} & \textbf{Vertices} & \textbf{Edges} & $dg_{avg}$ & $dg_{max}$ & $k_{max}$ \\ \hline
  San Francisco (SF) & 175K & 223K & 2.55 & 8 & - \\ \hline
  Florida (FL) & 1.1M & 1.4M & 2.53 & 12 & - \\ \hline
  \hline
  Slashdot & 79K & 0.5M & 13 & 2,507 & 85 \\ \hline
  Delicious & 536K & 1.4M & 5 & 3,216 & 34 \\ \hline
  Lastfm & 1.2M & 4.5M & 7 & 5,150 & 71 \\ \hline
  Flixster & 2.5M & 7.9M & 6 & 1,474 & 69 \\ \hline
  Yelp & 3.6M & 9.0M & 5 & 10,433 & 129 \\ \hline
\end{tabular}
\vspace{-15pt}
\end{table}

Comprehensive experiments are conducted to evaluate the proposed model and four algorithms, named \texttt{GS-T}, \texttt{GS-NC}, \texttt{LS-T} and \texttt{LS-NC} respectively. \texttt{GS-T} and \texttt{GS-NC} (resp. \texttt{LS-T} and \texttt{LS-NC}) are global search algorithms (resp. local search algorithms) used to compute the top-$j$ MACs and the non-contained MACs, respectively. 
Note that either of the two candidate selection strategies in Section~\ref{section:candidateGeneration} can be adopted in \texttt{LS-T} or \texttt{LS-NC}, and we just give the results by using Eq.~\ref{eq:li} with $\zeta \!=\! 100$ and $\lambda \!=\! 10$ (results by using Eq.~\ref{eq:lc} are similar and omitted to save space). All algorithms were implemented in C++, and all experiments were conducted on an Ubuntu server with 2GHz Intel Xeon E7-4820 CPU and 1TB memory.


\noindent
\textbf{Datasets.} We use five real-world social networks\footnote{http://networkrepository.com}\textsuperscript{,}\footnote{https://www.yelp.com/dataset} and two road networks (SF\footnote{https://www.cs.utah.edu/\textasciitilde lifeifei/SpatialDataset.htm}/FL\footnote{http://www.dis.uniroma1.it/challenge9/index.shtml}) in our experiments. Table~\ref{tab:table2} summarizes the statistics of datasets, of which $dg_{avg}$, $dg_{max}$ and $k_{max}$ denote the average degree, the maximal degree and the maximal core number, respectively. Note that numerical attributes are not contained in the first four original social networks\textsuperscript{2}, for which we employ a widely used method in \cite{borzsony2001skyline} to generate three different types of numerical attributes, i.e., \emph{independence}, \emph{correlation} and \emph{anti-correlation}. 
Due to space limit, we report the results obtained from datasets with independent and real attributes. In addition, we map each user $v$ to a spatial point $p$ in the road network that matches the scale of his/her social network as follows: we project SF/FL into range $[0,1]$ in each dimension and generate normalized $L(v)$ by drawing from a list of recent check-ins. Assume that $p$ is the current location of $v$ if it has the smallest Euclidean distance to $L(v)$ in the projection space.

\begin{table}[t]
\caption{Parameters}
\label{tab:table3}
\centering
\vspace{-6pt}
\begin{tabular}{|c|c|}
  \hline
  \textbf{Parameter} & \textbf{Tested values} \\ \hline
  $k$ & 4, 8, \textbf{16}, 32, 64 \\ \hline
  $t$ (SF/FL) & 600/800, \textbf{800/1000}, 1000/1200, 1200/1400, 1400/1600 \\ \hline
  $d$ & 2, \textbf{3}, 4, 5, 6 \\ \hline
  $|Q|$ & 1, \textbf{4}, 8, 16, 32 \\ \hline
  $j$ & 5, 10, \textbf{20}, 40, 60 \\ \hline
  $\sigma$ & 0.1\%, 0.5\%, \textbf{1\%}, 5\%, 10\% \\ \hline
\end{tabular}
\vspace{-15pt}
\end{table}

\noindent
\textbf{Parameters.} We vary $6$ parameters: structural cohesiveness $k$, query distance $t$, dimensionality $d$, number of query users $|Q|$, number $j$ of top MACs, and percentage $\sigma$ of axis length (i.e., side-length of $R$). Table~\ref{tab:table3} shows the range of parameters and their default values (in bold). For each value of $|Q|$, we randomly select $100$ sets of query vertices, that satisfy $t$ and can ensure the existence of the maximal $(k, t)$-core, from the $k$-core of each social network. In each experiment, only one parameter varies and the rest remains at the default. 
Every reported measurement is the average of $1,000$ MAC searches for ten randomly generated axis-parallel hypercubes $R$ in the preference domain. 

\subsection{Performance Evaluation}
\begin{figure*}[t]
\centering
\subfigcapskip=-4pt
\subfigure[Varying $k$] {\label{fig:slashdot-k}
\includegraphics[width=0.146\textwidth]{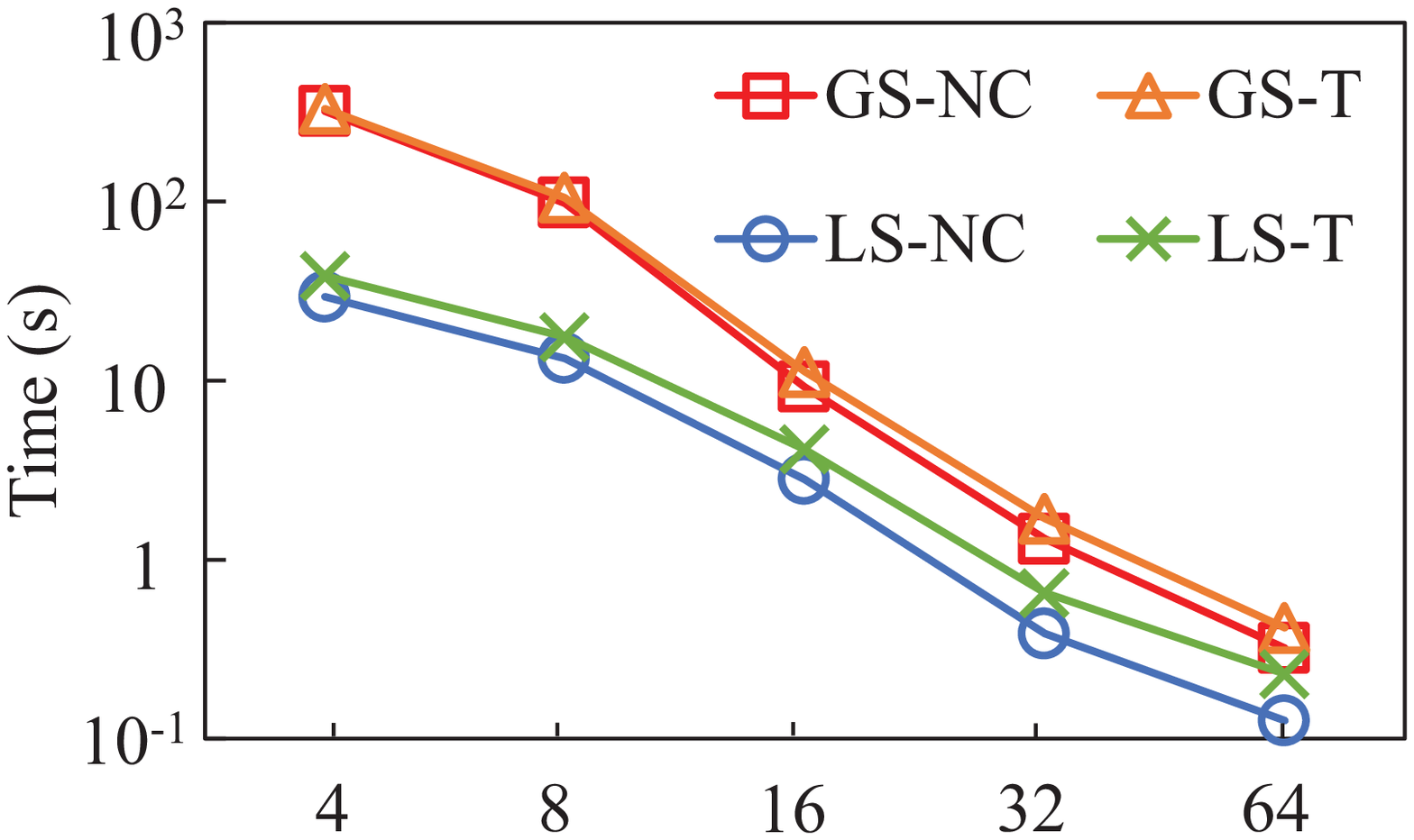}
}
\subfigure[Varying $t$] {\label{fig:slashdot-t}
\includegraphics[width=0.146\textwidth]{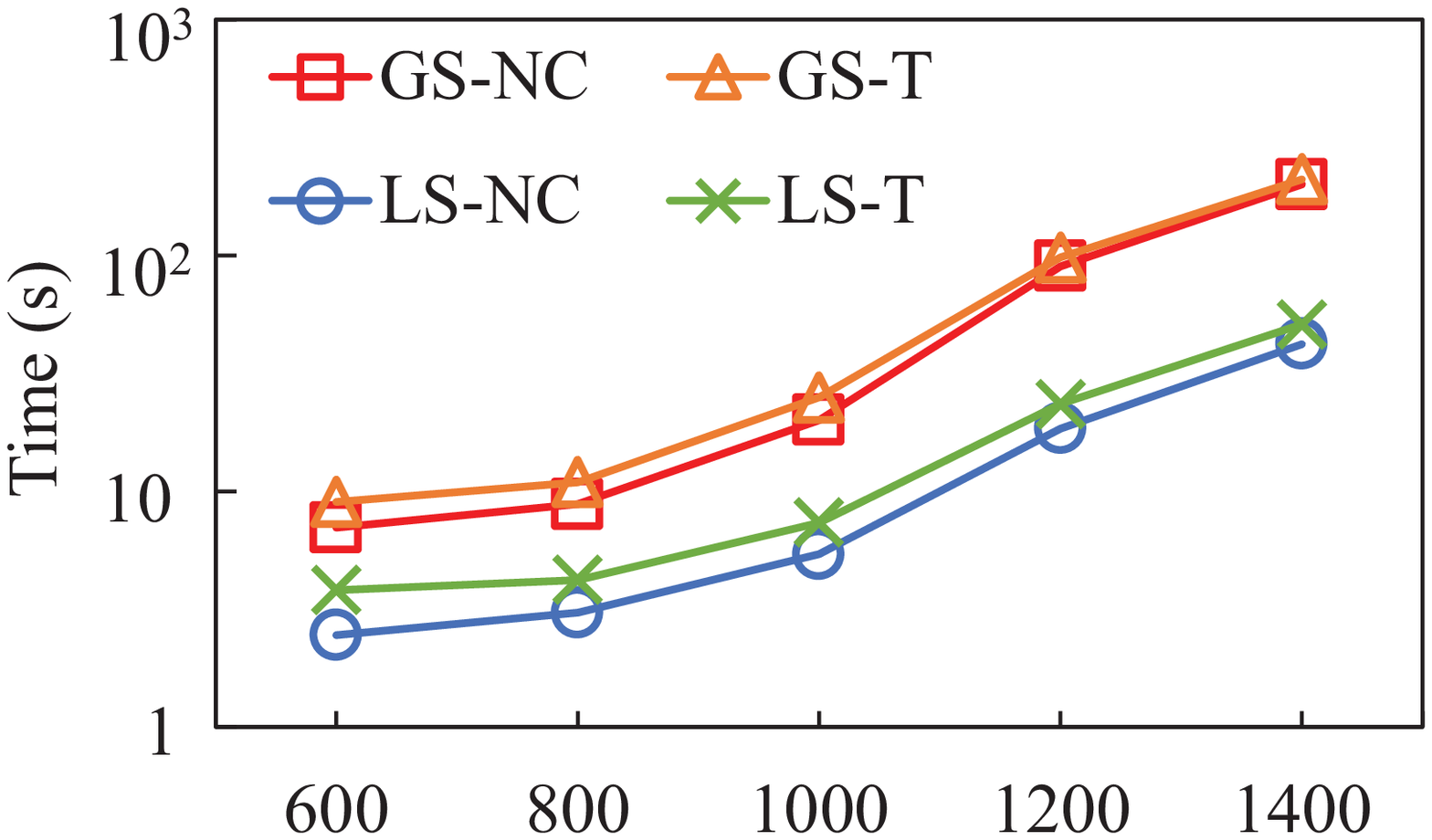}
}
\subfigure[Varying $d$] {\label{fig:slashdot-d}
\includegraphics[width=0.146\textwidth]{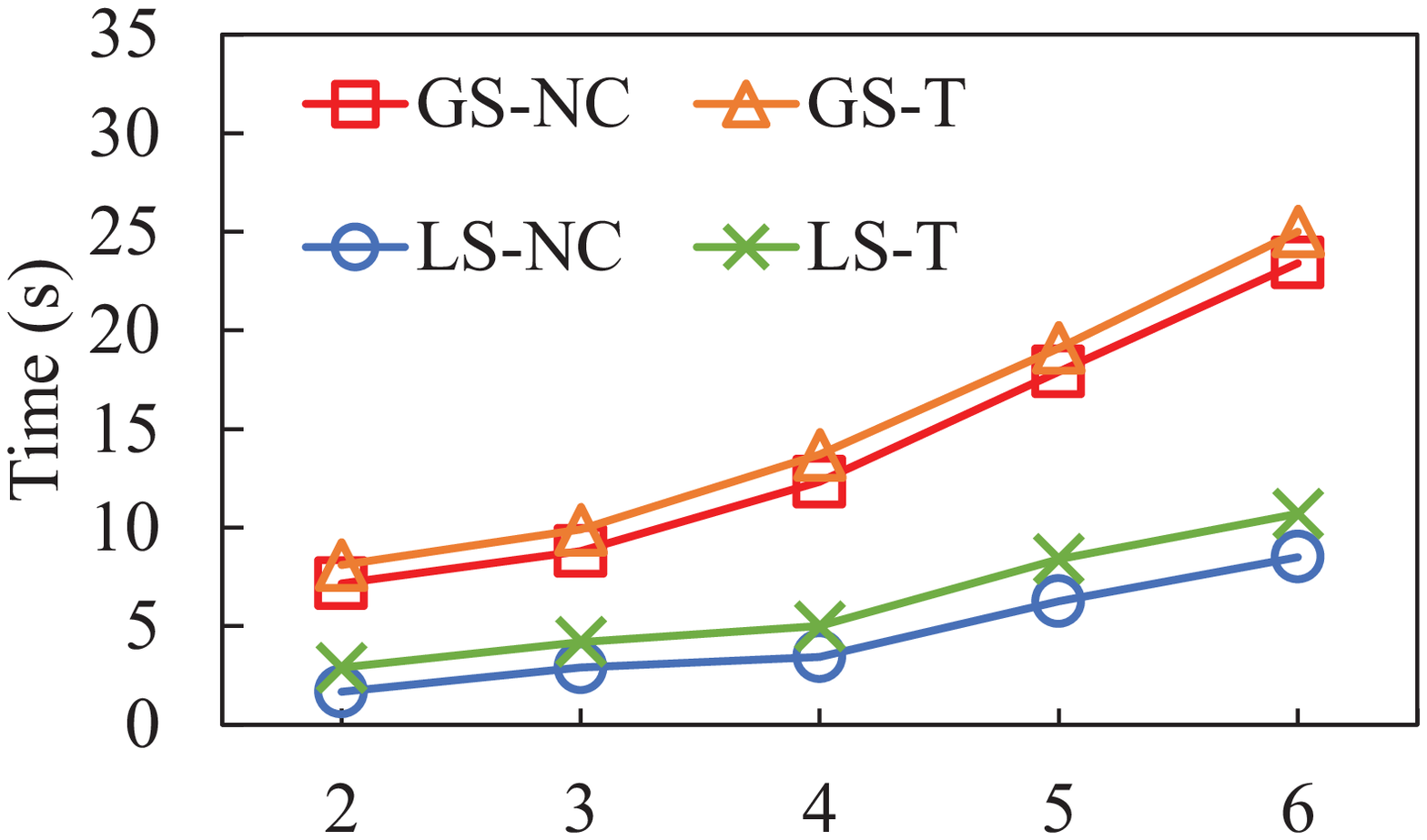}
}
\subfigure[Varying $|Q|$] {\label{fig:slashdot-q}
\includegraphics[width=0.146\textwidth]{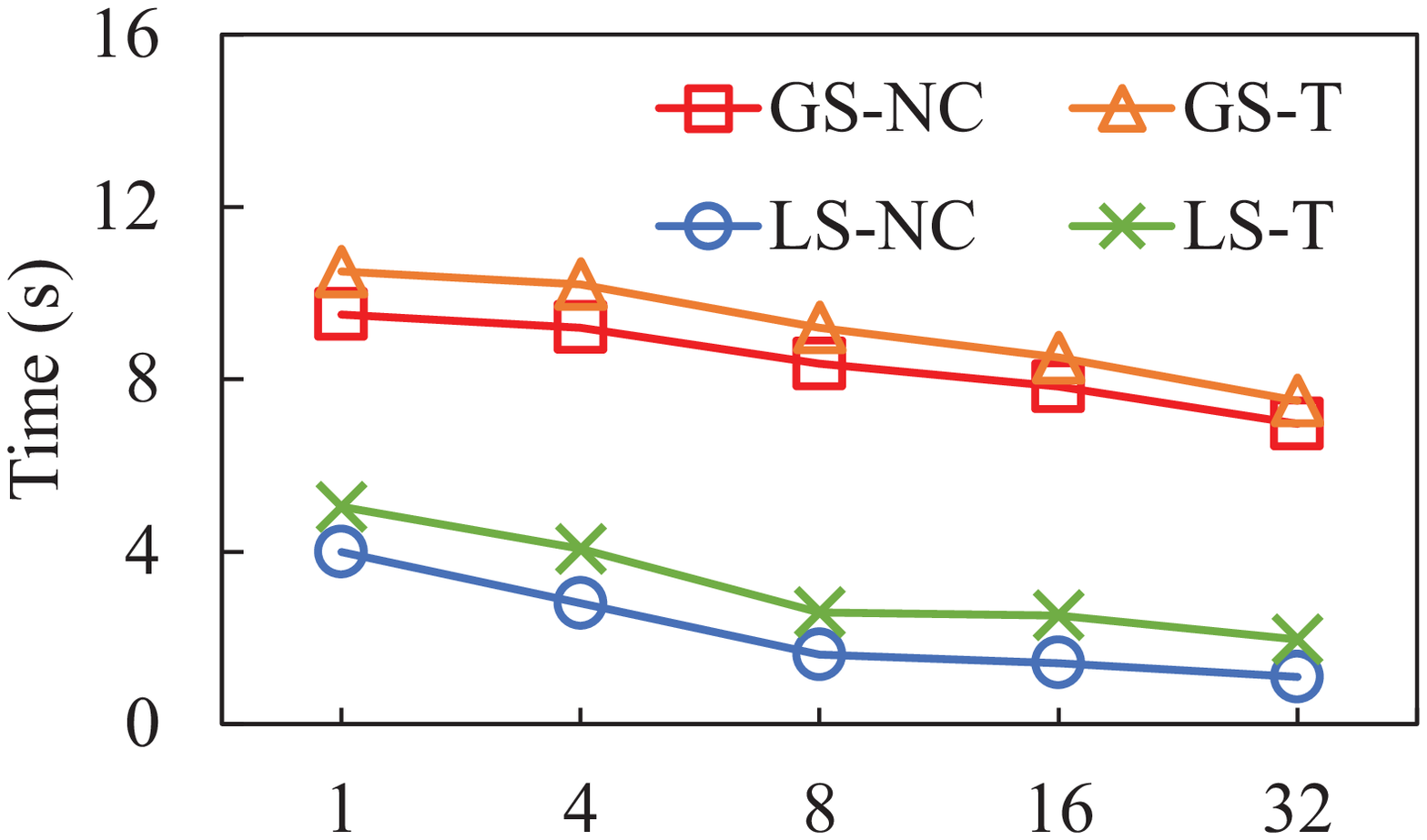}
}
\subfigure[Varying $j$] {\label{fig:slashdot-j}
\includegraphics[width=0.146\textwidth]{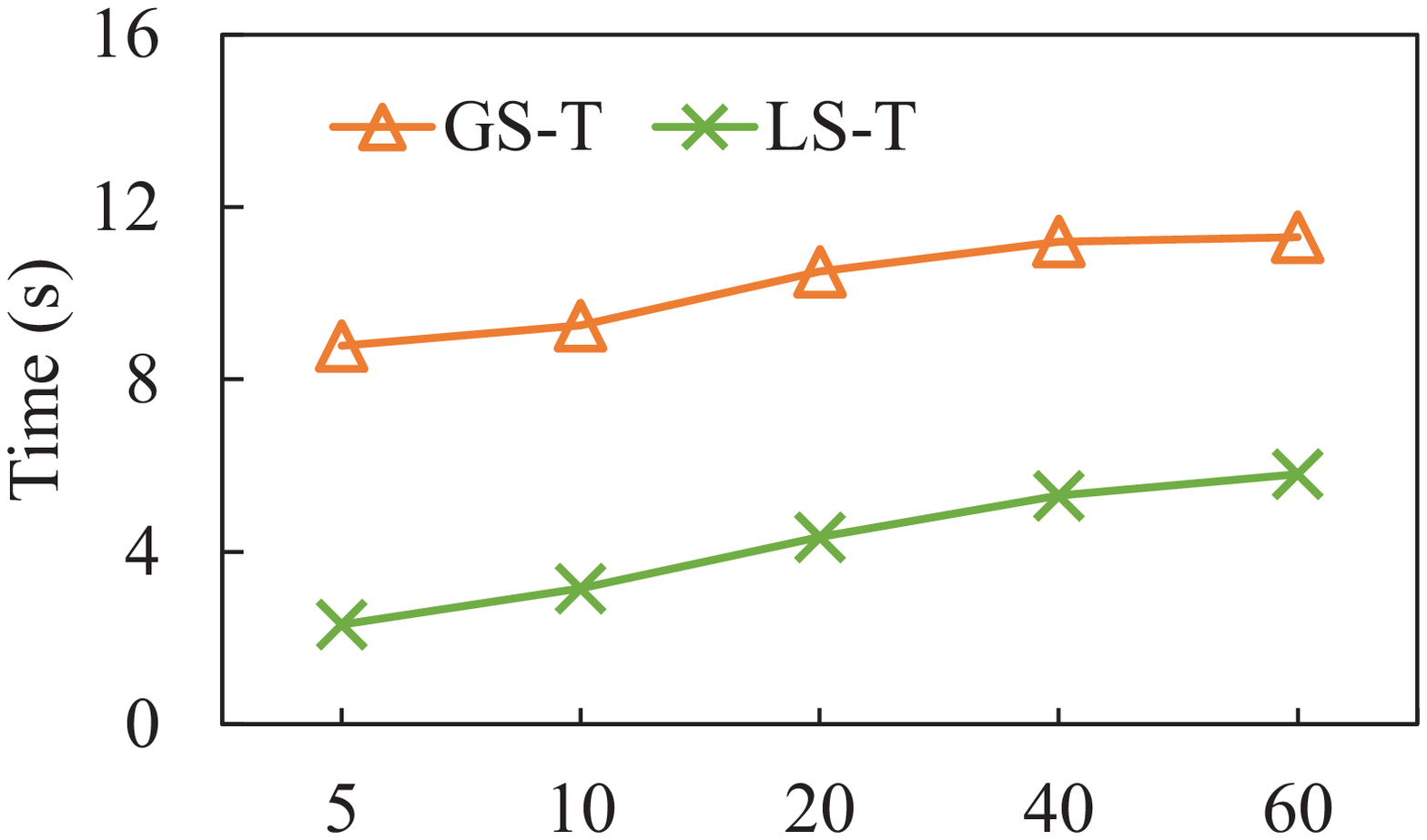}
}
\subfigure[Varying $\sigma$] {\label{fig:slashdot-sigma}
\includegraphics[width=0.146\textwidth]{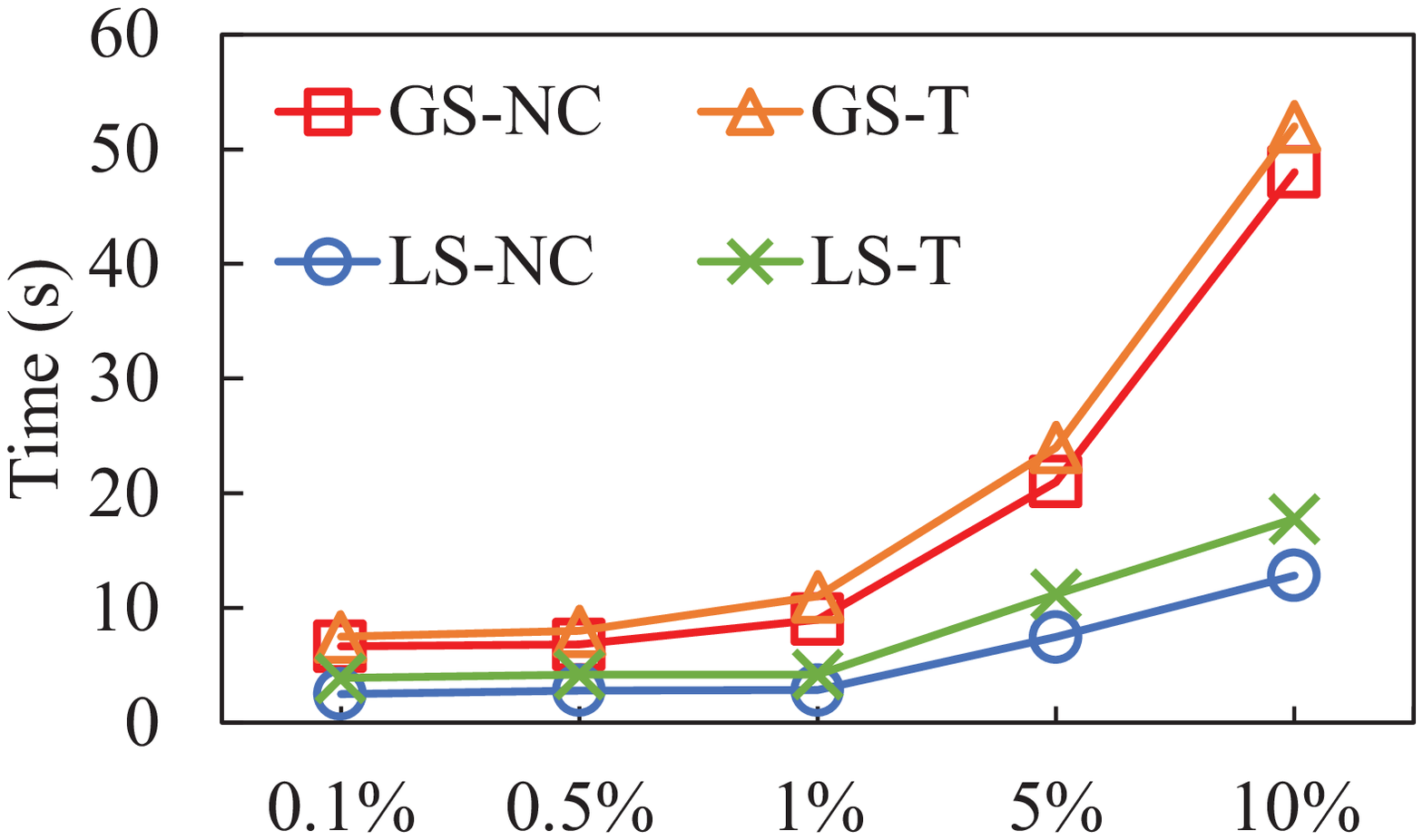}
}
\vspace{-7pt}
\caption{Efficiency and scalability of proposed algorithms in SF+Slashdot with independent attributes.}
\label{fig:slashdot}
\vspace{-8pt}
\end{figure*}

\begin{figure*}[t]
\centering
\subfigcapskip=-4pt
\subfigure[Varying $k$] {\label{fig:delicious-k}
\includegraphics[width=0.146\textwidth]{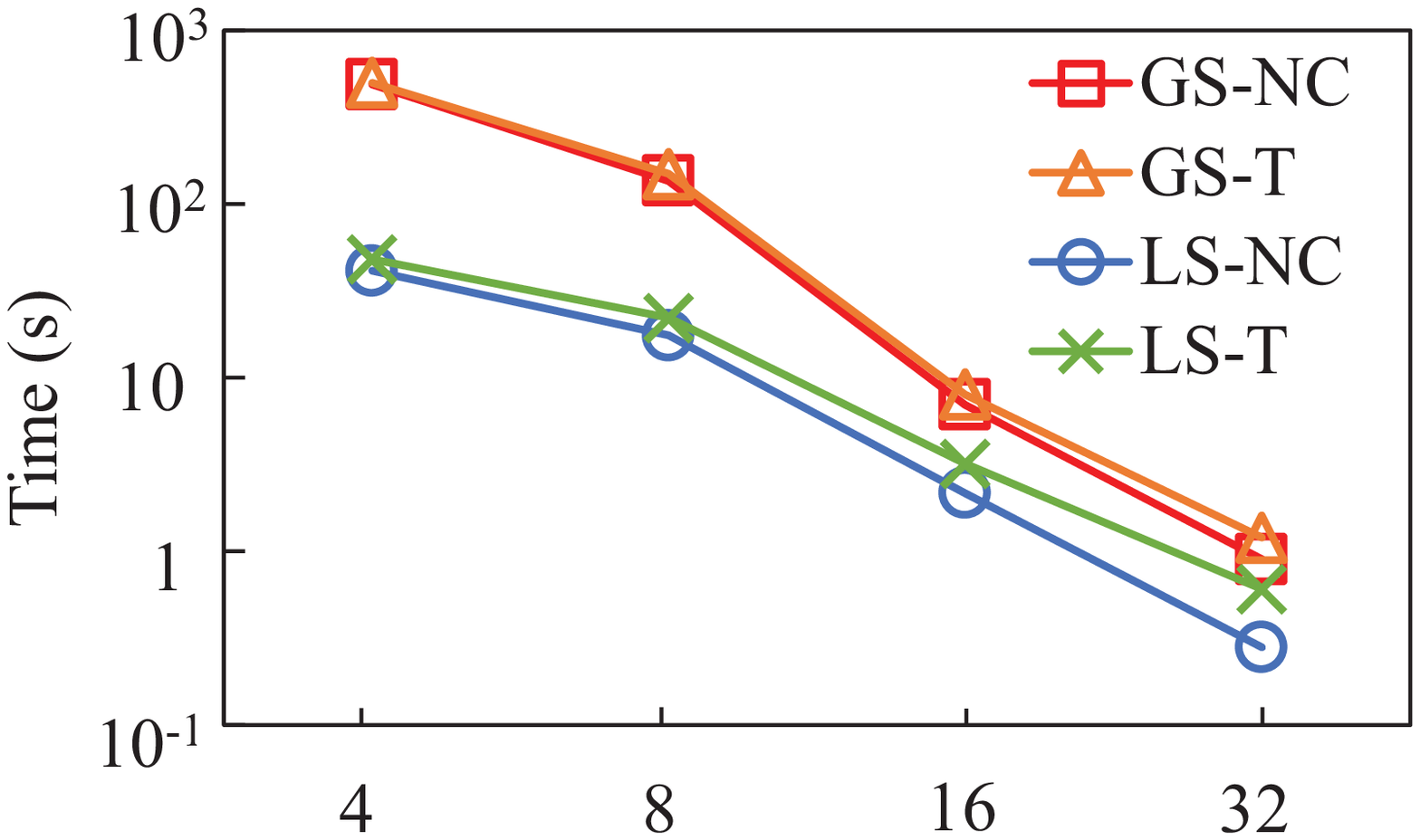}
}
\subfigure[Varying $t$] {\label{fig:delicious-t}
\includegraphics[width=0.146\textwidth]{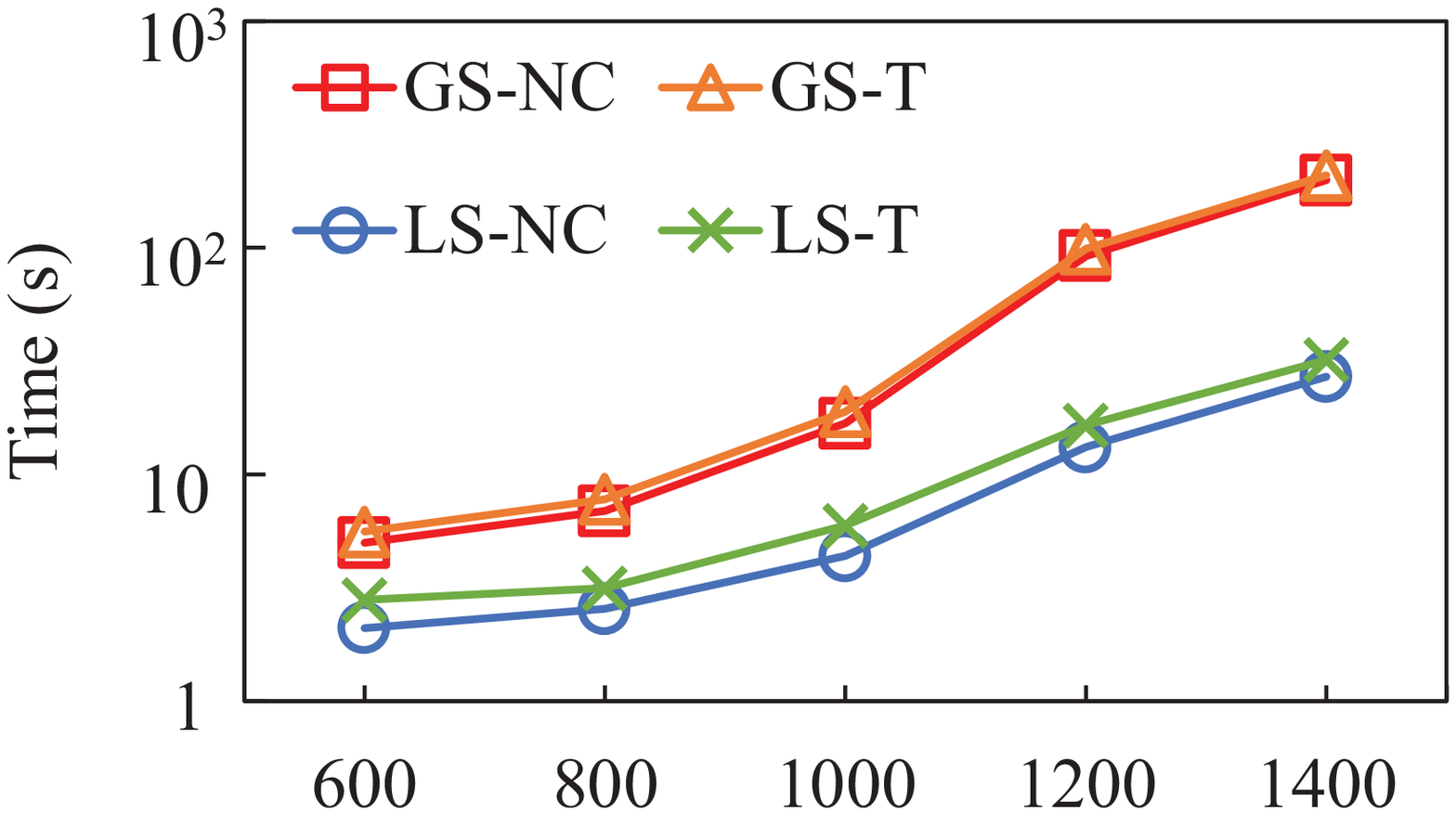}
}
\subfigure[Varying $d$] {\label{fig:delicious-d}
\includegraphics[width=0.146\textwidth]{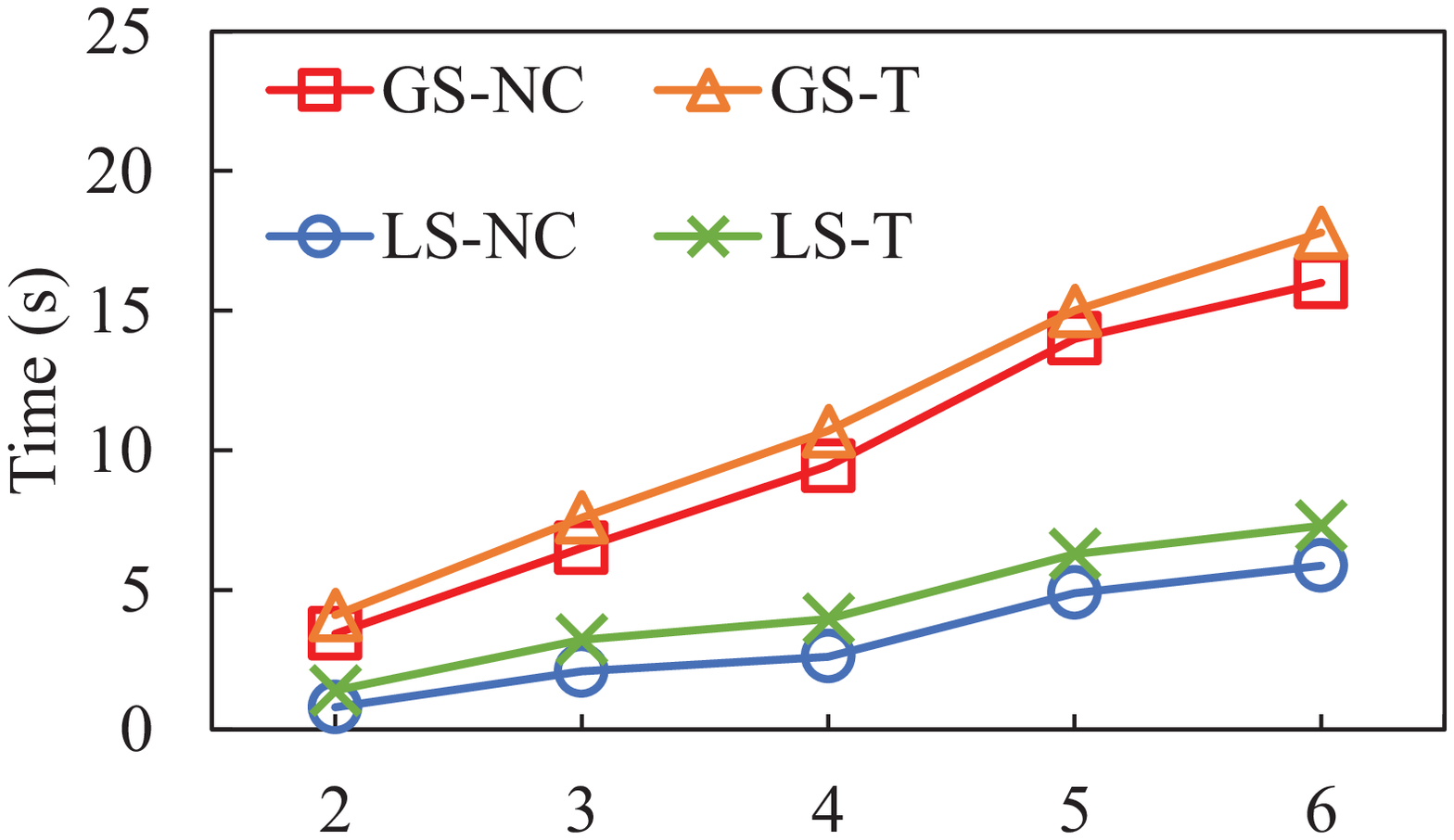}
}
\subfigure[Varying $|Q|$] {\label{fig:delicious-q}
\includegraphics[width=0.146\textwidth]{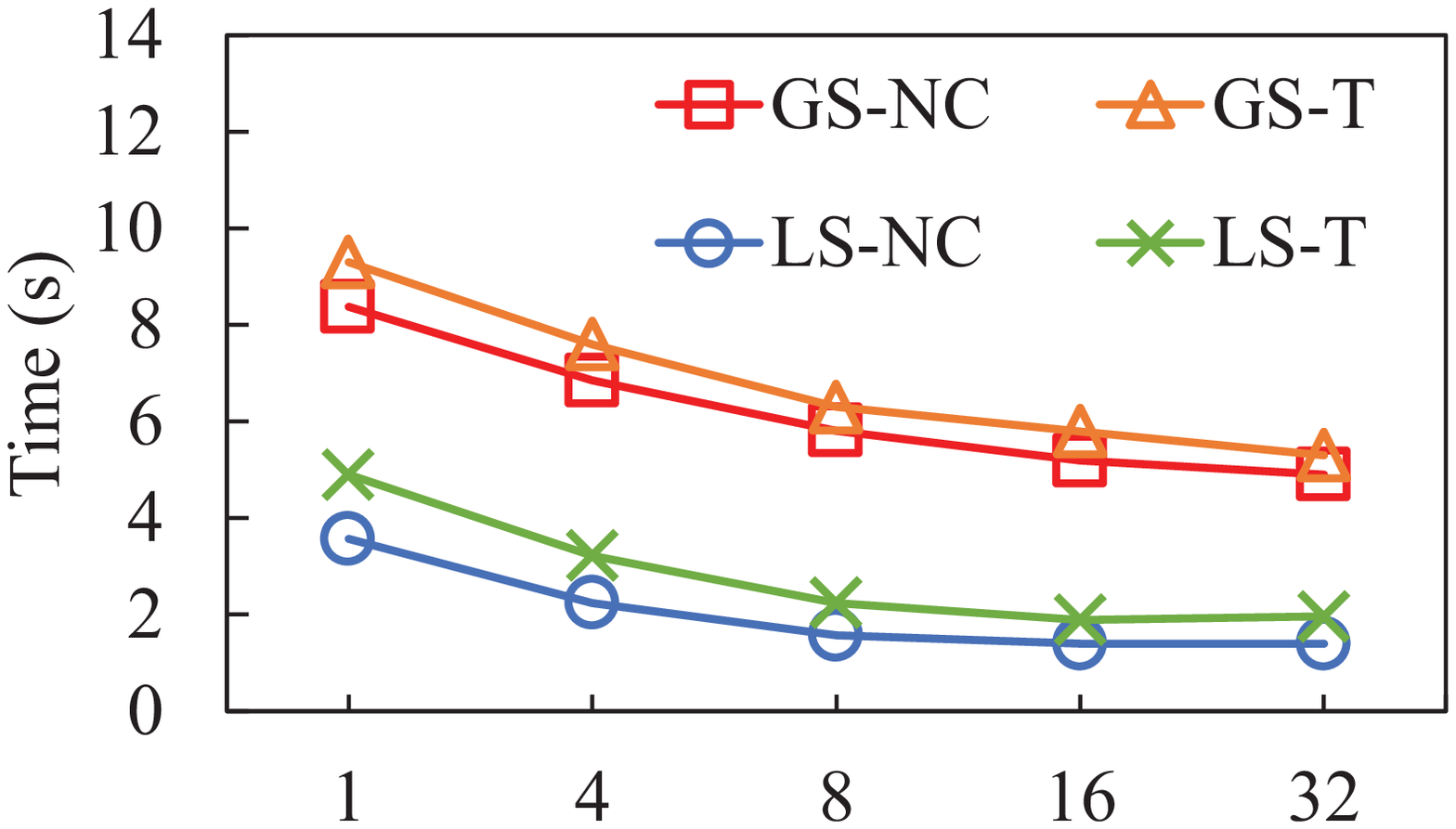}
}
\subfigure[Varying $j$] {\label{fig:delicious-j}
\includegraphics[width=0.146\textwidth]{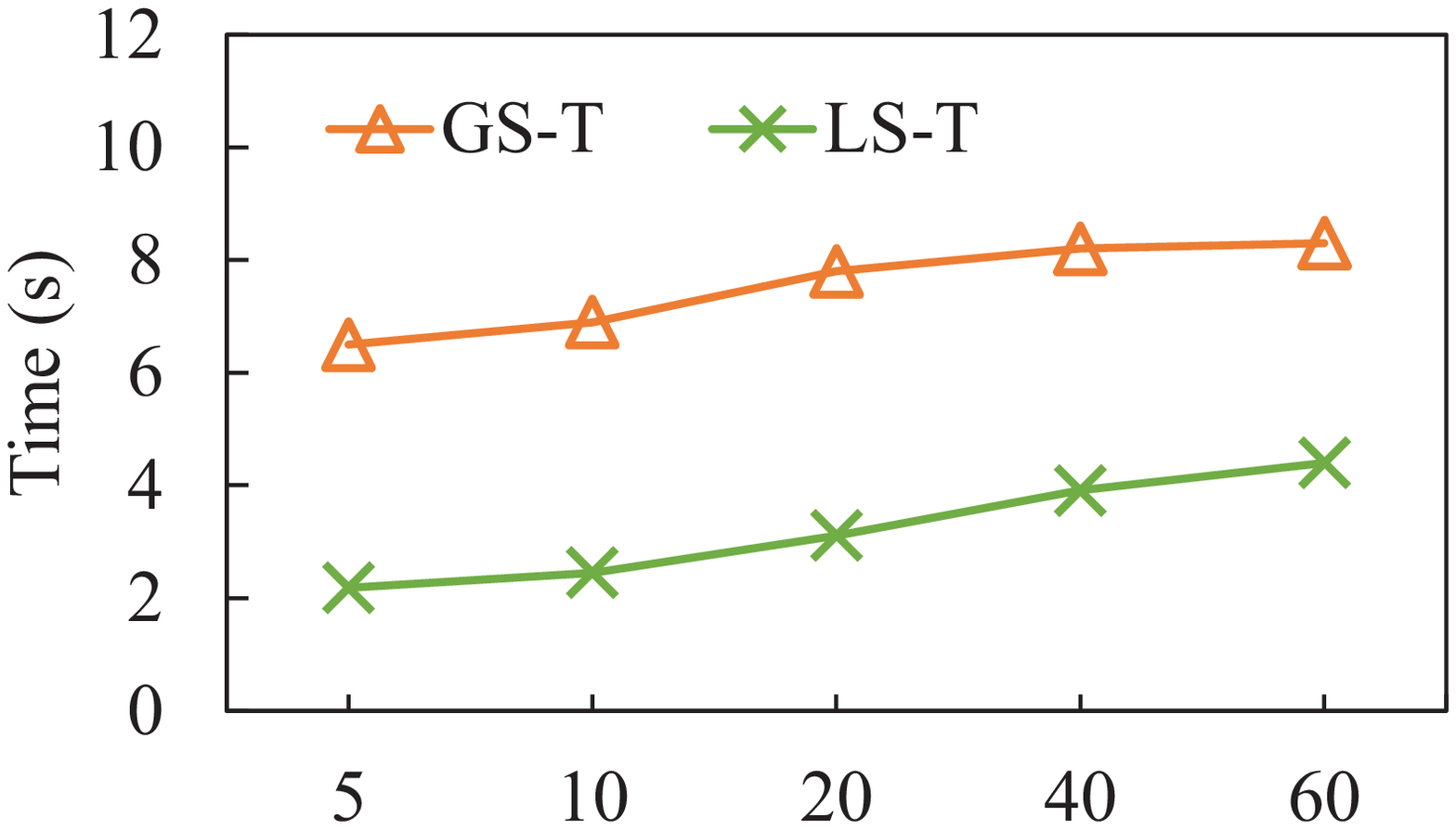}
}
\subfigure[Varying $\sigma$] {\label{fig:delicious-sigma}
\includegraphics[width=0.146\textwidth]{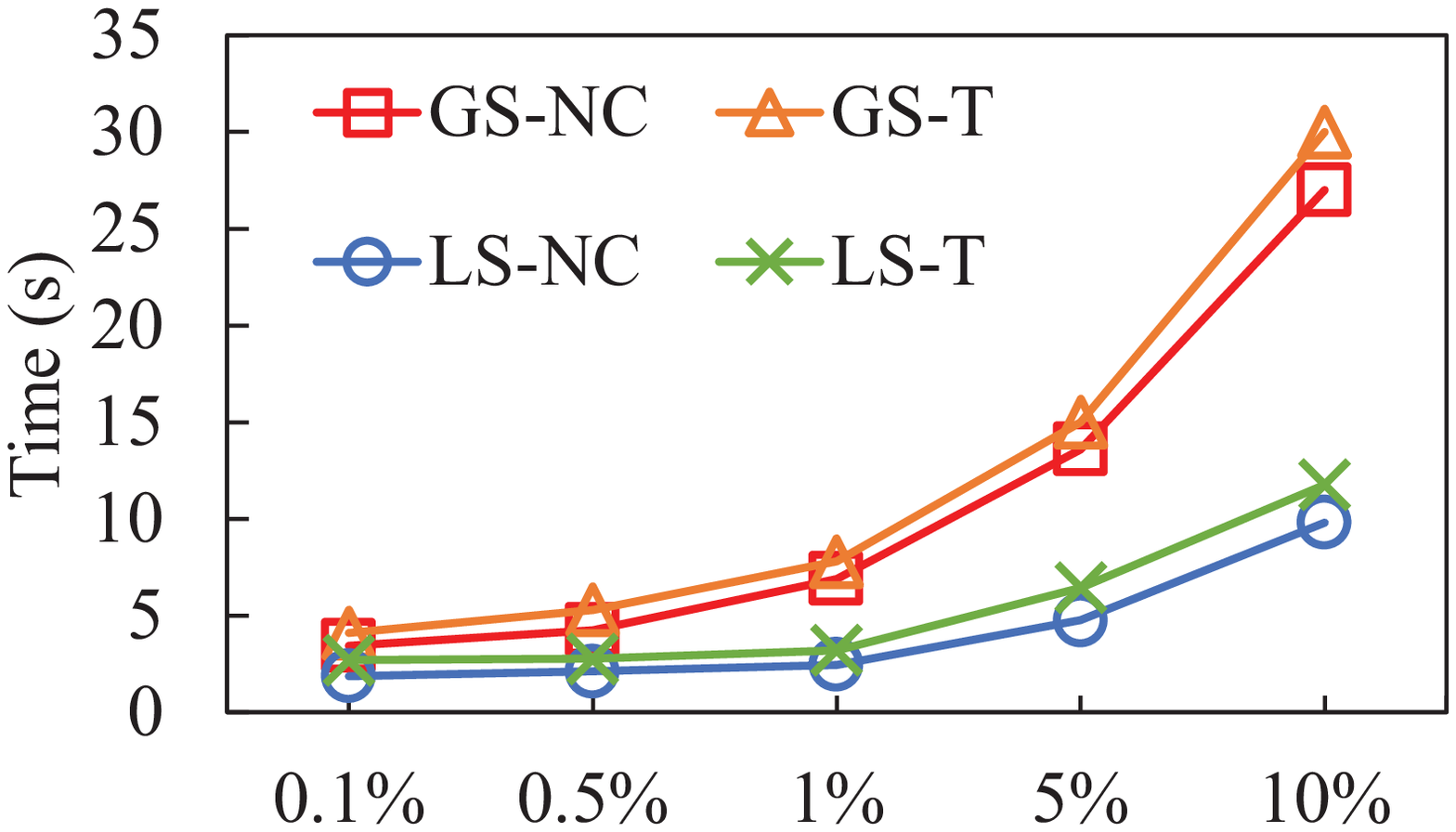}
}
\vspace{-7pt}
\caption{Efficiency and scalability of proposed algorithms in SF+Delicious with independent attributes.}
\label{fig:delicious}
\vspace{-8pt}
\end{figure*}

\begin{figure*}[t]
\centering
\subfigcapskip=-4pt
\subfigure[Varying $k$] {\label{fig:lastfm-k}
\includegraphics[width=0.146\textwidth]{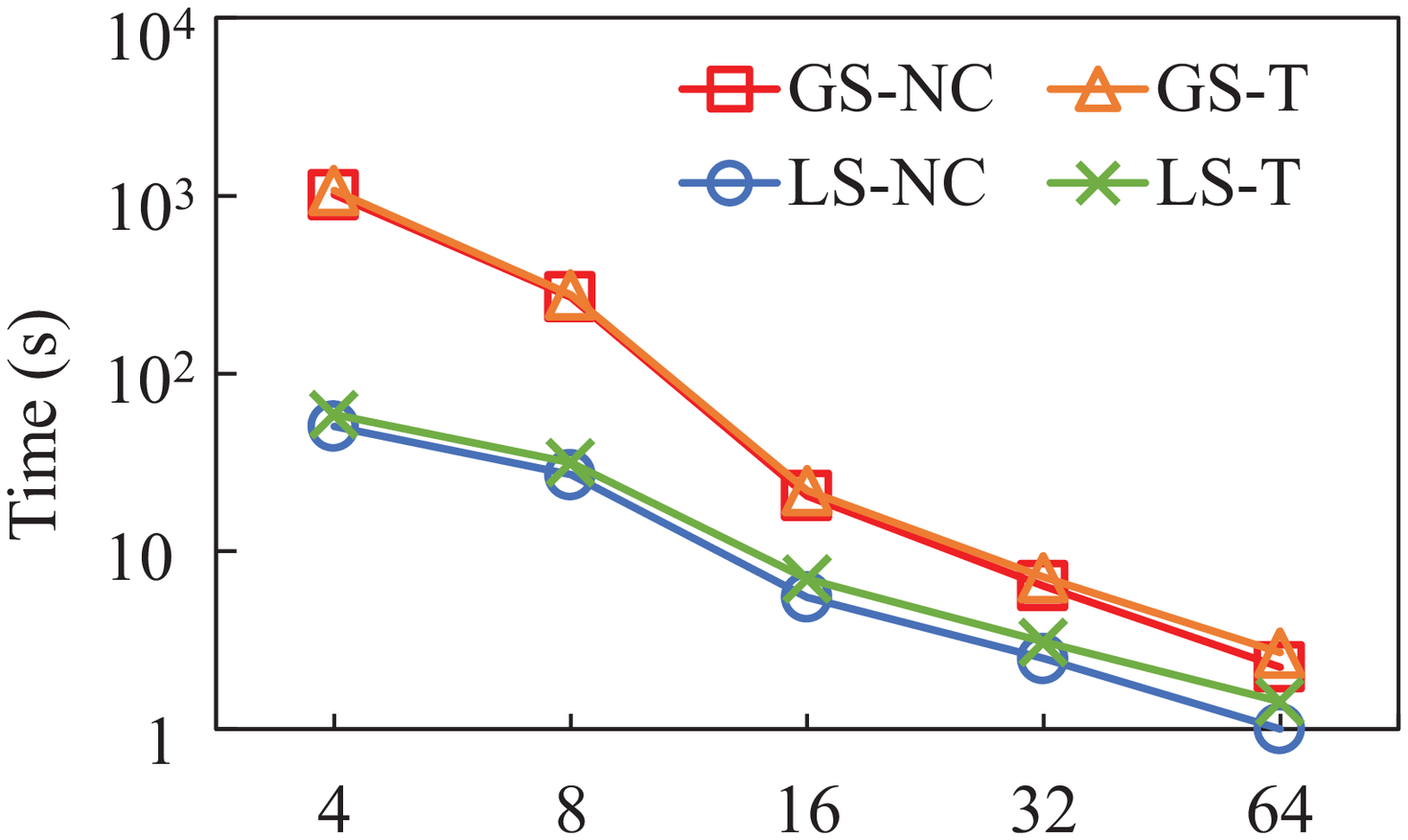}
}
\subfigure[Varying $t$] {\label{fig:lastfm-t}
\includegraphics[width=0.146\textwidth]{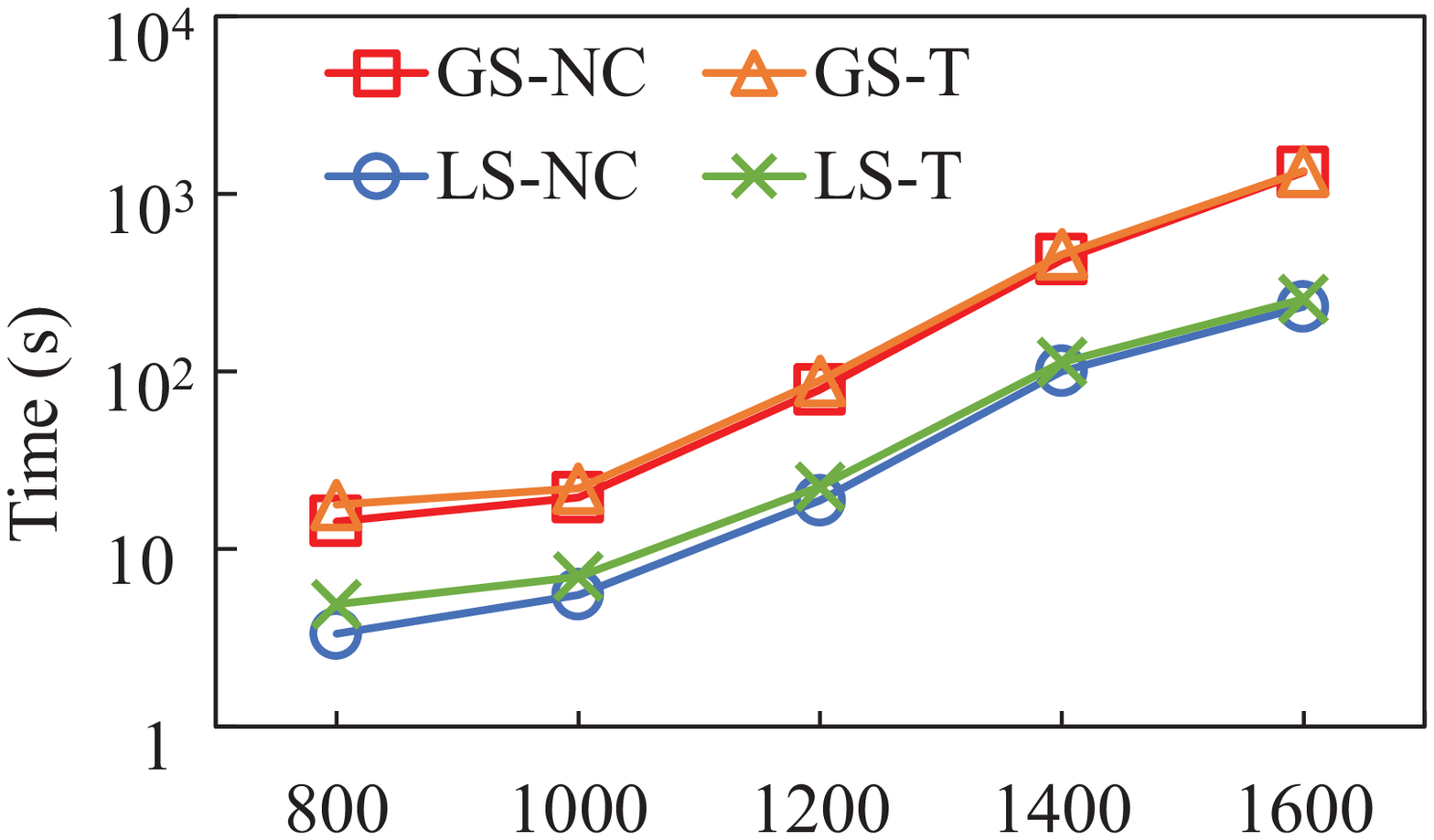}
}
\subfigure[Varying $d$] {\label{fig:lastfm-d}
\includegraphics[width=0.146\textwidth]{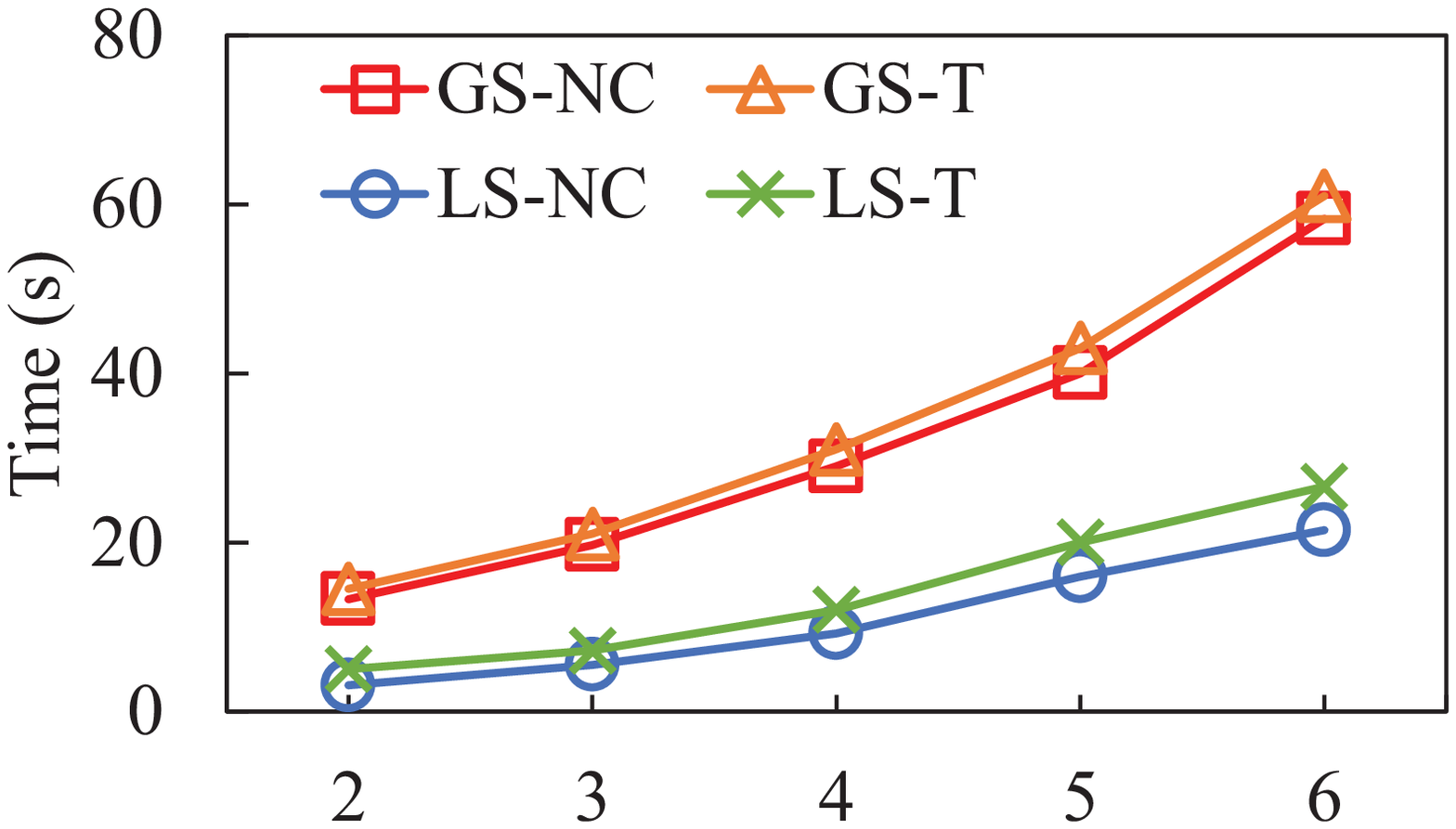}
}
\subfigure[Varying $|Q|$] {\label{fig:lastfm-q}
\includegraphics[width=0.146\textwidth]{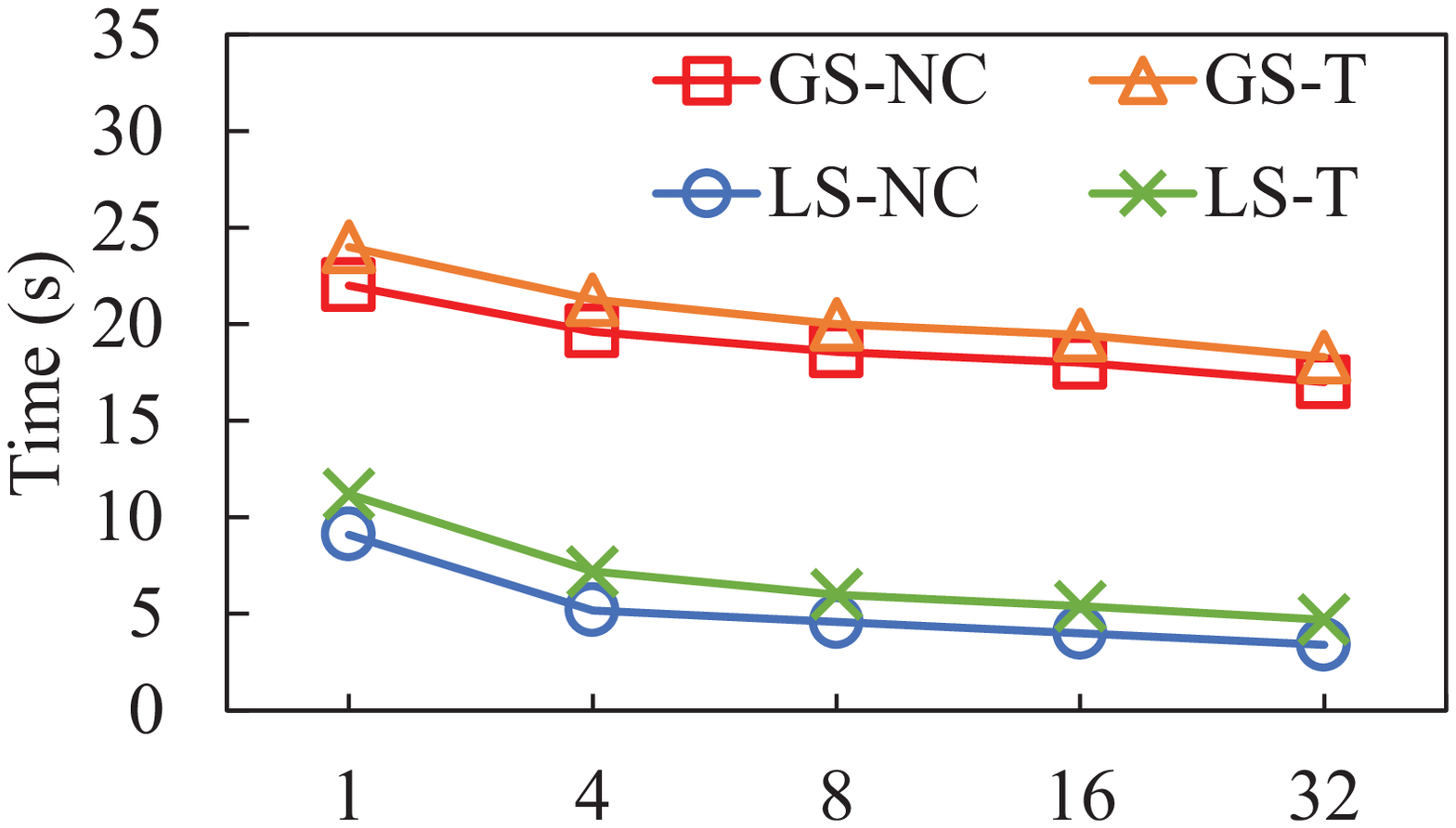}
}
\subfigure[Varying $j$] {\label{fig:lastfm-j}
\includegraphics[width=0.146\textwidth]{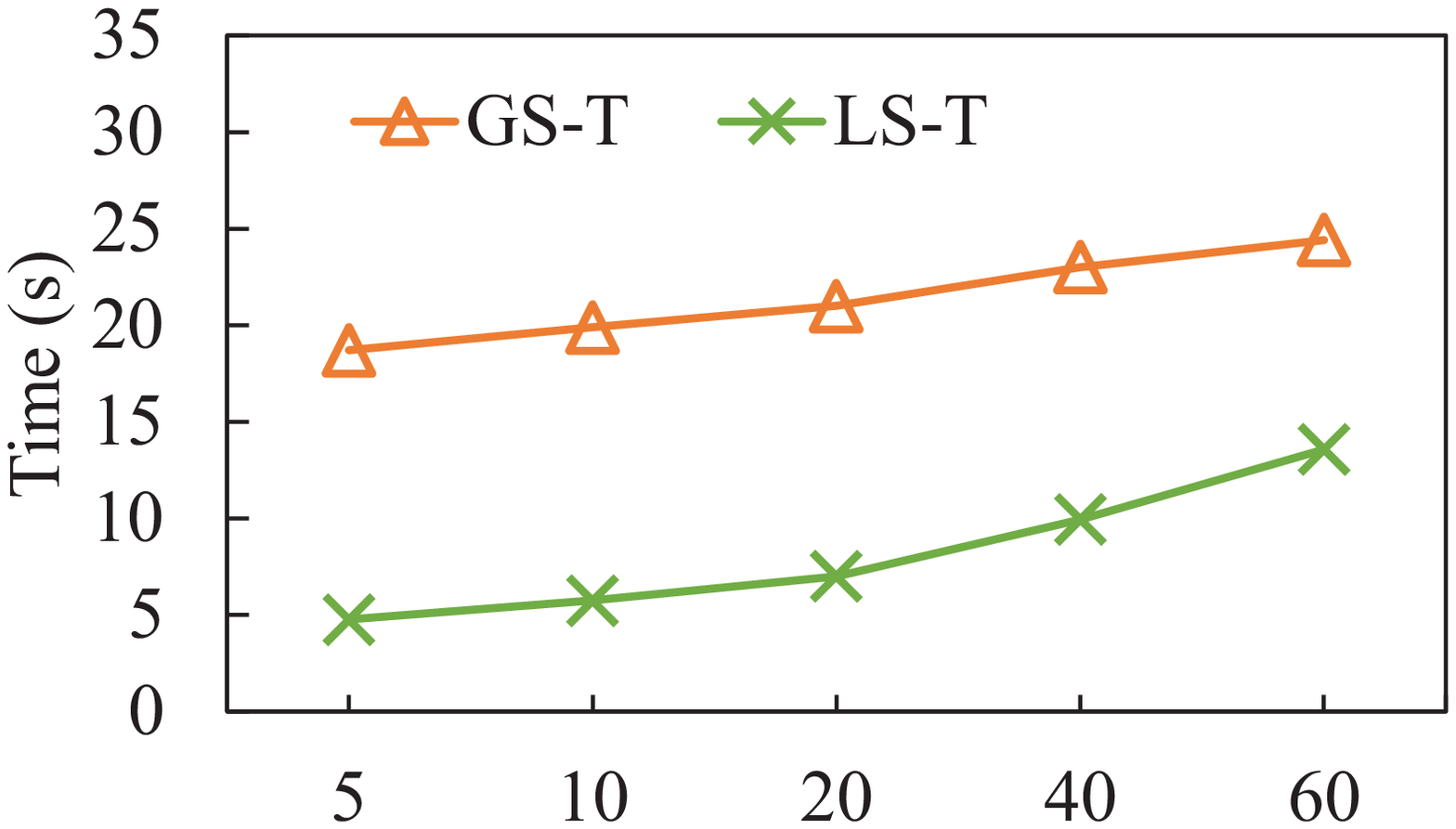}
}
\subfigure[Varying $\sigma$] {\label{fig:lastfm-sigma}
\includegraphics[width=0.146\textwidth]{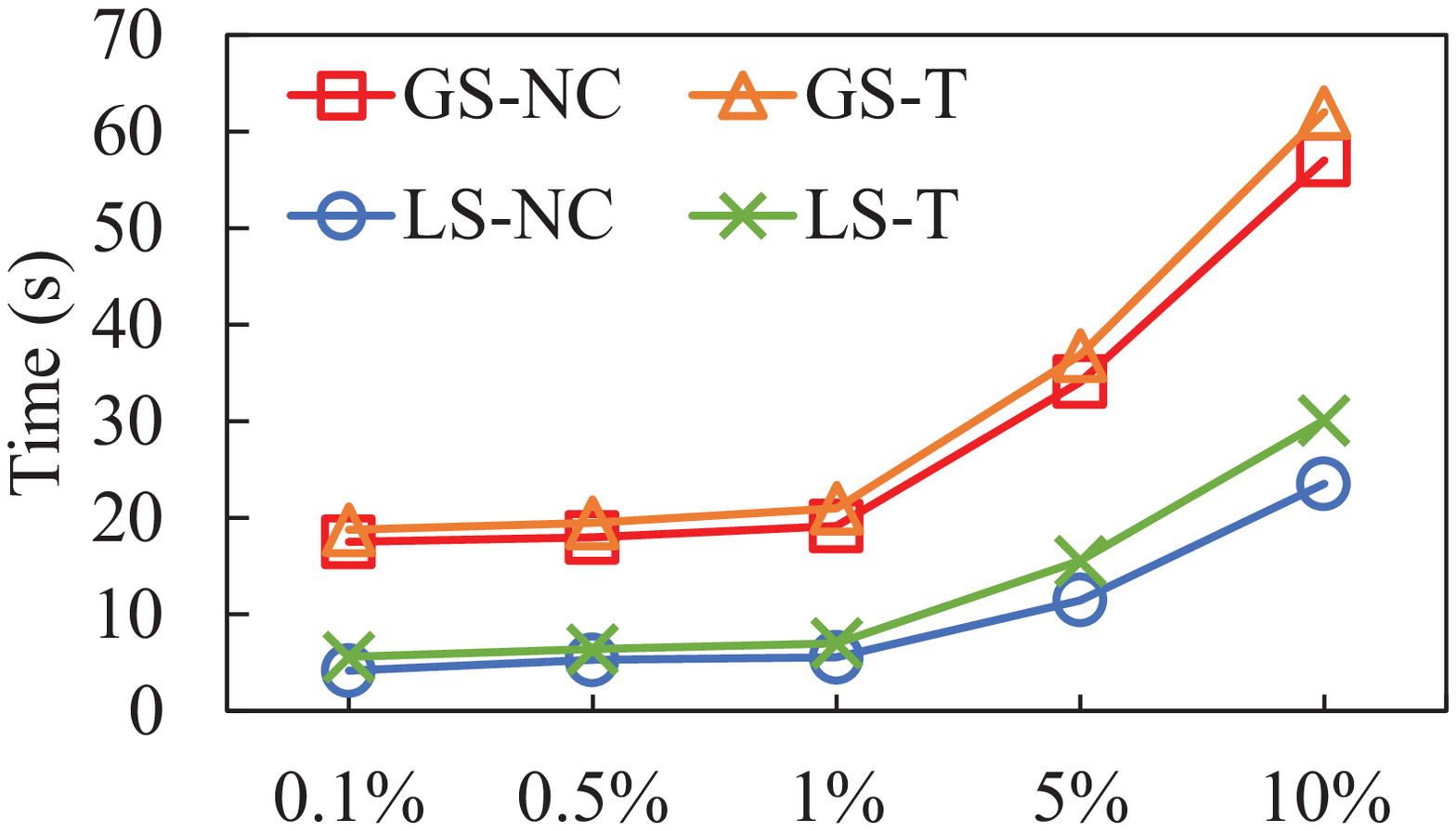}
}
\vspace{-7pt}
\caption{Efficiency and scalability of proposed algorithms in FL+Lastfm with independent attributes.}
\label{fig:lastfm}
\vspace{-8pt}
\end{figure*}

\begin{figure*}[t]
\centering
\subfigcapskip=-4pt
\subfigure[Varying $k$] {\label{fig:flixster-k}
\includegraphics[width=0.146\textwidth]{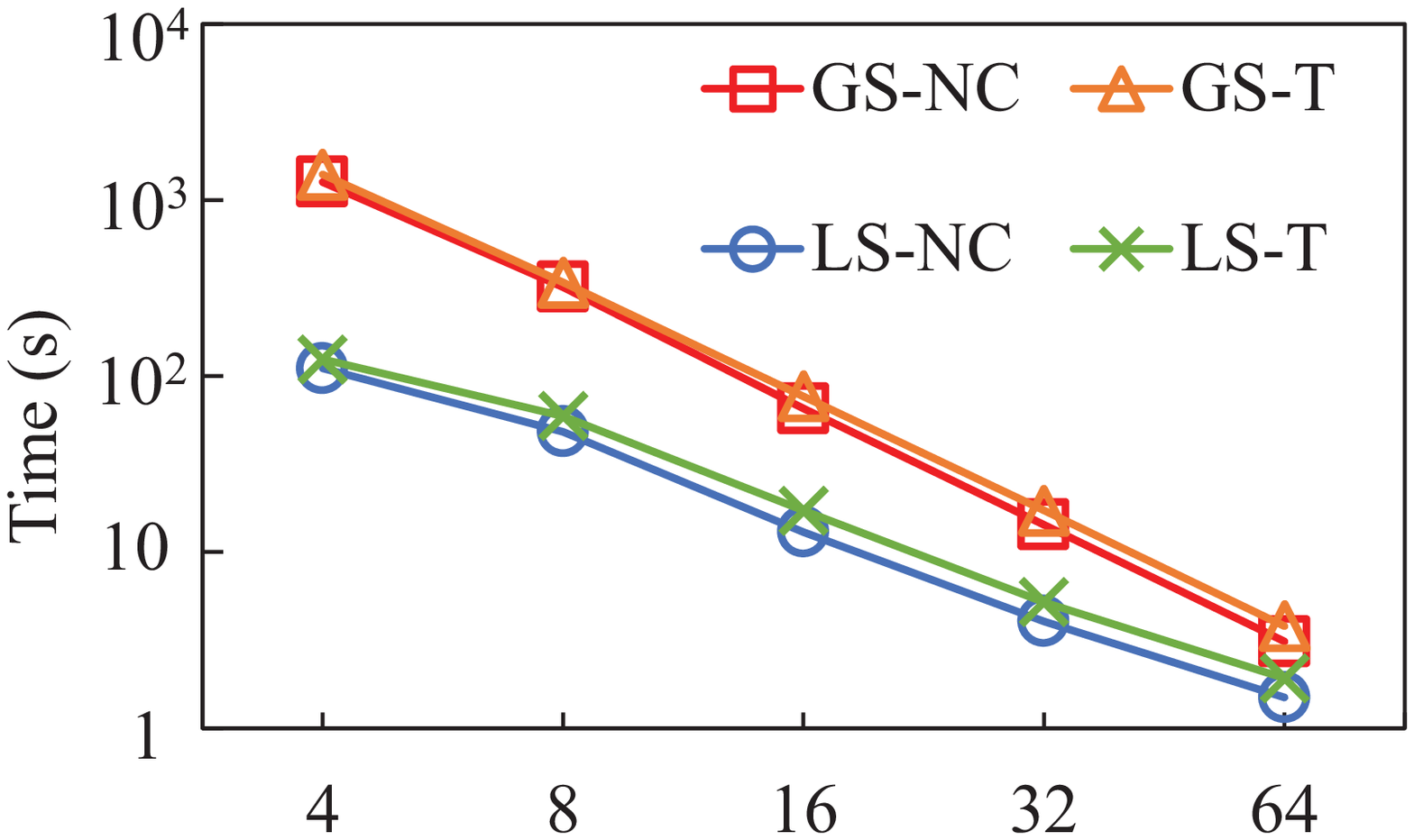}
}
\subfigure[Varying $t$] {\label{fig:flixster-t}
\includegraphics[width=0.146\textwidth]{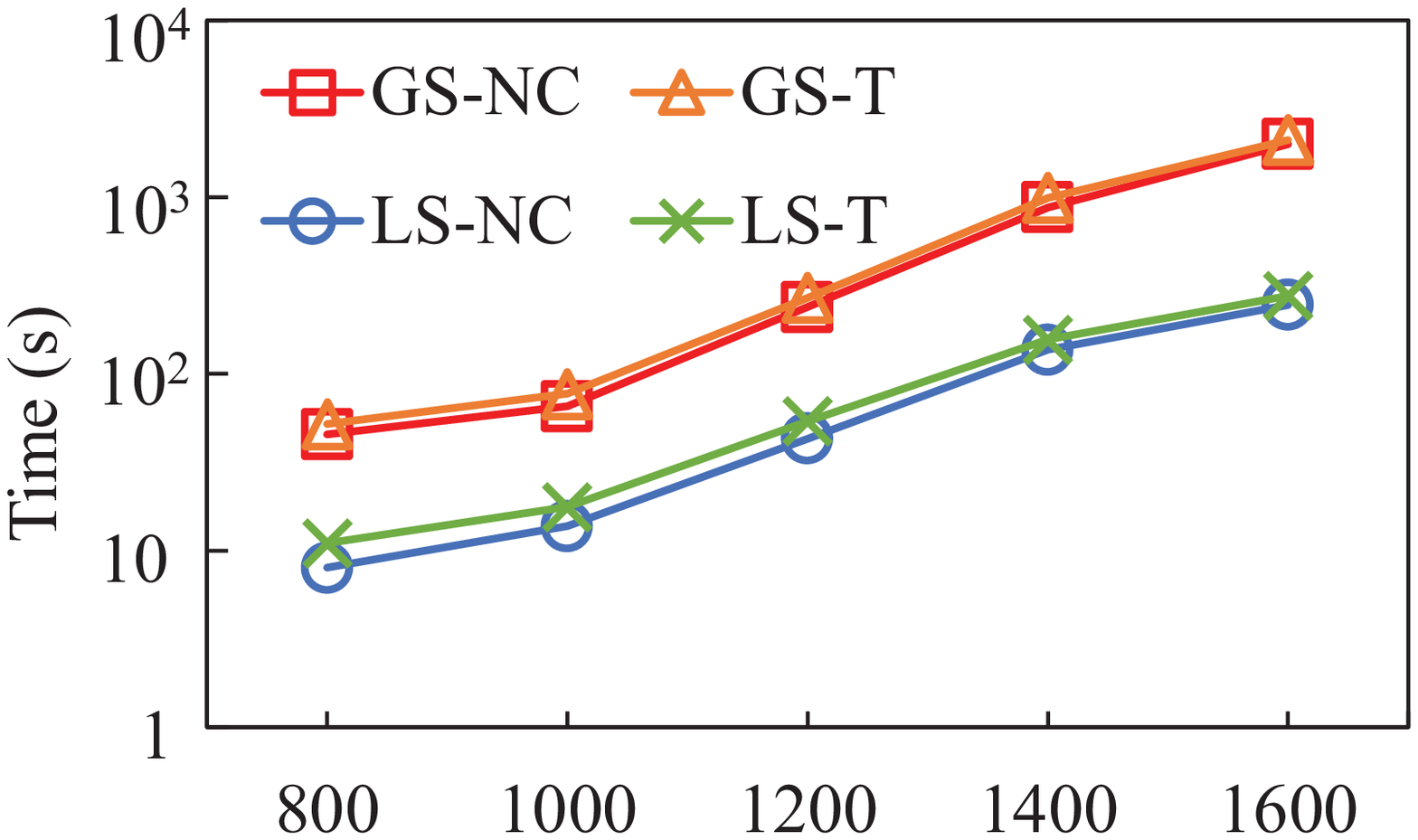}
}
\subfigure[Varying $d$] {\label{fig:flixster-d}
\includegraphics[width=0.146\textwidth]{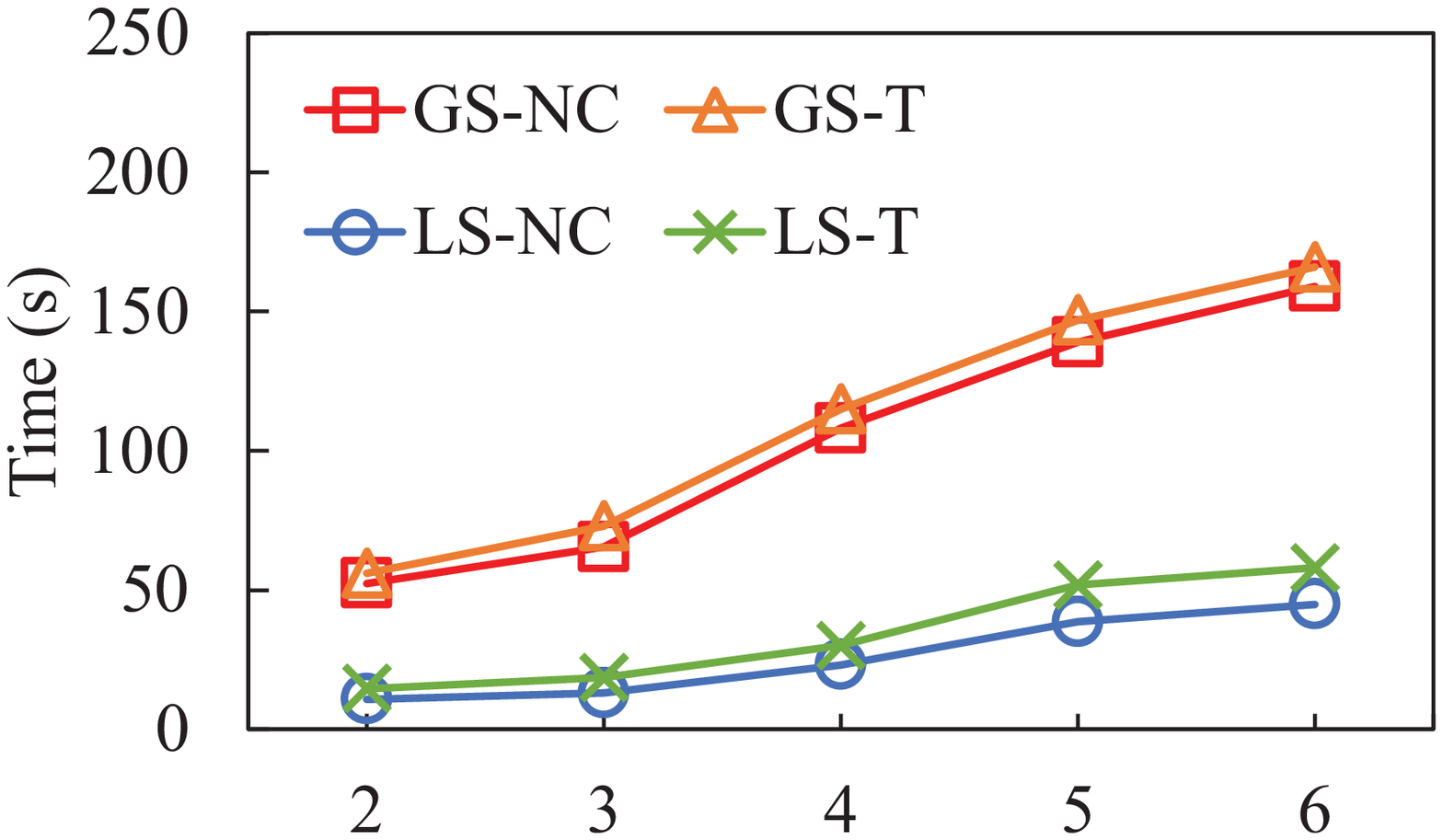}
}
\subfigure[Varying $|Q|$] {\label{fig:flixster-q}
\includegraphics[width=0.146\textwidth]{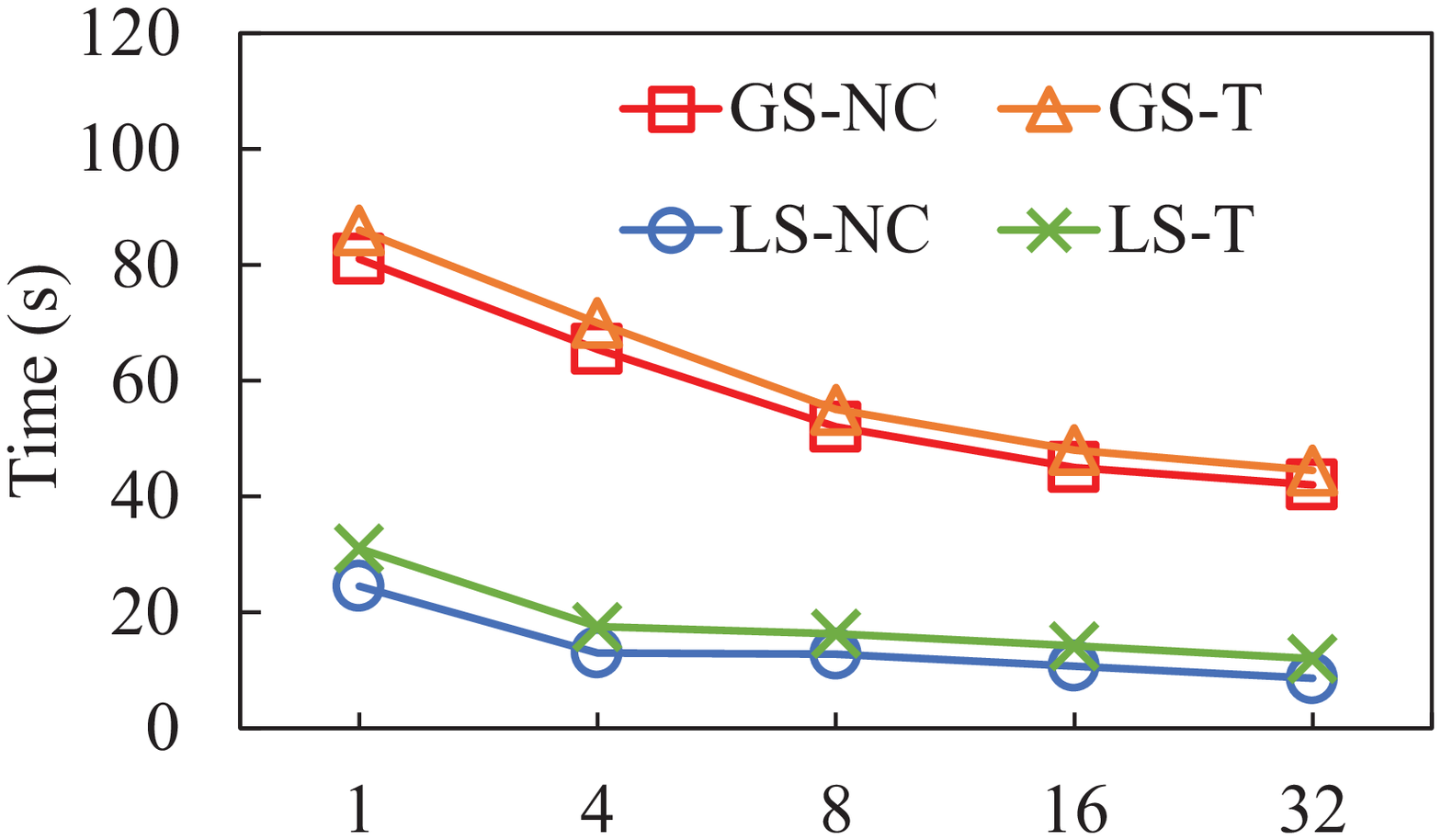}
}
\subfigure[Varying $j$] {\label{fig:flixster-j}
\includegraphics[width=0.146\textwidth]{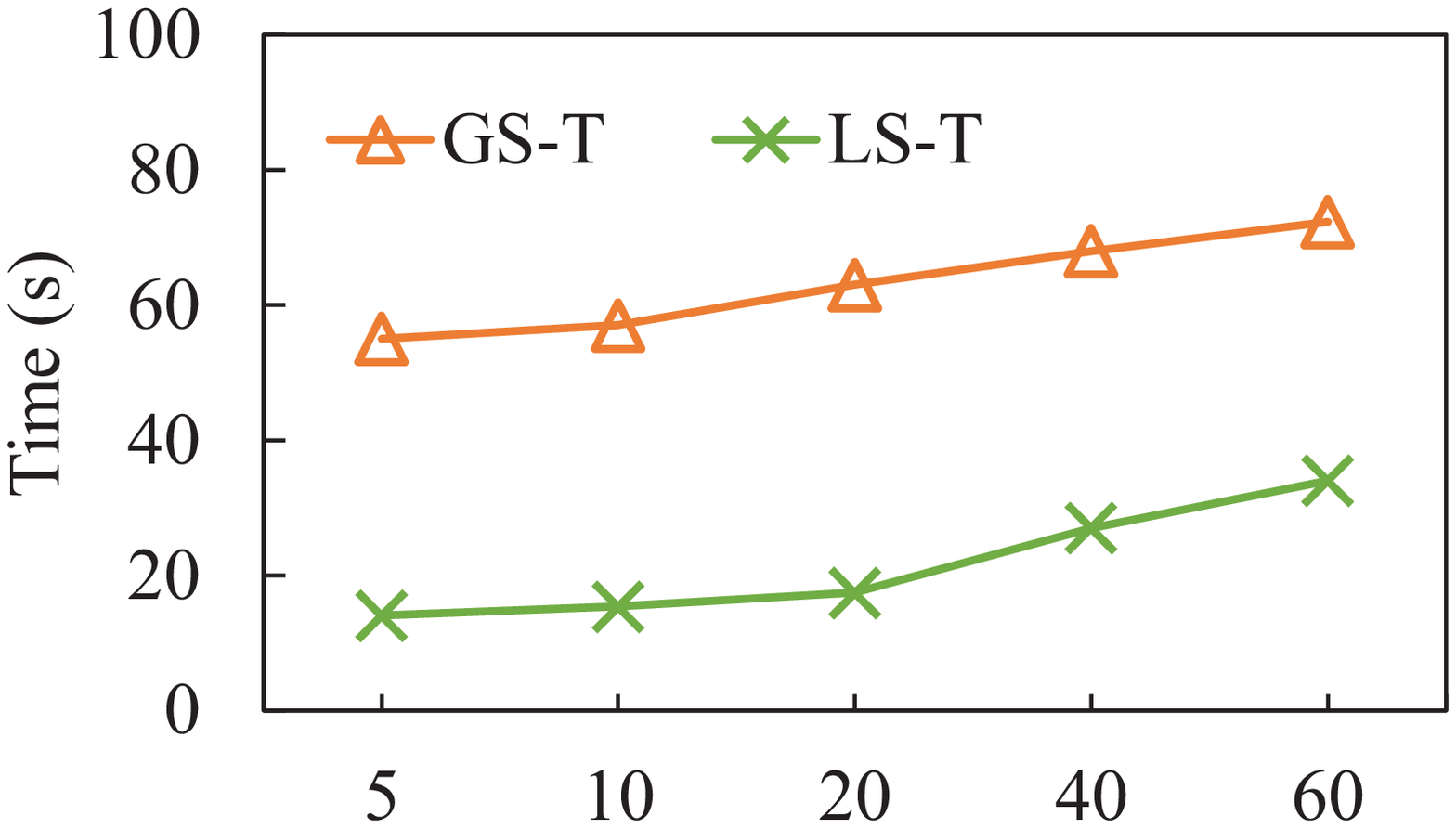}
}
\subfigure[Varying $\sigma$] {\label{fig:flixster-sigma}
\includegraphics[width=0.146\textwidth]{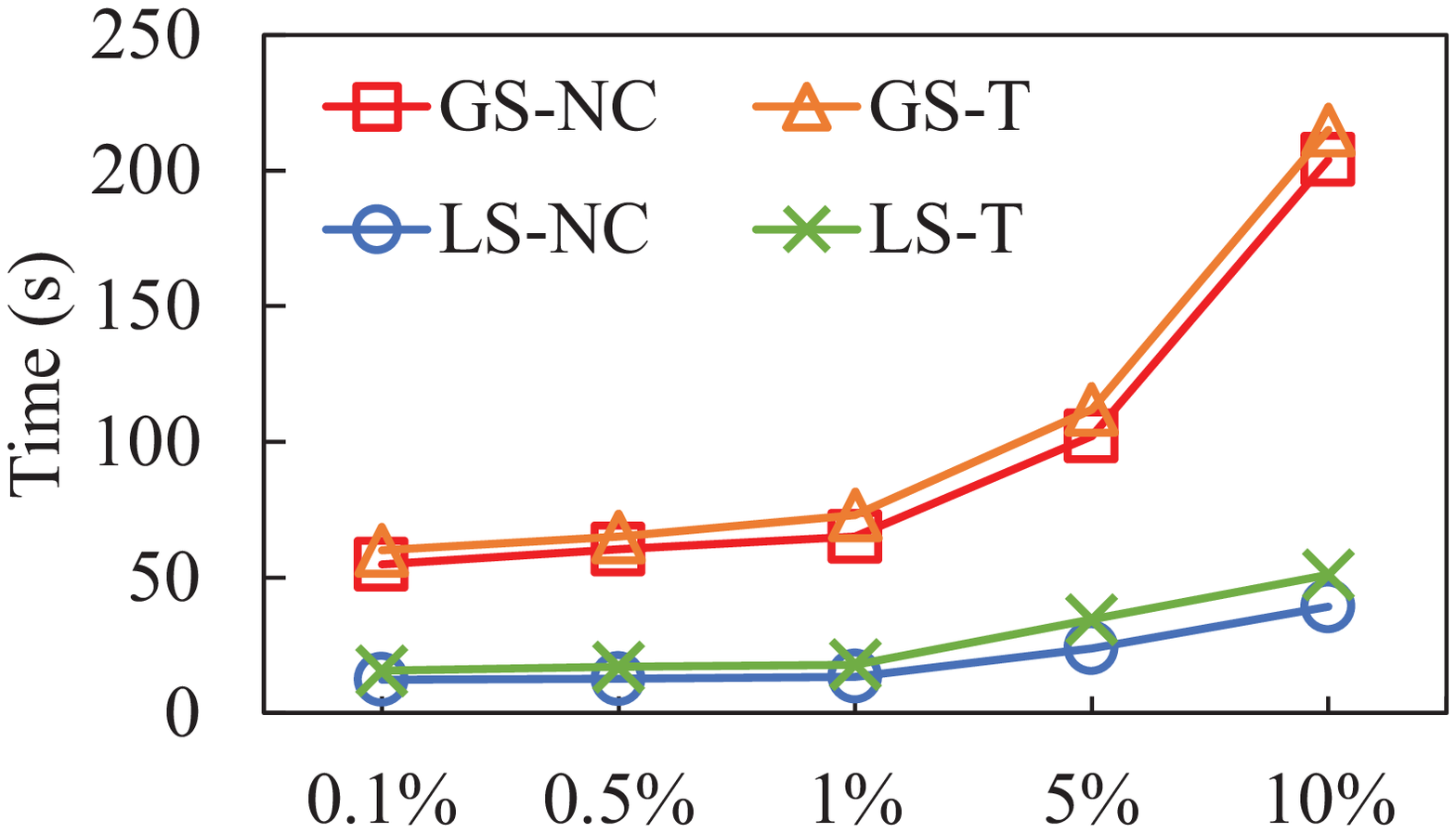}
}
\vspace{-7pt}
\caption{Efficiency and scalability of proposed algorithms in FL+Flixster with independent attributes.}
\label{fig:flixster}
\vspace{-15pt}
\end{figure*}

\begin{figure*}[t]
\centering
\subfigcapskip=-4pt
\subfigure[Varying $k$] {\label{fig:yelp-k}
\includegraphics[width=0.146\textwidth]{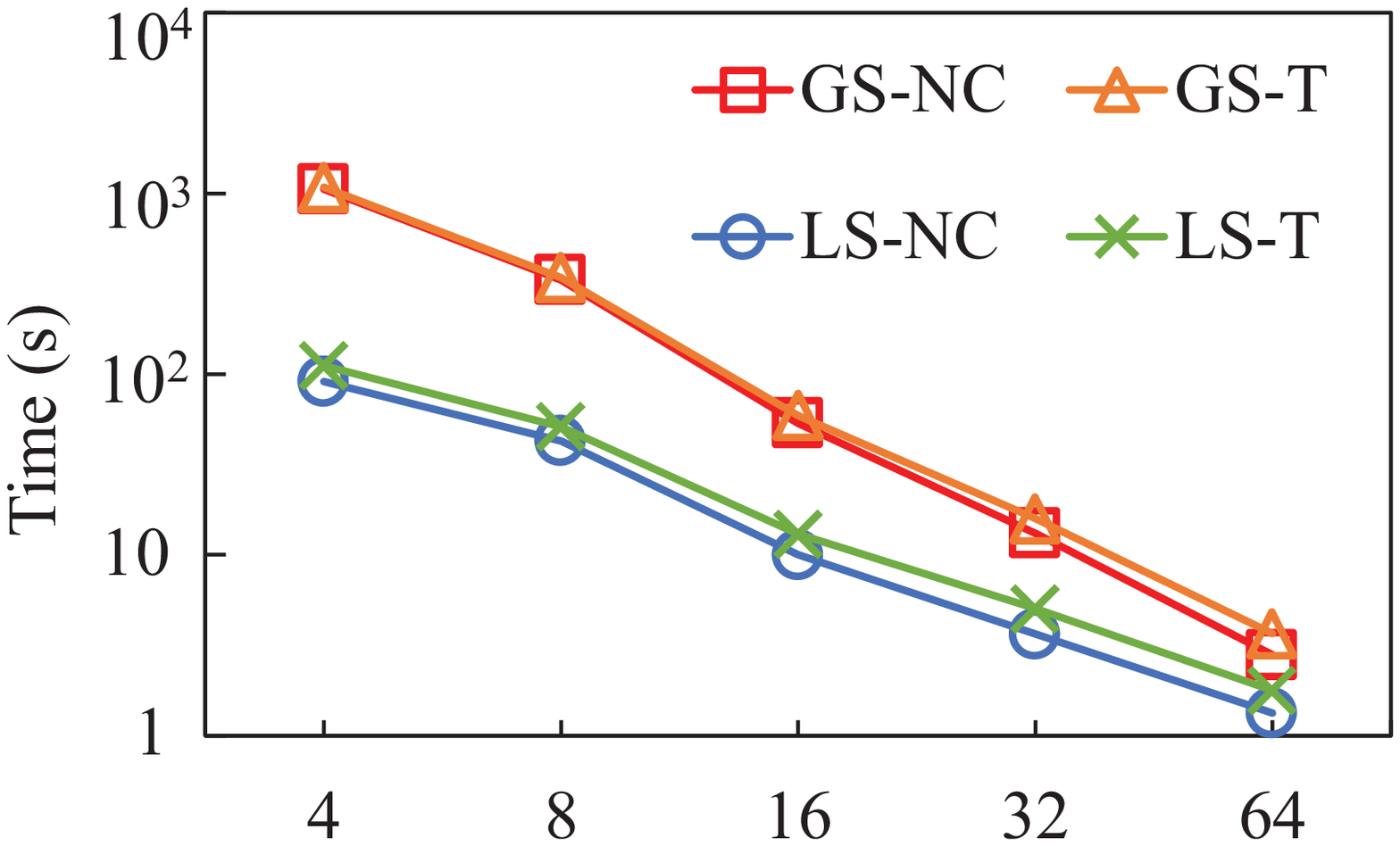}
}
\subfigure[Varying $t$] {\label{fig:yelp-t}
\includegraphics[width=0.146\textwidth]{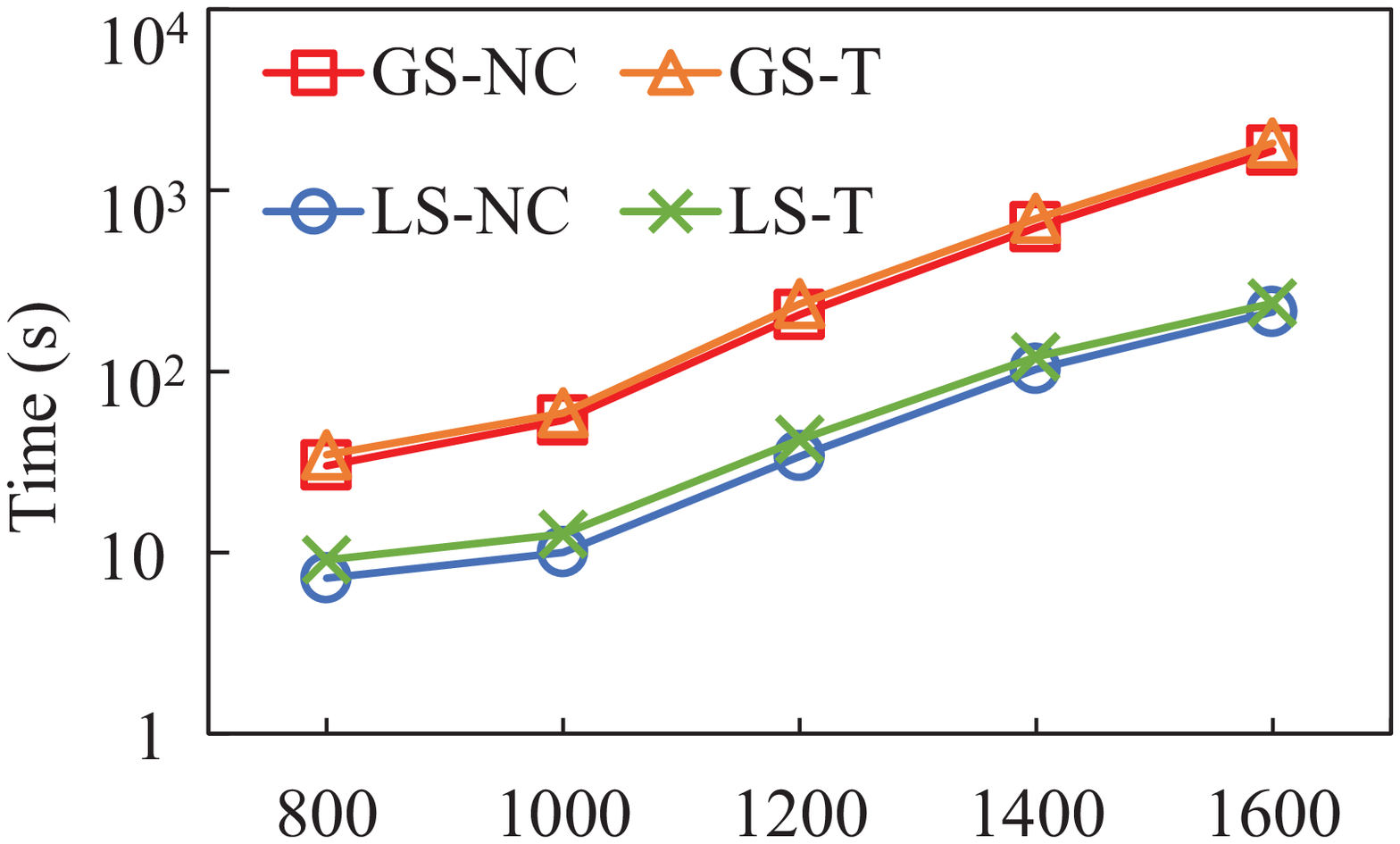}
}
\subfigure[Varying $d$] {\label{fig:yelp-d}
\includegraphics[width=0.146\textwidth]{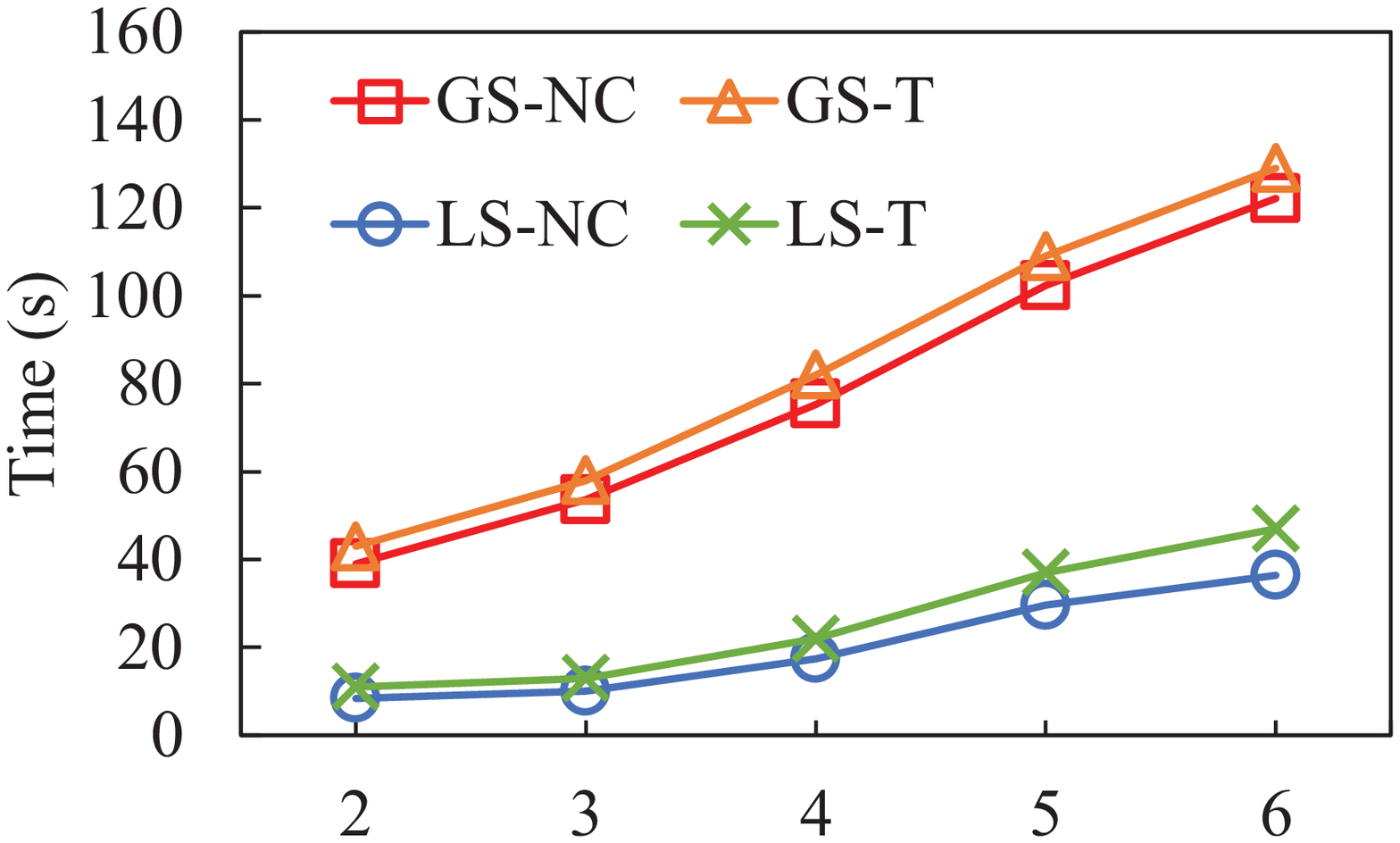}
}
\subfigure[Varying $|Q|$] {\label{fig:yelp-q}
\includegraphics[width=0.146\textwidth]{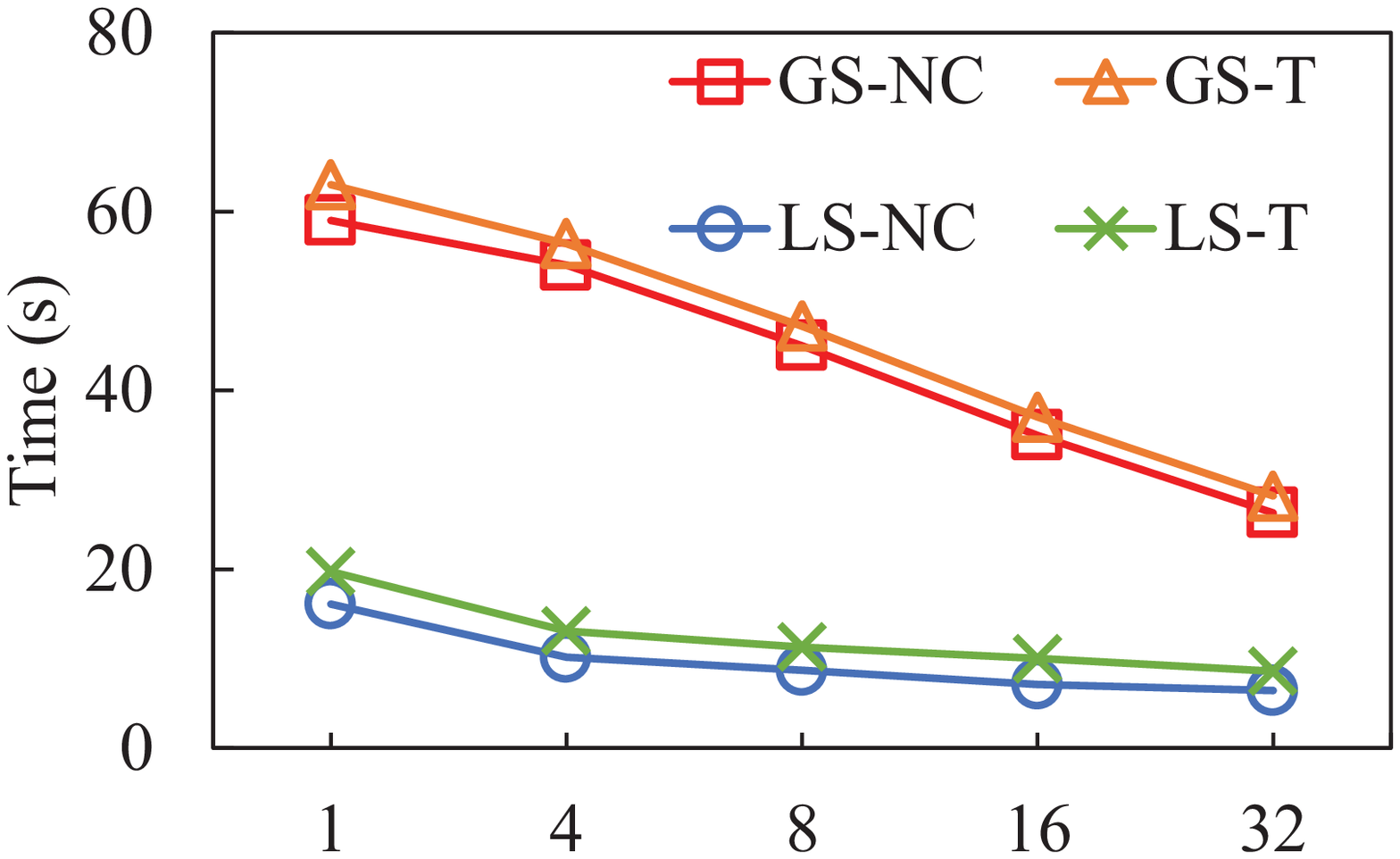}
}
\subfigure[Varying $j$] {\label{fig:yelp-j}
\includegraphics[width=0.146\textwidth]{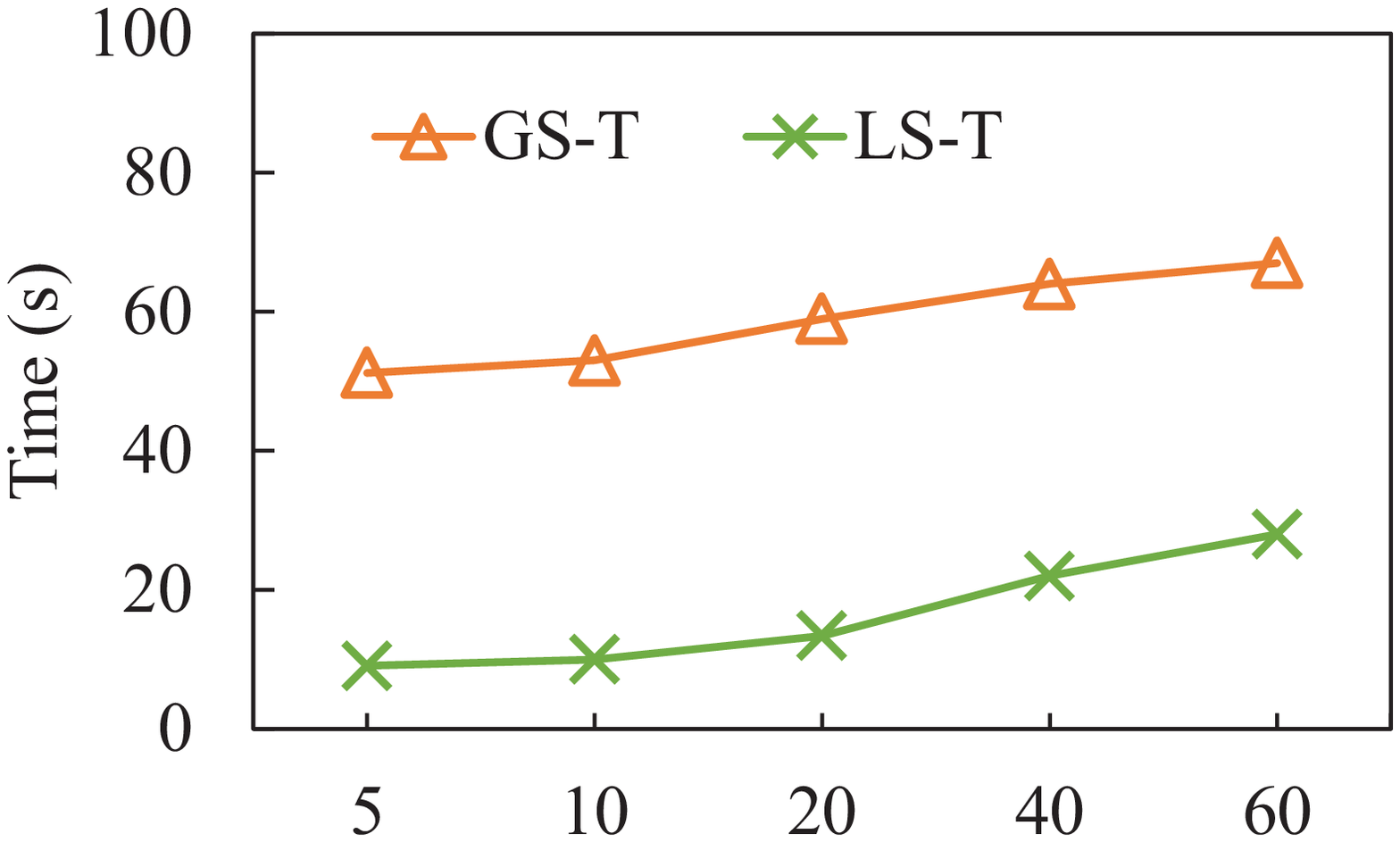}
}
\subfigure[Varying $\sigma$] {\label{fig:yelp-sigma}
\includegraphics[width=0.146\textwidth]{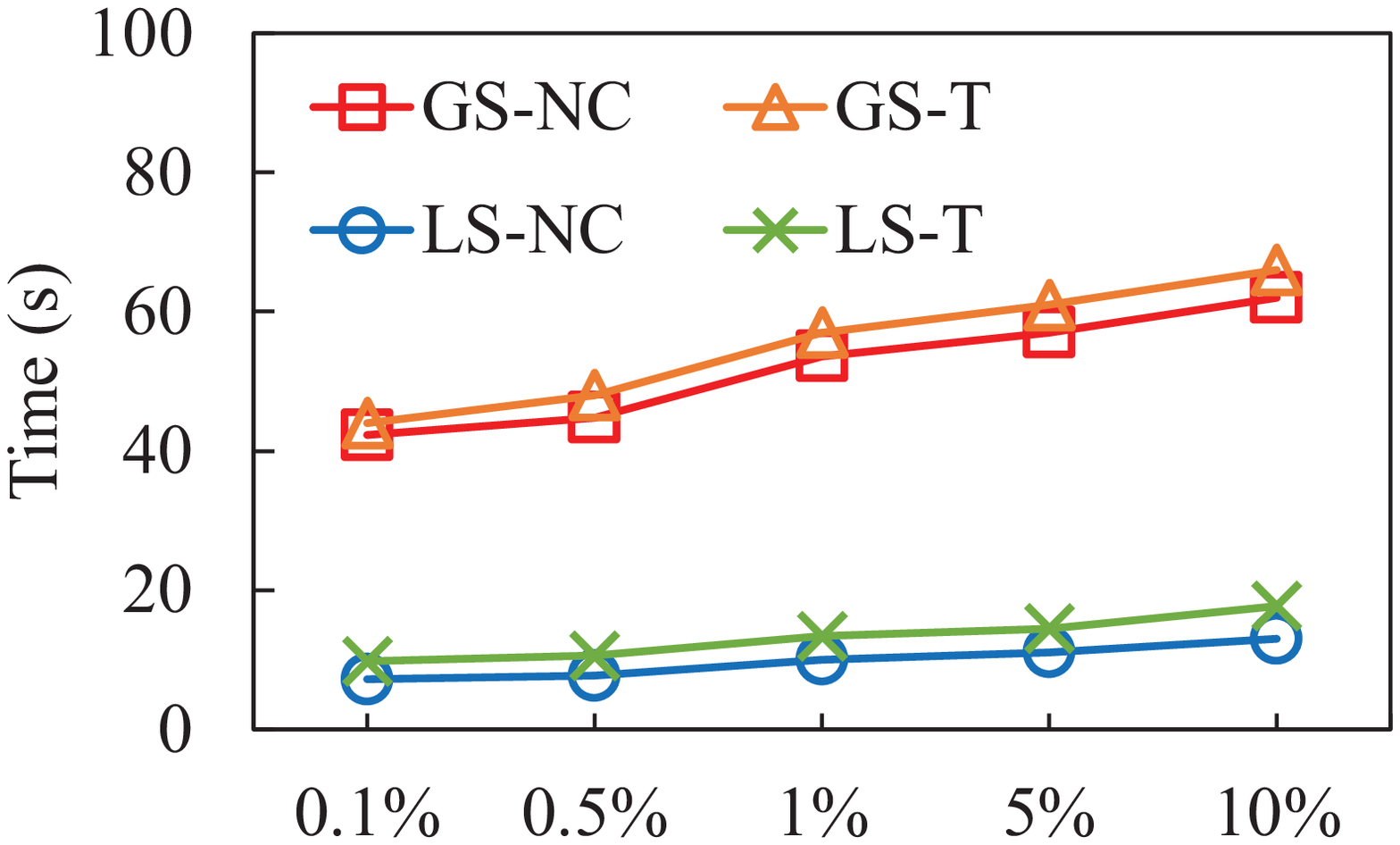}
}
\vspace{-7pt}
\caption{Efficiency and scalability of proposed algorithms in FL+Yelp with independent attributes.}
\label{fig:yelp}
\vspace{-15pt}
\end{figure*}

\begin{figure}[t]
\centering
\subfigbottomskip=3pt
\subfigcapskip=-4pt
\subfigure[No. of partitions (during search)] {\label{fig:no-partition}
\includegraphics[width=0.465\columnwidth]{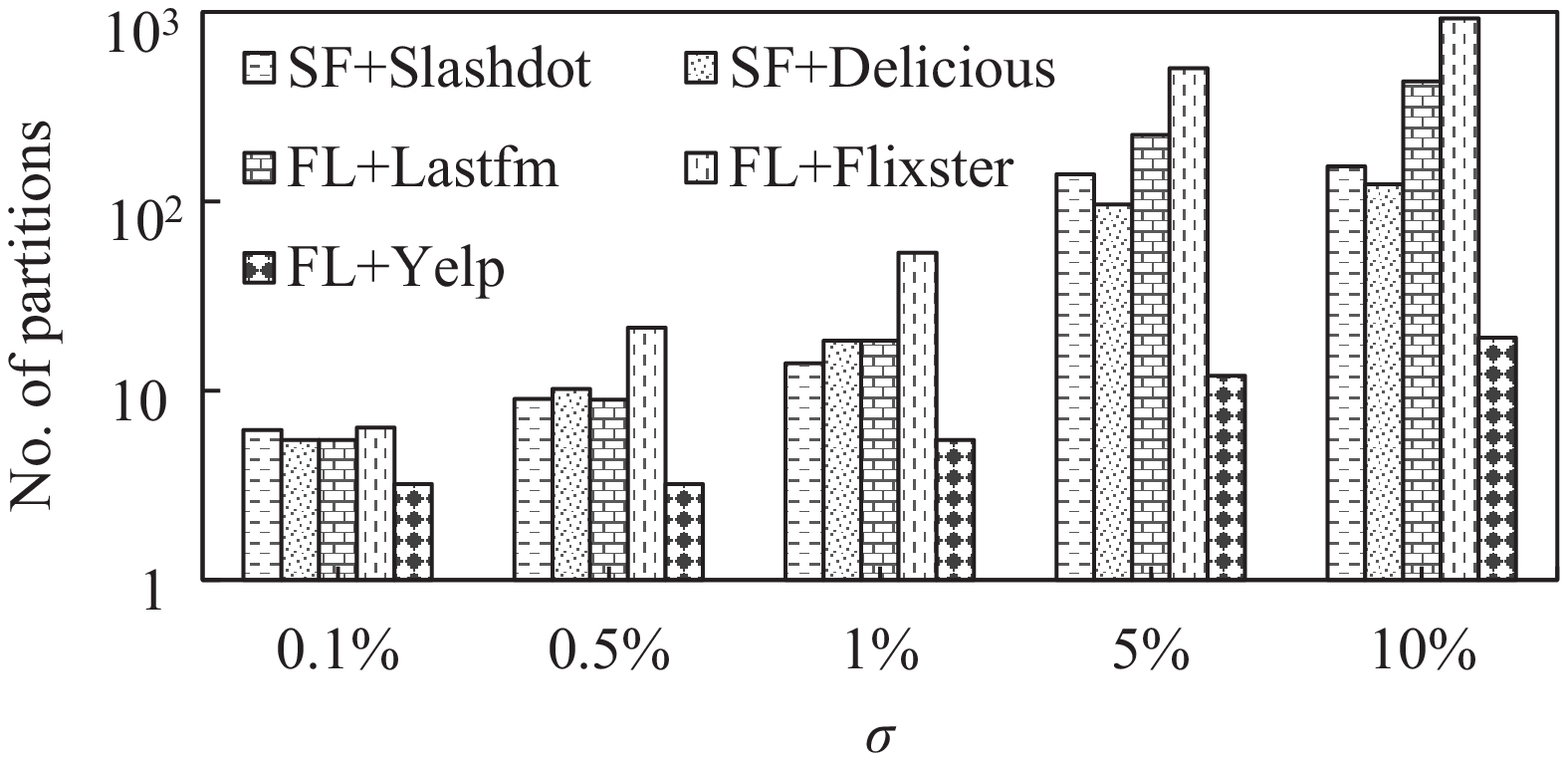}
}
\subfigure[No. of non-contained MACs] { \label{fig:no-nc-mac}
\includegraphics[width=0.465\columnwidth]{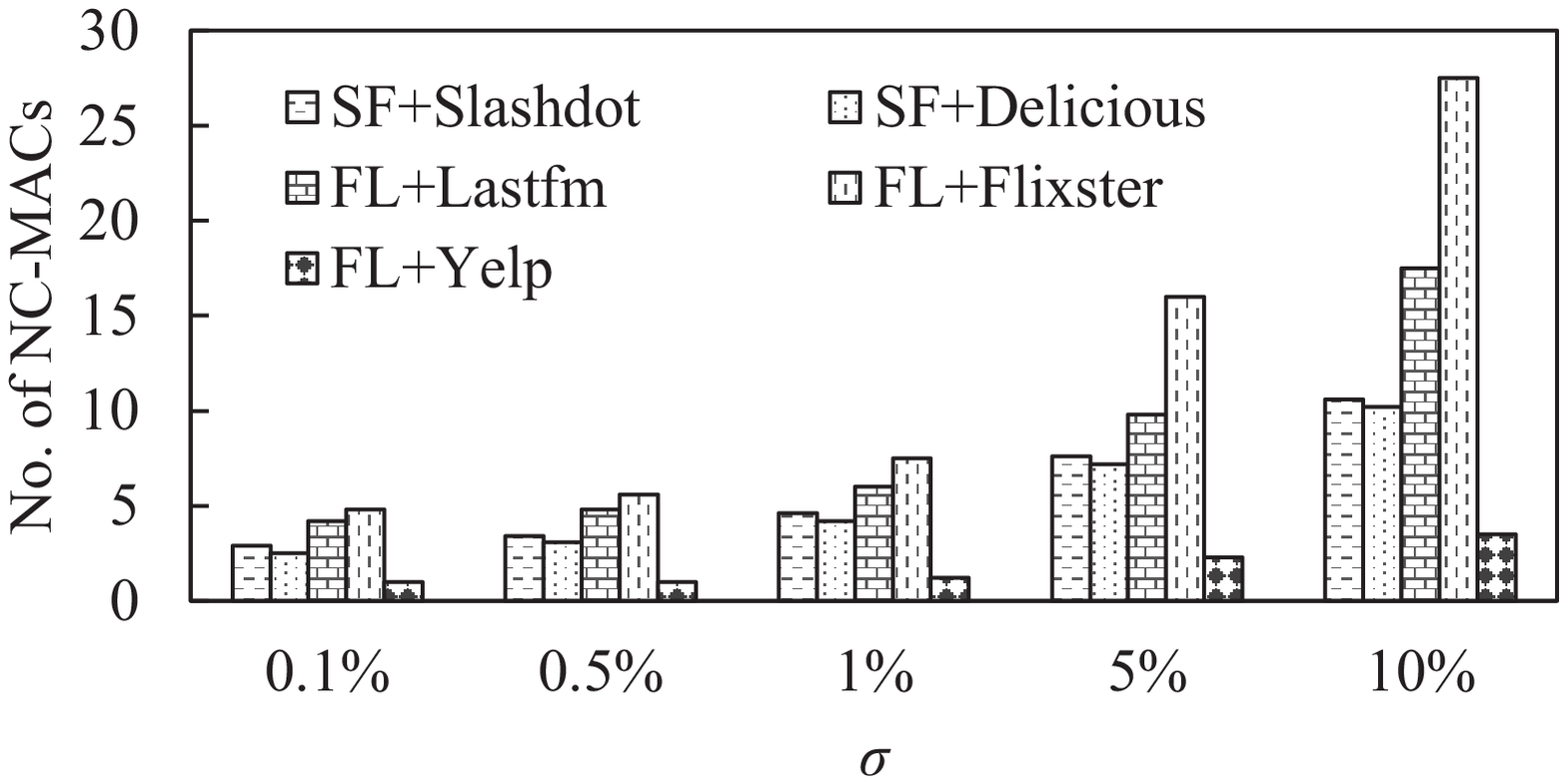}
}
\subfigure[\#Vertices of $H_k^t$] {\label{fig:vertices-in-Hkt}
\includegraphics[width=0.465\columnwidth]{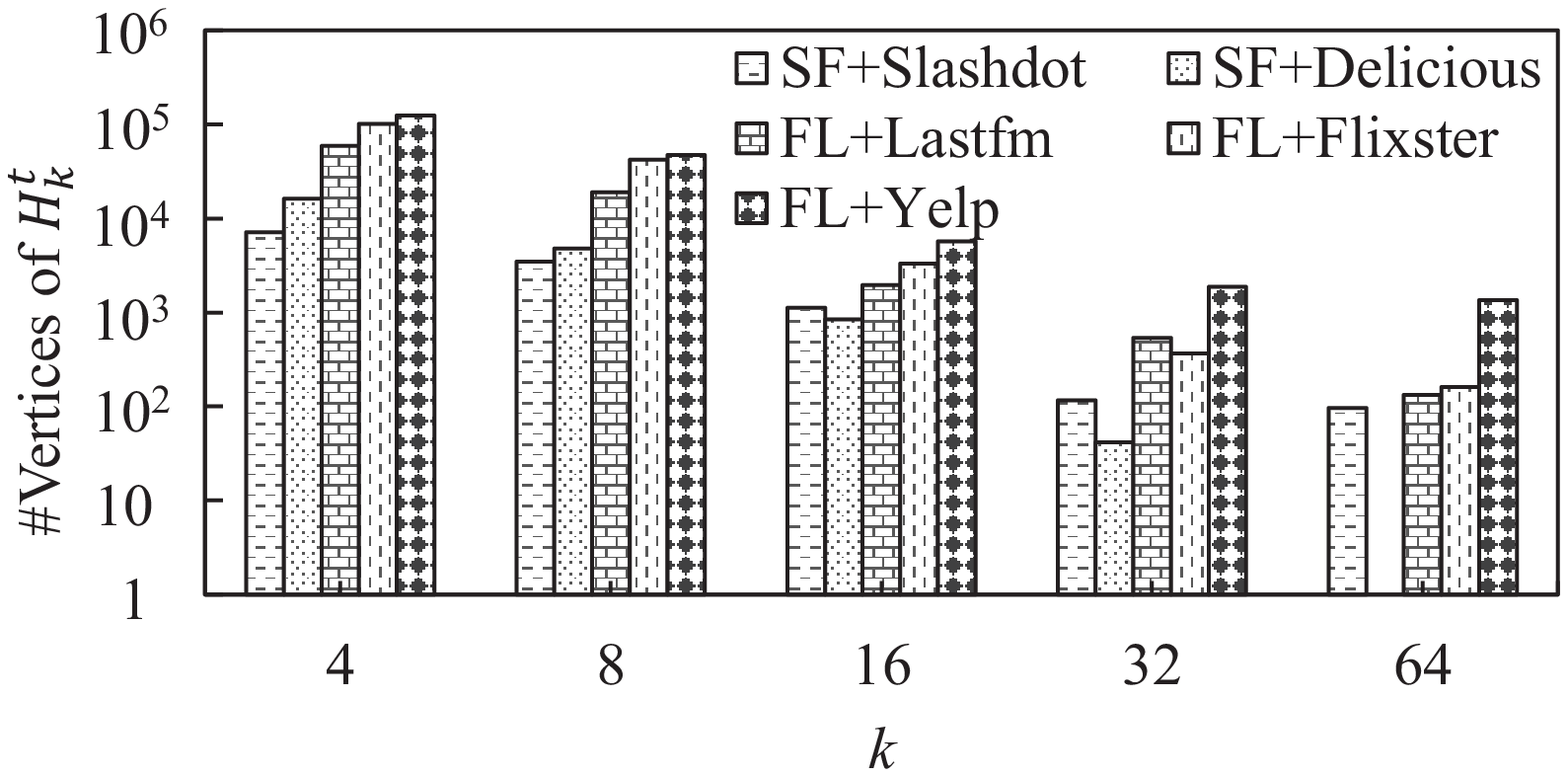}
}
\subfigure[Memory overhead (FL+Lastfm)] { \label{fig:spacecost}
\includegraphics[width=0.465\columnwidth]{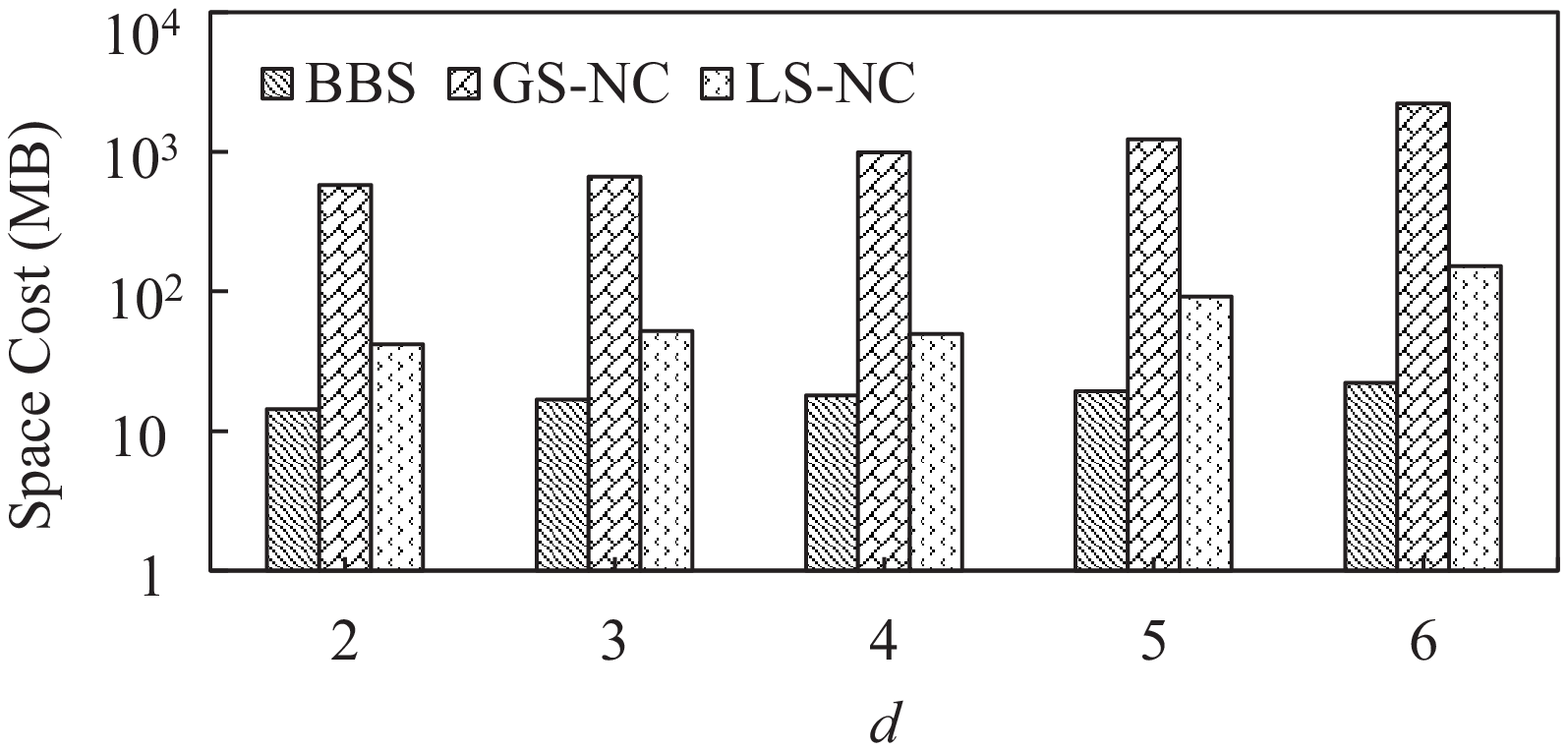}
}
\vspace{-12pt}
\caption{Scalability of proposed algorithms.}
\vspace{-8pt}
\label{fig:scalability}
\end{figure}

\noindent
\textbf{Exp-1: Varying $k$.} We evaluate the query processing time of all algorithms and the number of vertices of $H_k^t$ by varying $k$. In each road-social network, we can see that local search performs better than global search, but the advantage becomes less obvious when $k$ increases. In the best case, e.g., $k\!=\!4$, \texttt{LS-T} and \texttt{LS-NC} are more than one order of magnitude faster than \texttt{GS-T} and \texttt{GS-NC} in Fig.~\ref{fig:delicious-k} and~\ref{fig:flixster-k}. Only when $k\!=\!64$, global search is comparable to local search since $H_k^t$ size shrinks when $k$ is large (see Fig.~\ref{fig:vertices-in-Hkt}), resulting in a reduction in the time complexity of global search and in the number of promising vertices involved in local search. Although Delicious is larger than Slashdot, but algorithms run faster at $k\!=\!16$ and $k\!=\!32$ since $H_k^t$ contains fewer vertices (as $k_{max} \!=\! 34$ in Table~\ref{tab:table2}). Note that when $k$ increases from $4$ to $8$, local search is merely more than twice as fast, e.g., \texttt{LS-NC} takes $41$s and $17$s respectively in Fig.~\ref{fig:delicious-k}. This is because when $k$ is relatively small, candidate selection strategies tend to find more smaller candidates. However, algorithms run faster in Yelp than in Flixster, yet its $H_k^t$ size is larger. This will be explained in Exp-6. In short, across a wide range of $k$ local search is consistently better than global search.

\noindent
\textbf{Exp-2: Varying $t$.} We evaluate the query processing time of all algorithms by varying $t$. For each road-social network, local search outperforms global search significantly, with advantage becoming more obvious as $t$ increases. For example, in Fig.~\ref{fig:flixster-t}, \texttt{GS-NC} takes $1,992$s while \texttt{LS-NC} takes $245$s for $t\!=\!1600$. Note that the results are obtained in the case of $k\!=\!16$, thus generally \texttt{LS-T} and \texttt{LS-NC} are almost one order of magnitude faster than \texttt{GS-T} and \texttt{GS-NC} in terms of $t$. Because when $t$ is large, more users are retained via range query accelerated by G-tree \cite{zhong2015g} or G*-tree \cite{li2019g}. This favors local search radiating outward from $Q$, equivalent to increasing the expansion radius.

\begin{figure}[t]
\centering
\subfigbottomskip=1pt
\subfigcapskip=-4pt
\subfigure{\label{fig:legend1}
\includegraphics[width=0.65\columnwidth]{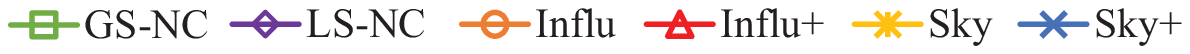}
}
\subfigure[Varying $k$] {\label{fig:comparison-delicious-k}
\includegraphics[width=0.465\columnwidth]{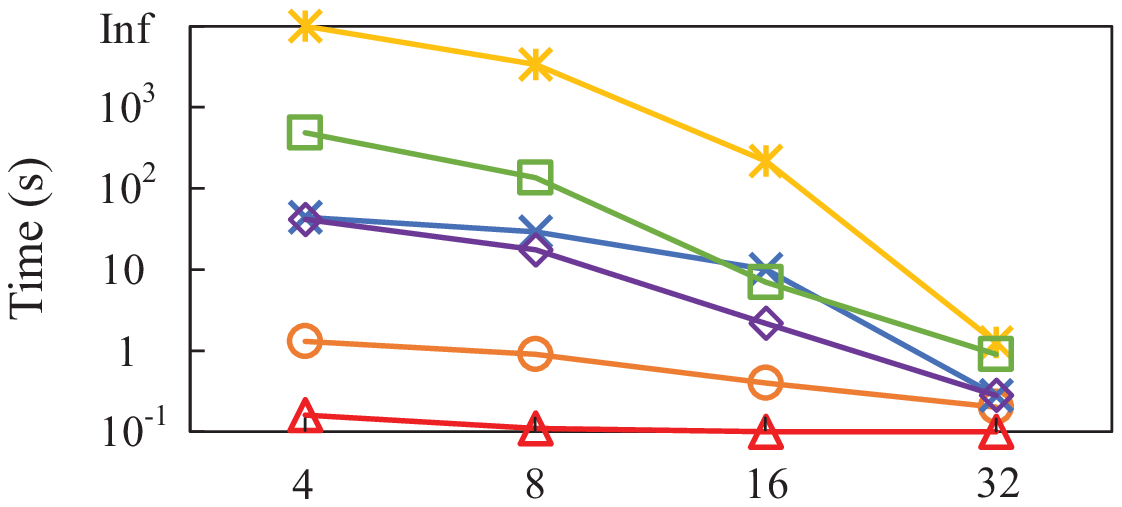}
}
\subfigure[Varying $d$] {\label{fig:comparison-delicious-d}
\includegraphics[width=0.465\columnwidth]{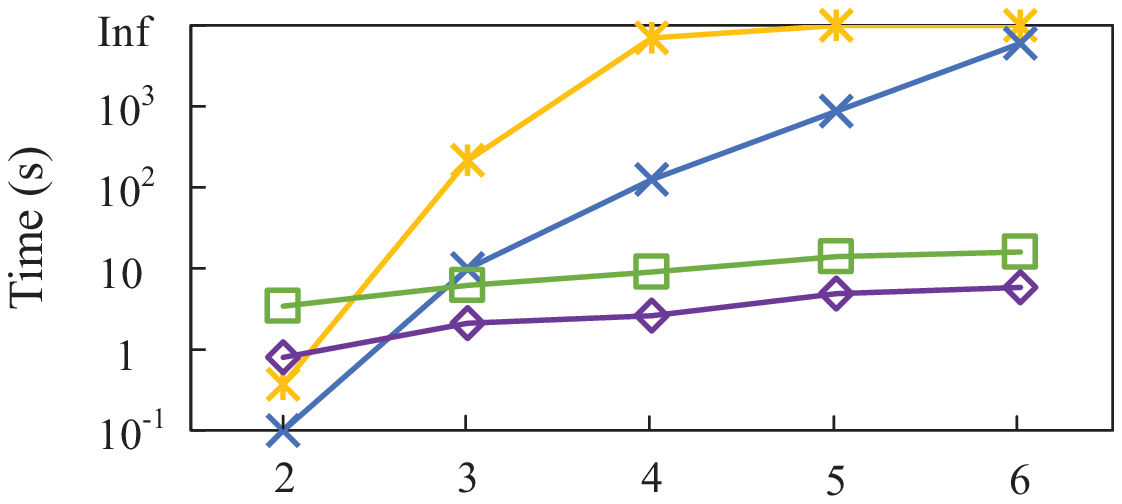}
}
\vspace{-12pt}
\caption{Comparison of different methods in SF+Delicious (independence).}
\vspace{-8pt}
\label{fig:comparison1}
\end{figure}

\begin{figure}[t]
\centering
\subfigbottomskip=1pt
\subfigcapskip=-4pt
\subfigure{\label{fig:legend2}
\includegraphics[width=0.65\columnwidth]{Exp/comparison/legend-1.eps}
}
\subfigure[Varying $k$] { \label{fig:comparison-flixster-k}
\includegraphics[width=0.465\columnwidth]{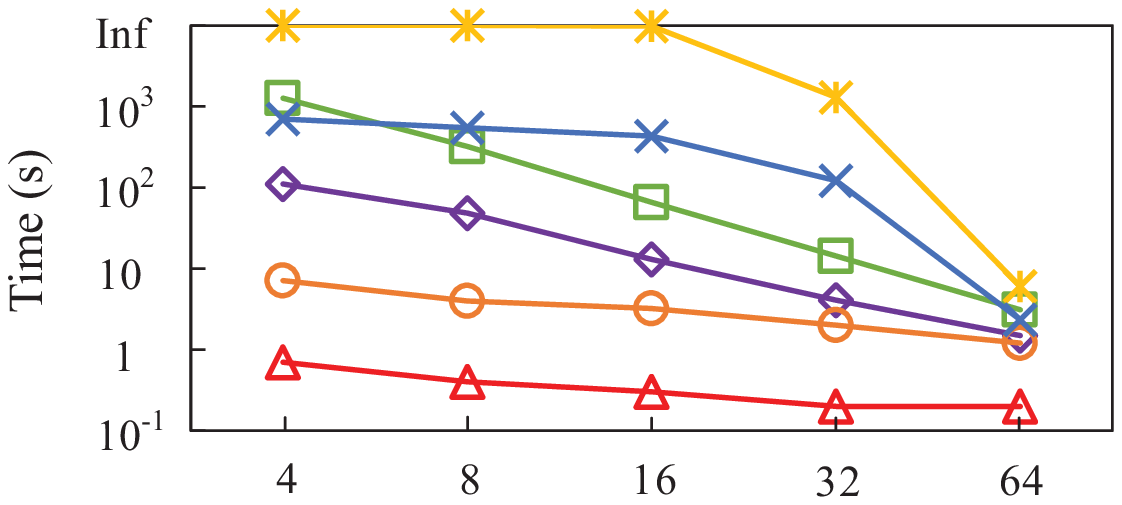}
}
\subfigure[Varying $d$] { \label{fig:comparison-flixster-d}
\includegraphics[width=0.465\columnwidth]{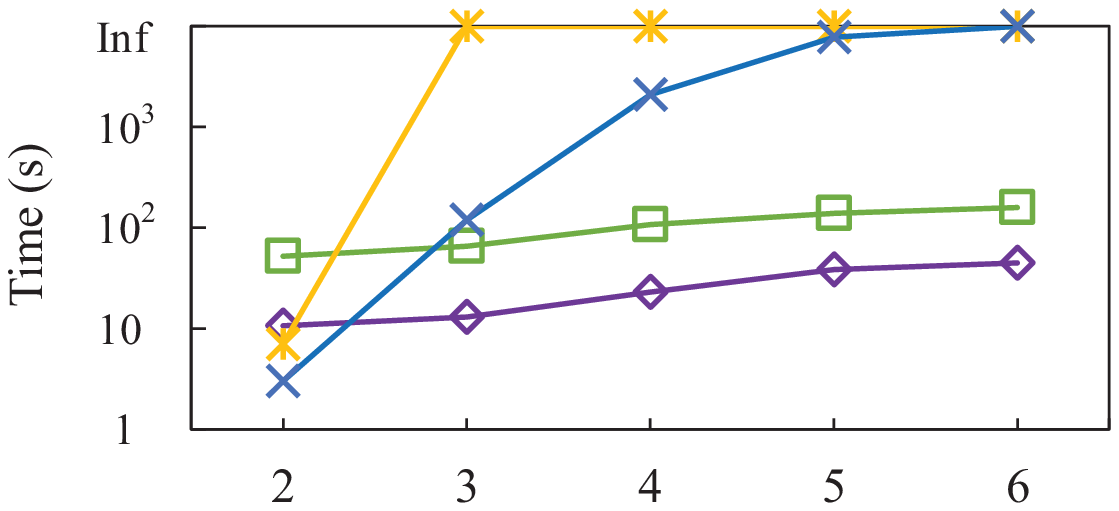}
}
\vspace{-12pt}
\caption{Comparison of different methods in FL+Flixster (independence).}
\vspace{-15pt}
\label{fig:comparison2}
\end{figure}

\noindent
\textbf{Exp-3: Varying $d$.} By varying $d$ we study the query processing time and the memory overhead of all algorithms, and comparison of different methods. The toughness of MAC search rises with $d$ due to its computational geometric nature. Nonetheless, all four algorithms offer practical processing time, taking respectively $159$s and $45$s for $d=6$ in Fig.~\ref{fig:flixster-d}. Furthermore, Fig.~\ref{fig:spacecost} shows the memory overhead of \texttt{GS-NC}/\texttt{LS-NC} and \emph{BBS} process ($X$ indexed and $G_d$ built). When $d$ increases, dimension of R-tree increases but memory overhead changes not much due to unchanged $G_d$ size; local search is very lightweight against global search. For example, \texttt{GS-NC} takes $664$MB and $2,219$MB while \texttt{LS-NC} takes only $52$MB and $152$MB for $d\!=\!3$ and $d\!=\!6$, respectively. The results confirm both the theoretical analysis and the claims on arrangement indexing in Section~\ref{section:findSmallestScoreVertex}. In addition, Fig.~\ref{fig:comparison1} and~\ref{fig:comparison2} show the comparison with methods in \cite{li2015influential} and \cite{li2018skyline}, where \texttt{Influ} (resp. \texttt{Influ+}) is the DFS-based (resp. ICP-index based) algorithm, and \texttt{Sky} (resp. \texttt{Sky+}) is the basic (resp. space-partition) algorithm. We implement \texttt{Influ} and \texttt{Influ+} by varying $k$ instead of $d$ since they can only capture $1$-dimensional attribute. For a fair comparison, $100$ weight vectors that fall anywhere in $R$ are randomly selected to respectively calculate the weighted sum of $d$ (at the default) numerical attributes as vertex influence (i.e., score), and the average processing time is reported in Fig.~\ref{fig:comparison-delicious-k} and~\ref{fig:comparison-flixster-k}. Since no r-dominance graph needs to be maintained and no half-space has to be computed nor inserted, \texttt{Influ} and \texttt{Influ+} are superior to \texttt{GS-NC} and \texttt{LS-NC} in terms of processing time, while \texttt{Sky} and \texttt{Sky+} are generally the most expensive due to their time complexity. On the other hand, in terms of $d$, \texttt{Sky} and \texttt{Sky+} are much costlier than ours and intractable when $d$ is relatively large, e.g., $d\!\geq\!3$ and $d\!\geq\!5$ respectively in Fig.~\ref{fig:comparison-flixster-d}. Here, ``Inf" means processing time exceeds $10,000$s. Therefore, our model and algorithms are tractable and scalable to handle real-world applications comprehensively and flexibly.

\begin{figure}[t]
\centering
\subfigbottomskip=3pt
\subfigcapskip=-4pt
\subfigure[Varying $k$] {\label{fig:ratio-k}
\includegraphics[width=0.465\columnwidth]{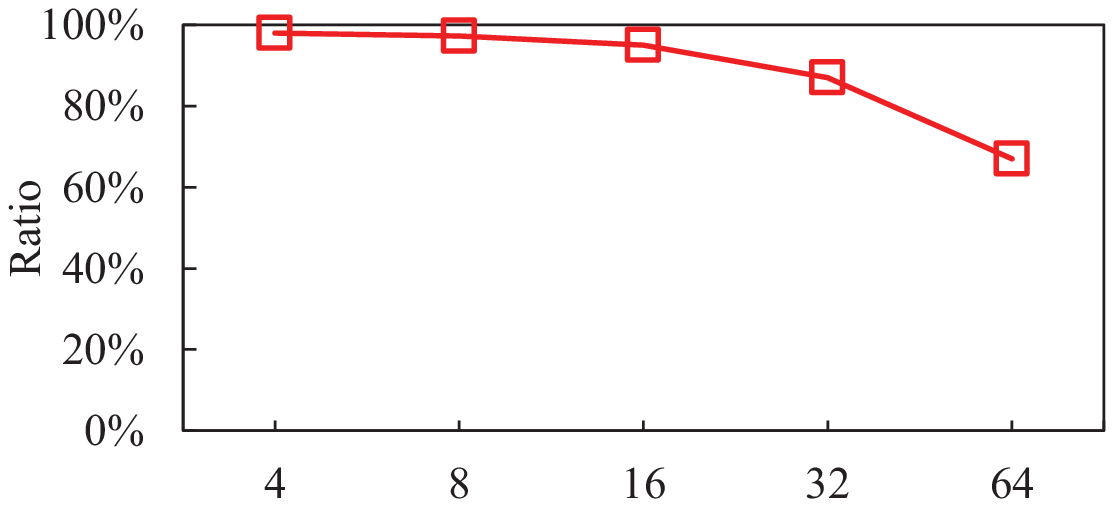}
}
\subfigure[Varying $|Q|$] { \label{fig:ratio-q}
\includegraphics[width=0.465\columnwidth]{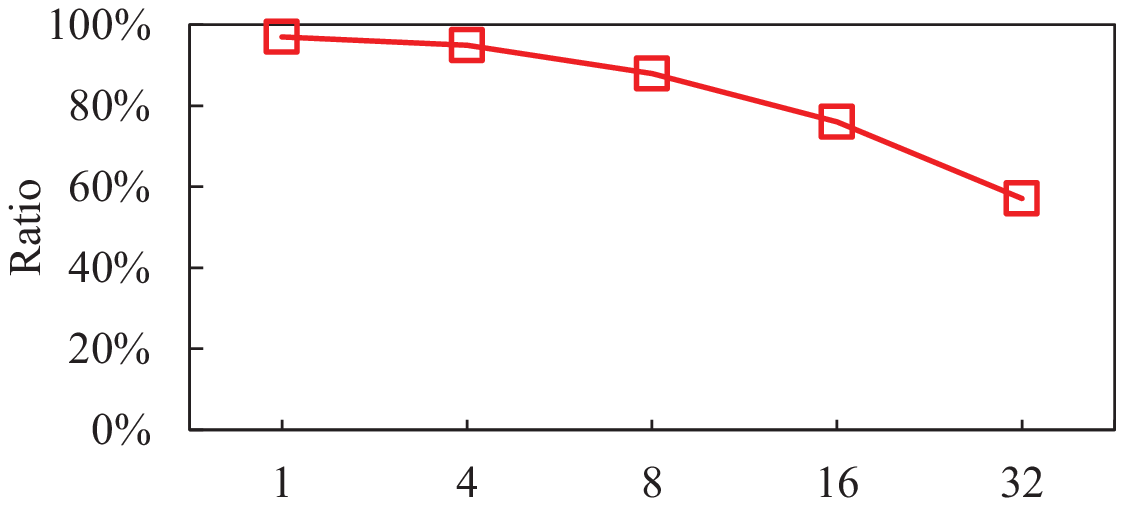}
}
\vspace{-12pt}
\caption{Ratio of NC-MACs found by LS-NC to GS-NC in FL+Lastfm.}
\vspace{-15pt}
\label{fig:ratio}
\end{figure}

\noindent
\textbf{Exp-4: Varying $|Q|$.} We evaluate the query processing time of all algorithms and the ratio of non-contained MACs (NC-MACs) found by \texttt{LS-NC} to \texttt{GS-NC} by varying $|Q|$. All processing time almost monotonically decreases with the growth of $|Q|$ since it accelerates the convergence of both global and local search. The reason is that increasing $|Q|$ means more vertices cannot be deleted in global search and selecting fewer vertices may find a candidate in local search. Note that in Fig.~\ref{fig:delicious-q}, as $k\!=\!16$, algorithms are not much faster at $Q\!=\!32$ than $Q\!=\!16$. In fact, we also find that global search terminates early if the generated query vertices $Q$ are located in the lower layer of $G_d$ because it is more likely to encounter $Q$ when deleting vertices, while local search is just the opposite. Fig.~\ref{fig:ratio-k} and~\ref{fig:ratio-q} show the ratio against $k$ and $|Q|$, respectively. We can see that both the ratios decrease with the growth of $k$ and $Q$, but it is still satisfactory. The reason is that the number of community candidates expanded by the \emph{Expand} procedure tends to decrease with the increase of $k$ or $|Q|$, but the \emph{Verify} procedure can always ensure the correctness of the corresponding partitions in $R$ of any (promising) community. On the other hand, this proves that local search is very useful for applications expecting only part of MACs. Note that, in practice, $|Q|$ is usually not very large and the ratio reaches $95\%$ at default $|Q|$, which confirms that local search can find all non-contained MACs in most cases.

\noindent
\textbf{Exp-5: Varying $j$.} We examine the query processing time of \texttt{GS-T} and \texttt{LS-T} by varying $j$. The curve of \texttt{GS-T} is rising slowly with increasing $j$ since the top-$j$ MACs can be directly obtained after executing global search. But for \texttt{LS-T}, after obtaining the non-contained MAC, its corresponding cell has to be divided again to find the top-$j$ MACs, resulting in an increase in processing time.

\noindent
\textbf{Exp-6: Varying $\sigma$.} We evaluate the query processing time of all algorithms, and the number of partitions and non-contained MACs by varying $\sigma$ (determining the size of region $R$). As anticipated, a larger $R$ means a larger output, thereby more computations needed. In Fig.~\ref{fig:no-partition}, we can see that the growth of $\sigma$ will lead to a significant increase in the number of partitions in $R$, which also explains the increase in processing time of all algorithms. Fig.~\ref{fig:no-nc-mac} records the relationship between the number of non-contained MACs obtained by \texttt{GS-NC} and $\sigma$. Similarly, as the number of partitions increases, the diversity of non-contained MACs w.r.t. $R$ will also increase. Note that both bars of Yelp in Fig.~\ref{fig:no-partition} and~\ref{fig:no-nc-mac} are much shorter than those of Flixster while its $H_k^t$ size is larger in Fig.~\ref{fig:vertices-in-Hkt}. Since attributes not only in Yelp but in real world are usually correlated or more, fewer (even unique) branches in DAG result in less processing time, that is, less half-space computation and insertion.

\subsection{Case Study}
\begin{figure}[t]
\centering
\subfigbottomskip=3pt
\subfigcapskip=-4pt
\subfigure[top-$1$ NC-MAC] {\label{fig:NC-MAC-1}
\includegraphics[width=0.265\columnwidth]{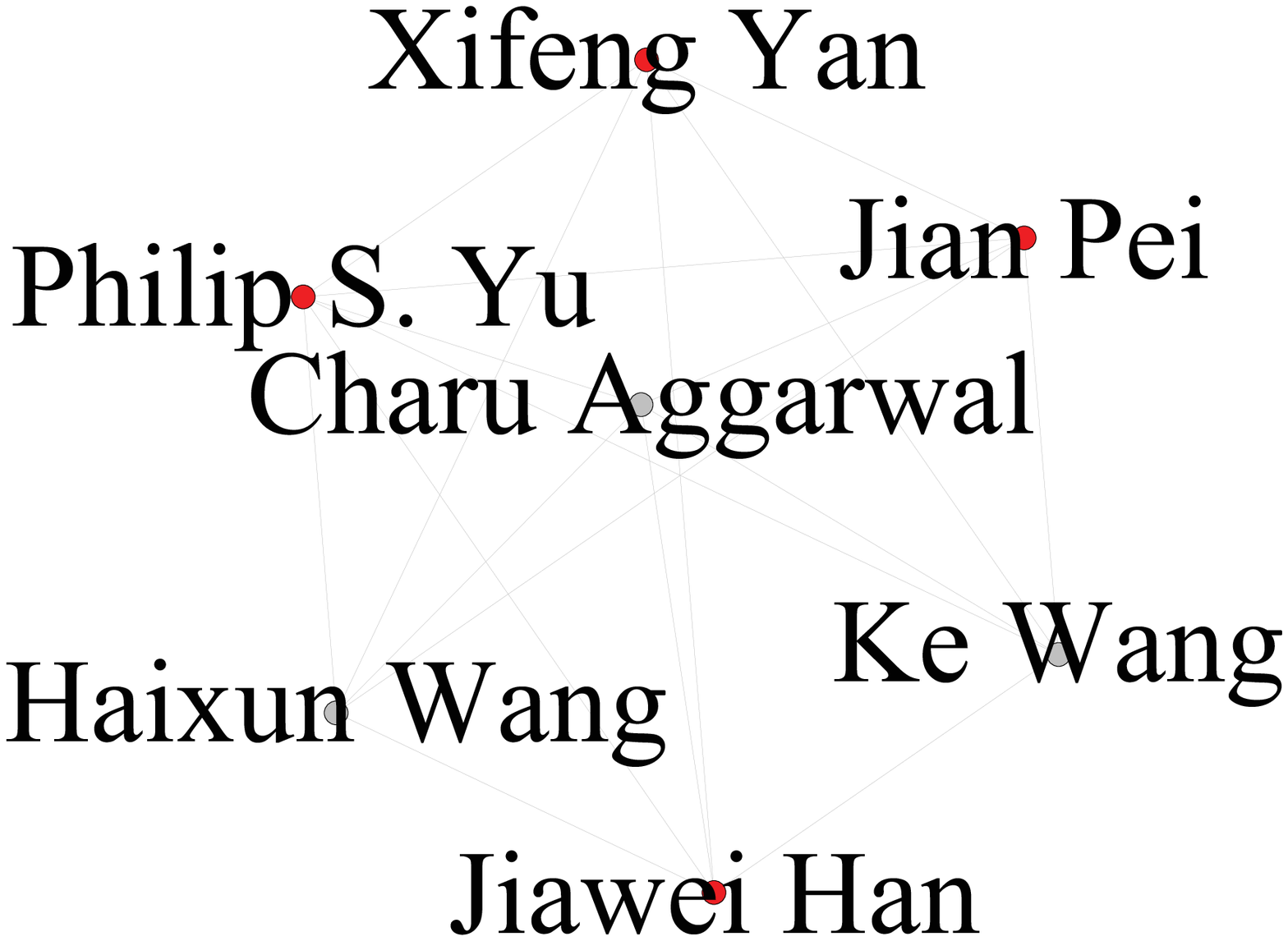}
}
\subfigure[top-$2$ MAC] { \label{fig:top2-MAC-1}
\includegraphics[width=0.3\columnwidth]{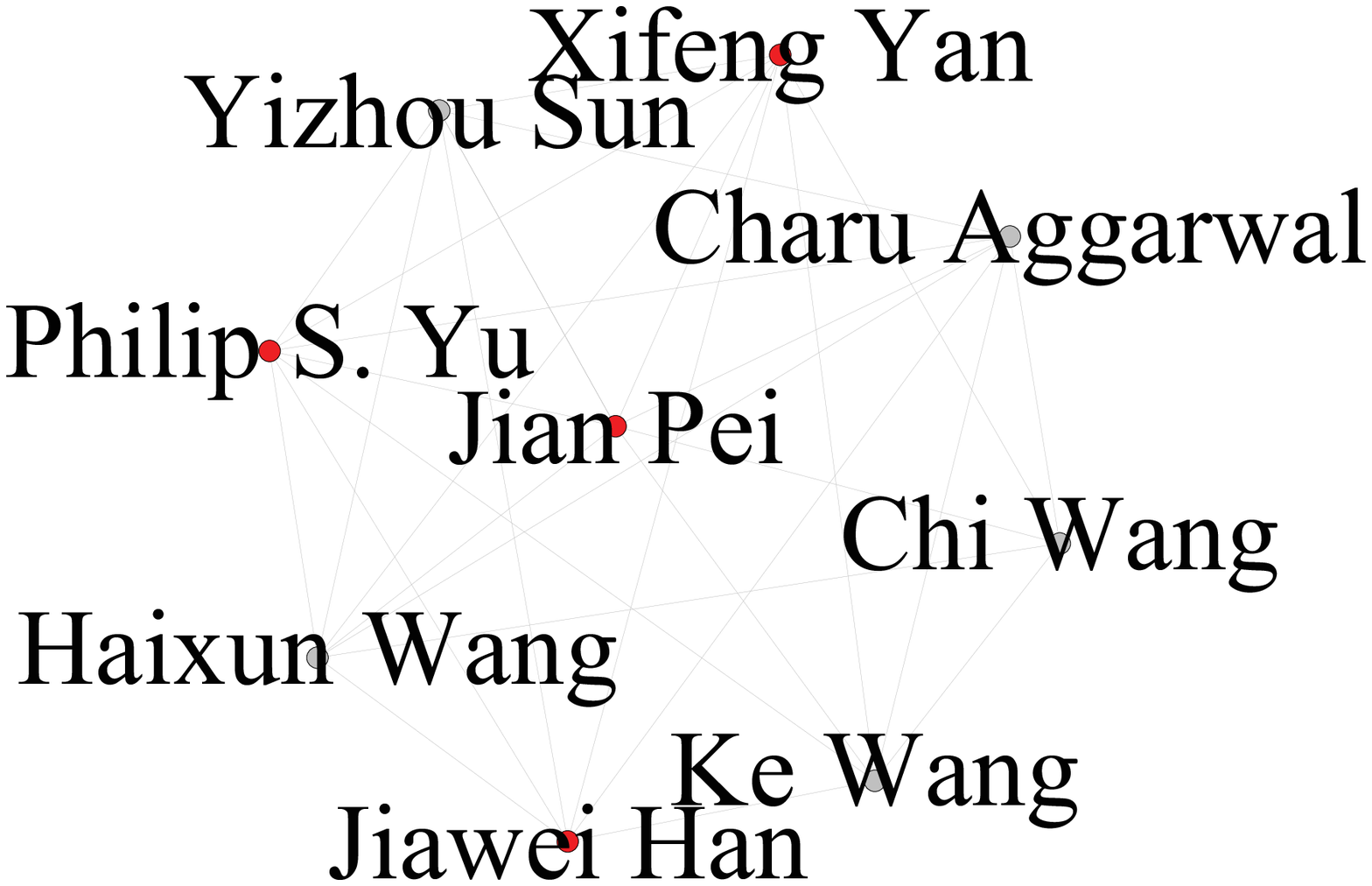}
}
\subfigure[top-$1$ NC-MAC] {\label{fig:NC-MAC-2}
\includegraphics[width=0.28\columnwidth]{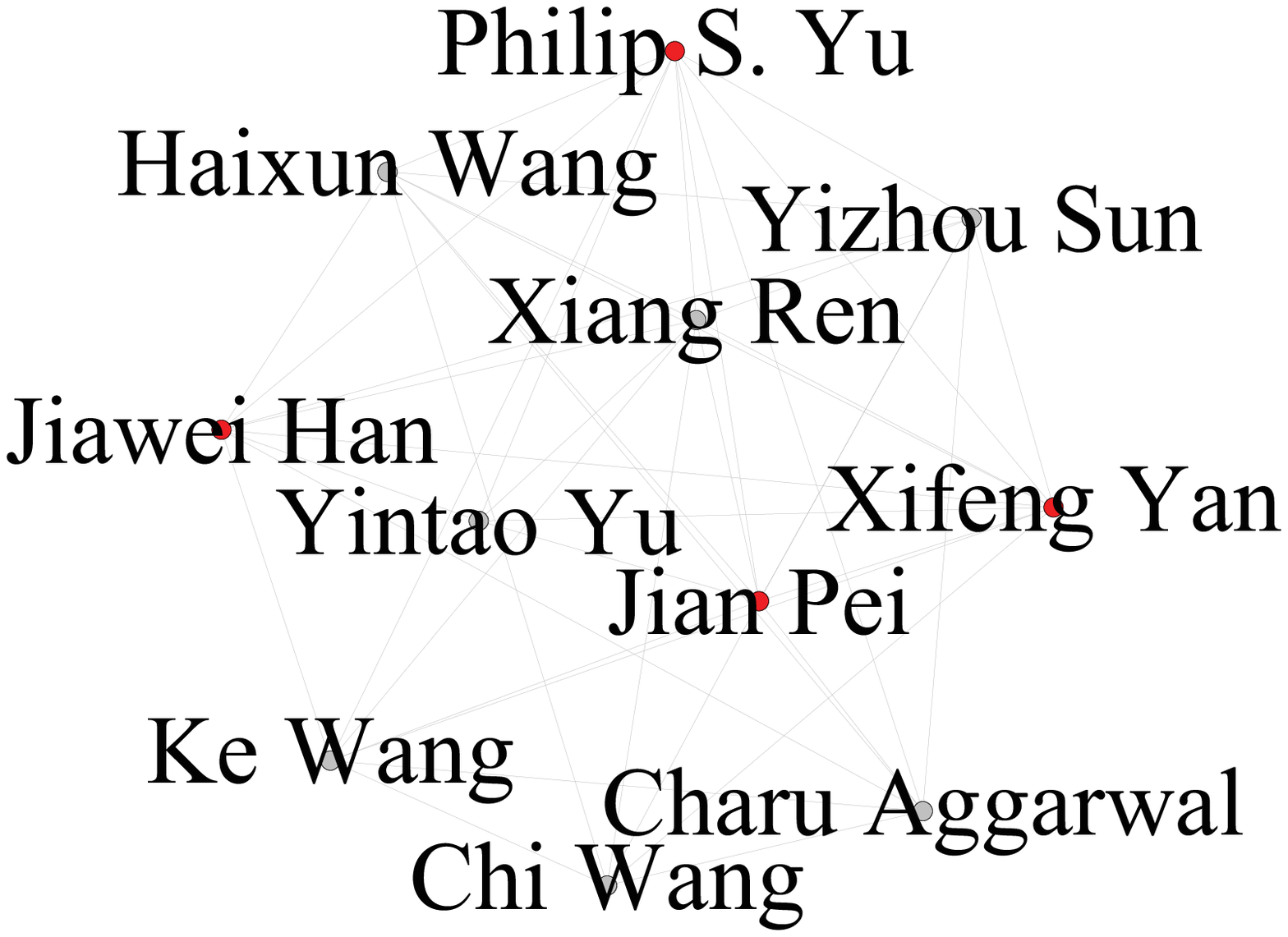}
}
\subfigure[top-$2$ MAC] { \label{fig:top2-MAC-2}
\includegraphics[width=0.28\columnwidth]{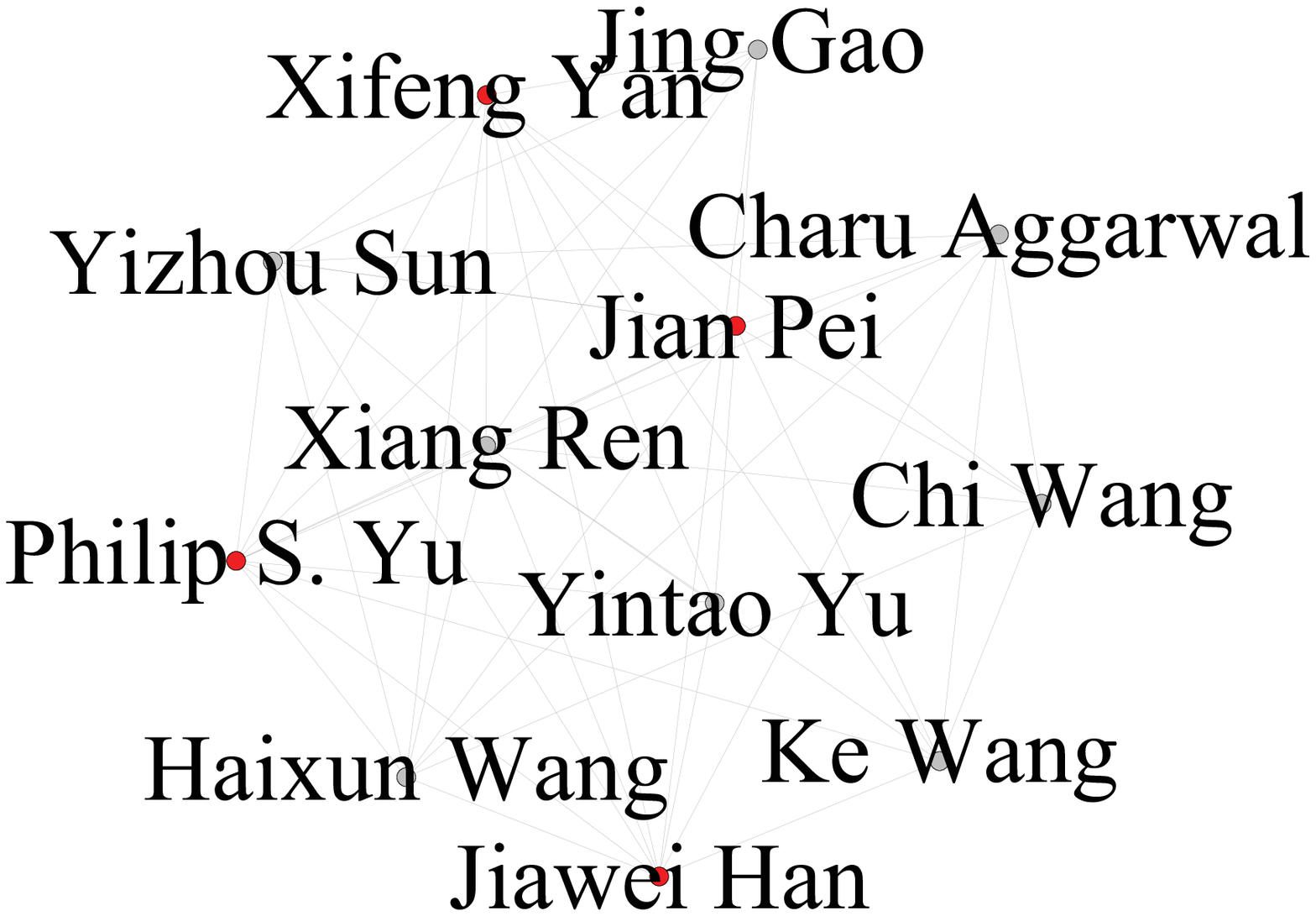}
}
\subfigure[SkyC] {\label{fig:sky-partialQ}
\includegraphics[width=0.28\columnwidth]{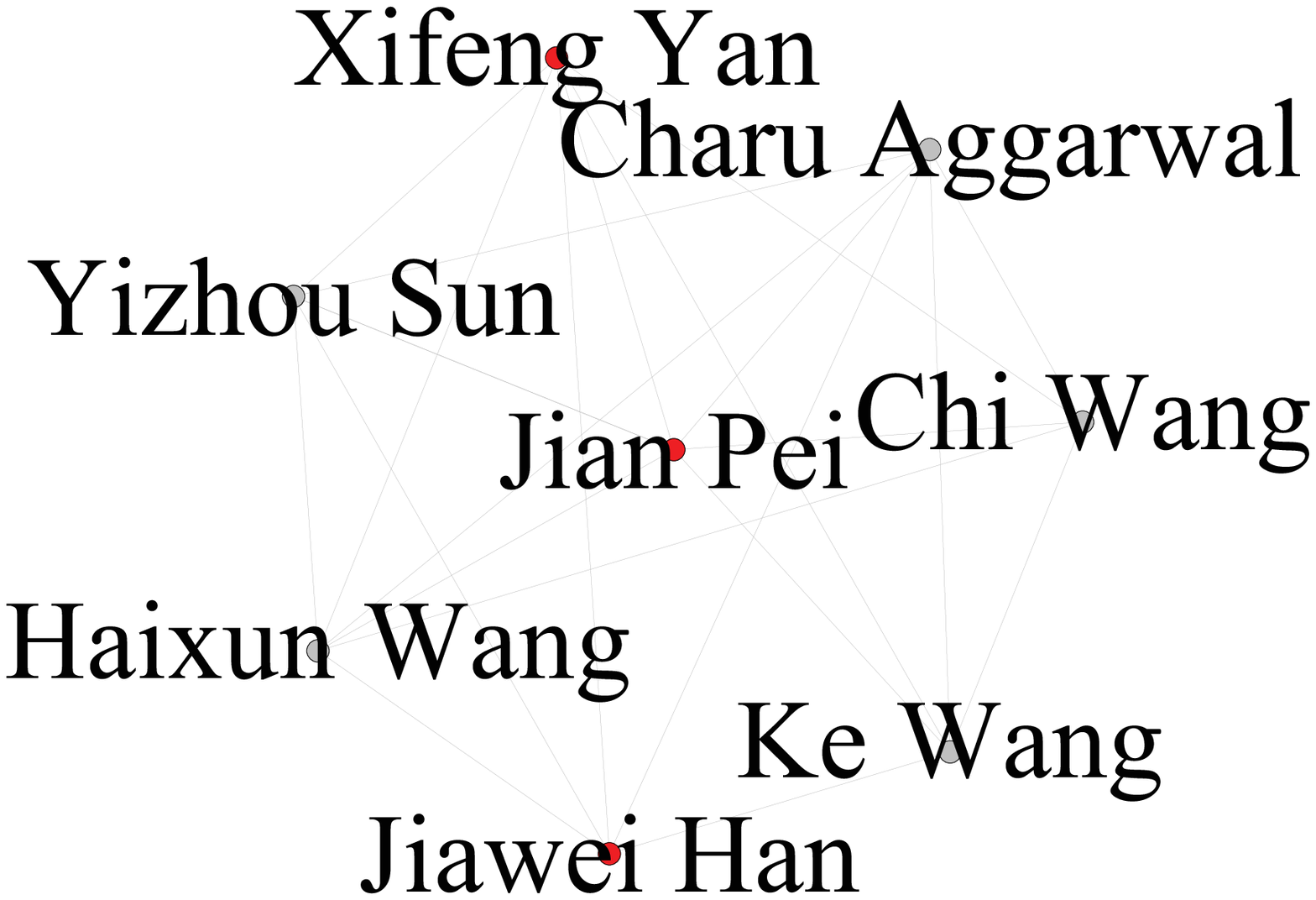}
}
\subfigure[InfC ($1$-D)] {\label{fig:influ-publication}
\includegraphics[width=0.28\columnwidth]{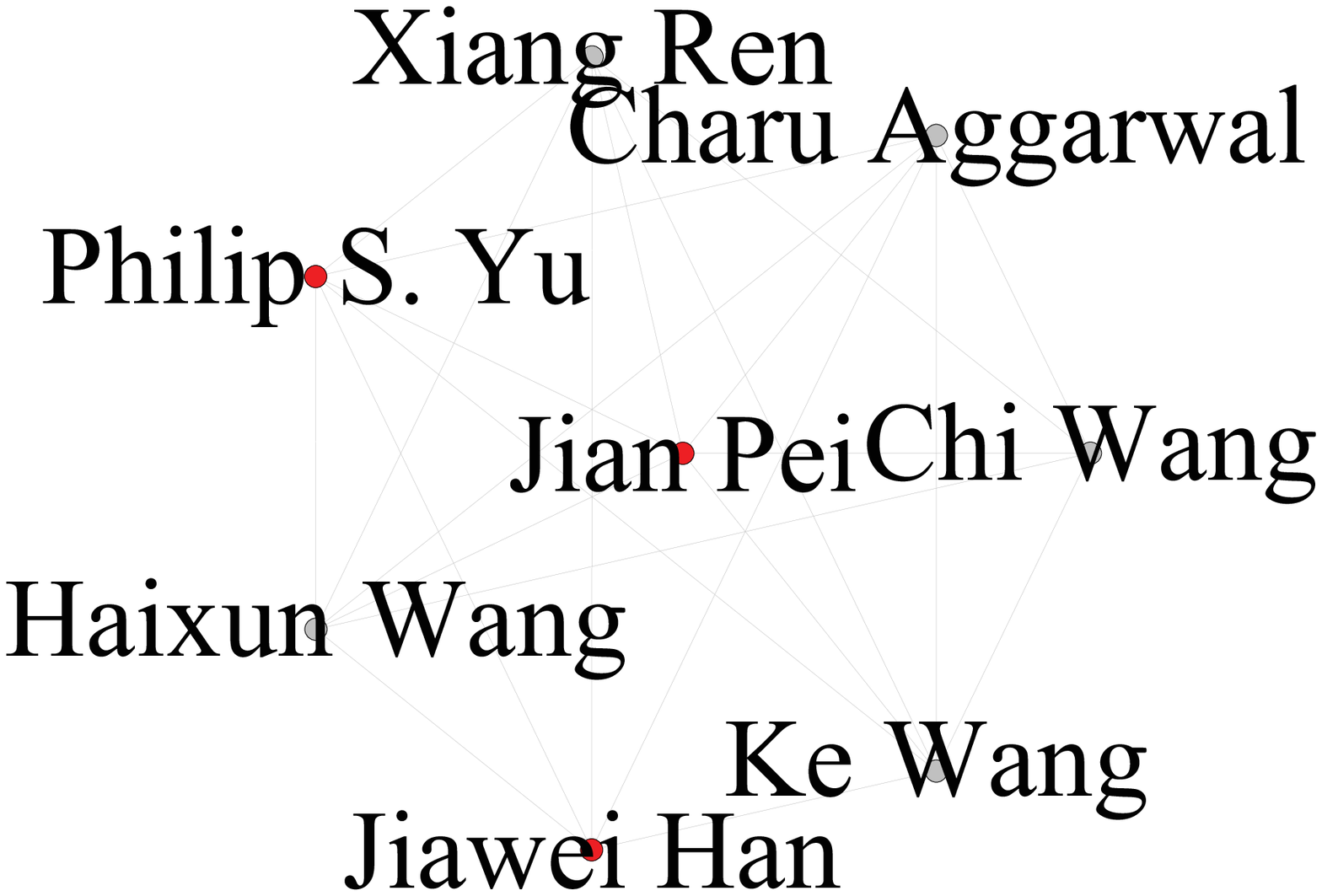}
}
\subfigure[InfC ($w \!\in\! R$)] { \label{fig:influ-partialQ}
\includegraphics[width=0.265\columnwidth]{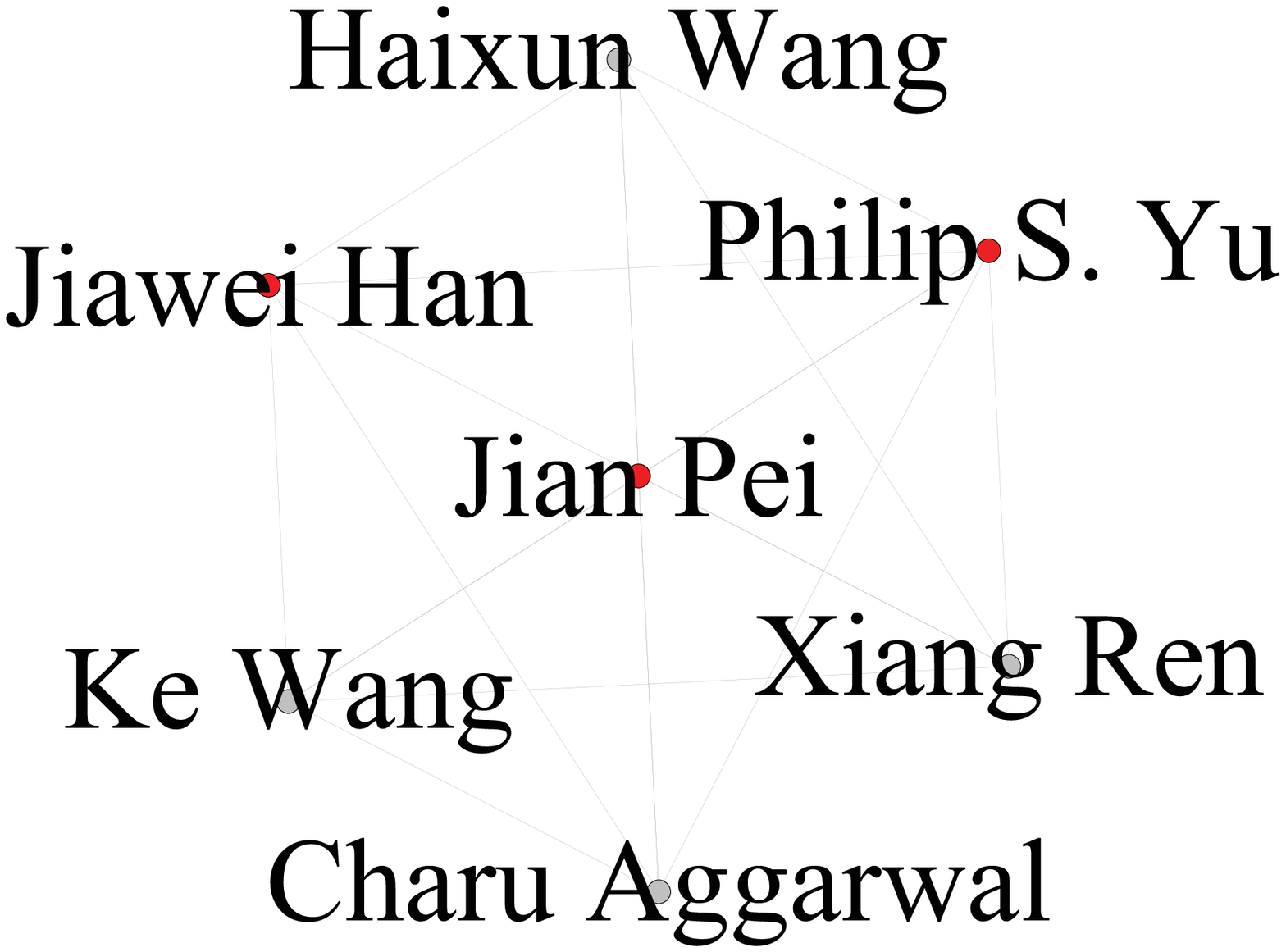}
}
\subfigure[ATC (``DM")] { \label{fig:truss}
\includegraphics[width=0.265\columnwidth]{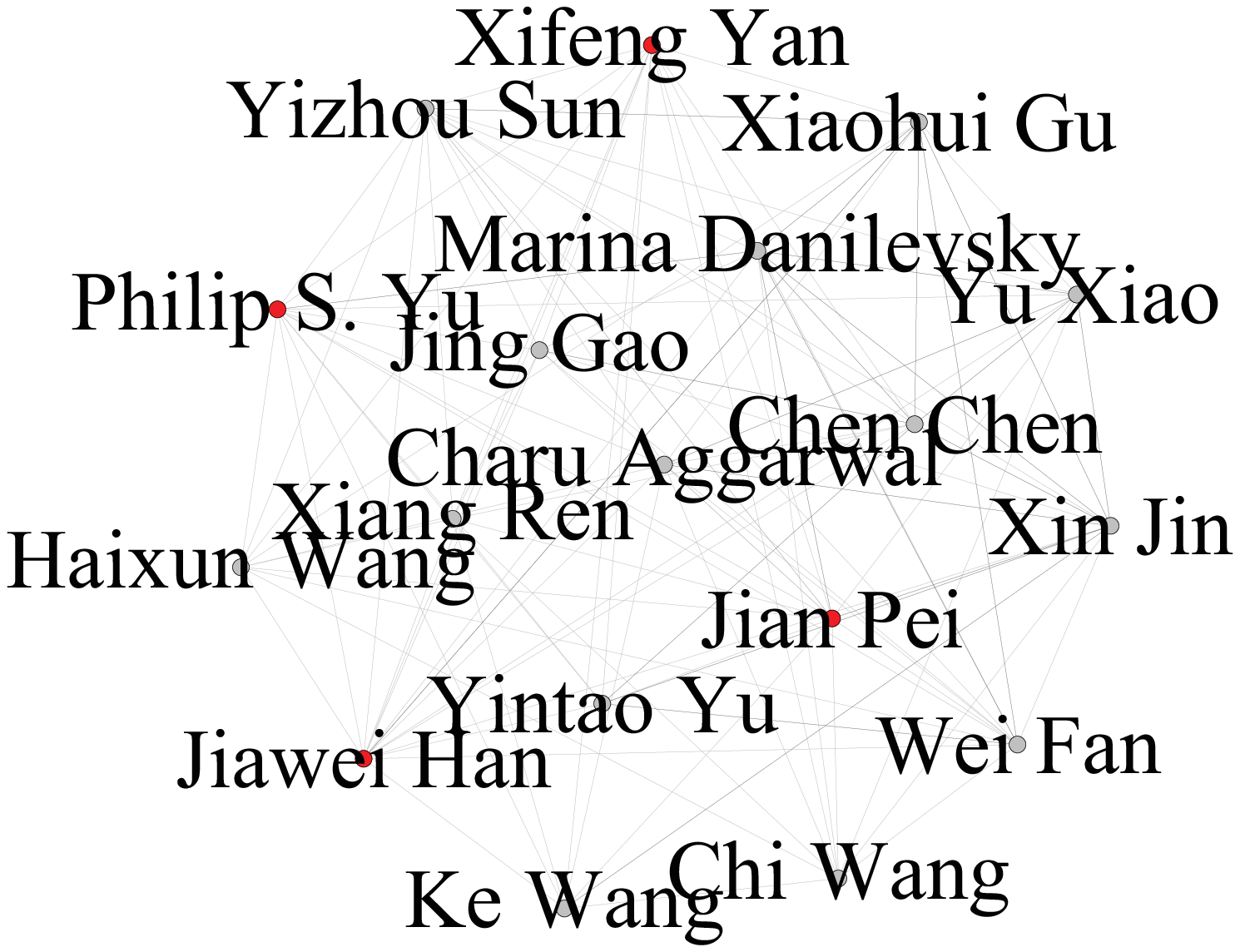}
}
\vspace{-5pt}
\caption{Case study of Aminer+NA: results for $k=5$.}
\vspace{-15pt}
\label{fig:casestudy}
\end{figure}

\noindent
\textbf{NA+Aminer.} We apply the road network of North America\textsuperscript{4} (NA) and the Aminer (aminer.org) for the first case study. The Aminer is a scientific collaboration network that incorporates authors in DB, DM, IR, and ML fields, comprised of $109,931$ vertices and $300,000$ edges. For each author we crawl four numerical attributes: \emph{h-index}, \emph{\#publications}, \emph{activeness}, and \emph{diverseness}. 
To evaluate the effectiveness of MAC model in real world, we map each author to the location in NA according to its affiliation and use $Q \!=\! \{$``Jiawei Han", ``Jian Pei", ``Philip S. Yu", ``Xifeng Yan"$\}$, who are renowned scientists in DM (i.e., relatively high scores), as query vertices. After setting $k\!=\!5, j\!=\!2$ and $R\!=\![0.1,0.3]\!\times\![0.3,0.5]\!\times\![0.05,0.1]$ (with $t$ large enough), Fig.~\ref{fig:casestudy}(a-d) show the top-$2$ MACs anywhere in $R$. Furthermore, we compare MAC with different models. For \texttt{InfC} \cite{li2015influential}, Fig.~\ref{fig:casestudy}(f, g) report the results involving $Q$, respectively taking \emph{\#publications} only and weighted sum (by $w\!=\!(0.3,0.4,0.1)$) as influence. In fact, \texttt{InfC} either cannot capture the characteristics of all attributes, or must be covered by a non-contained MAC (NC-MAC) if $w \!\in\! R$ (e.g., Fig.~\ref{fig:NC-MAC-2}). For \texttt{SkyC} \cite{li2018skyline}, there are two results, one is the same as NC-MAC in Fig.~\ref{fig:NC-MAC-1}, while the other shown in Fig.~\ref{fig:sky-partialQ} only contains partial $Q$ and is covered by NC-MAC in Fig.~\ref{fig:NC-MAC-2}. In effect, we find that \texttt{SkyC} is always contained in NC-MACs due to no query vertices and its skyline property. For \texttt{ATC} \cite{huang2017attribute}, Fig.~\ref{fig:truss} reports the $(6,2)$-truss w.r.t. $Q$ and keyword ``DM" as a $(k\!+\!1)$-truss is a $k$-core. Although communication cost is low (i.e., $2$), its size is still too large since it only considers maximum inclusion of keywords but ignores attributes. Therefore, the MAC model is very effective, comprehensive and flexible for applications.

\begin{figure}[t]
\centering
\subfigbottomskip=3pt
\subfigcapskip=-4pt
\subfigure[top-$1$ NC-MAC] {\label{fig:yelp-top1-MAC}
\includegraphics[width=0.265\columnwidth]{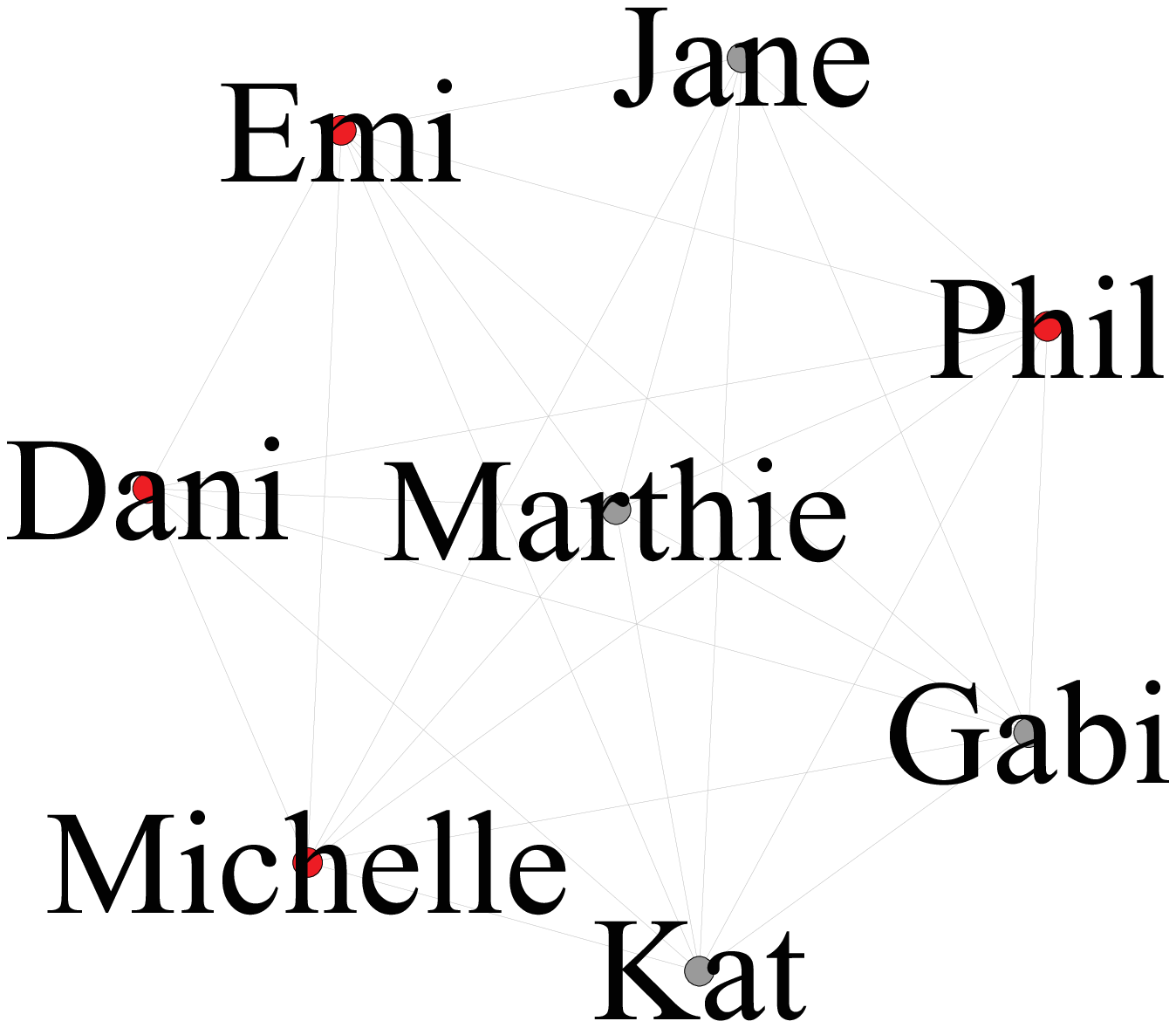}
}
\subfigure[top-$2$ MAC] { \label{fig:yelp-top2-MAC}
\includegraphics[width=0.28\columnwidth]{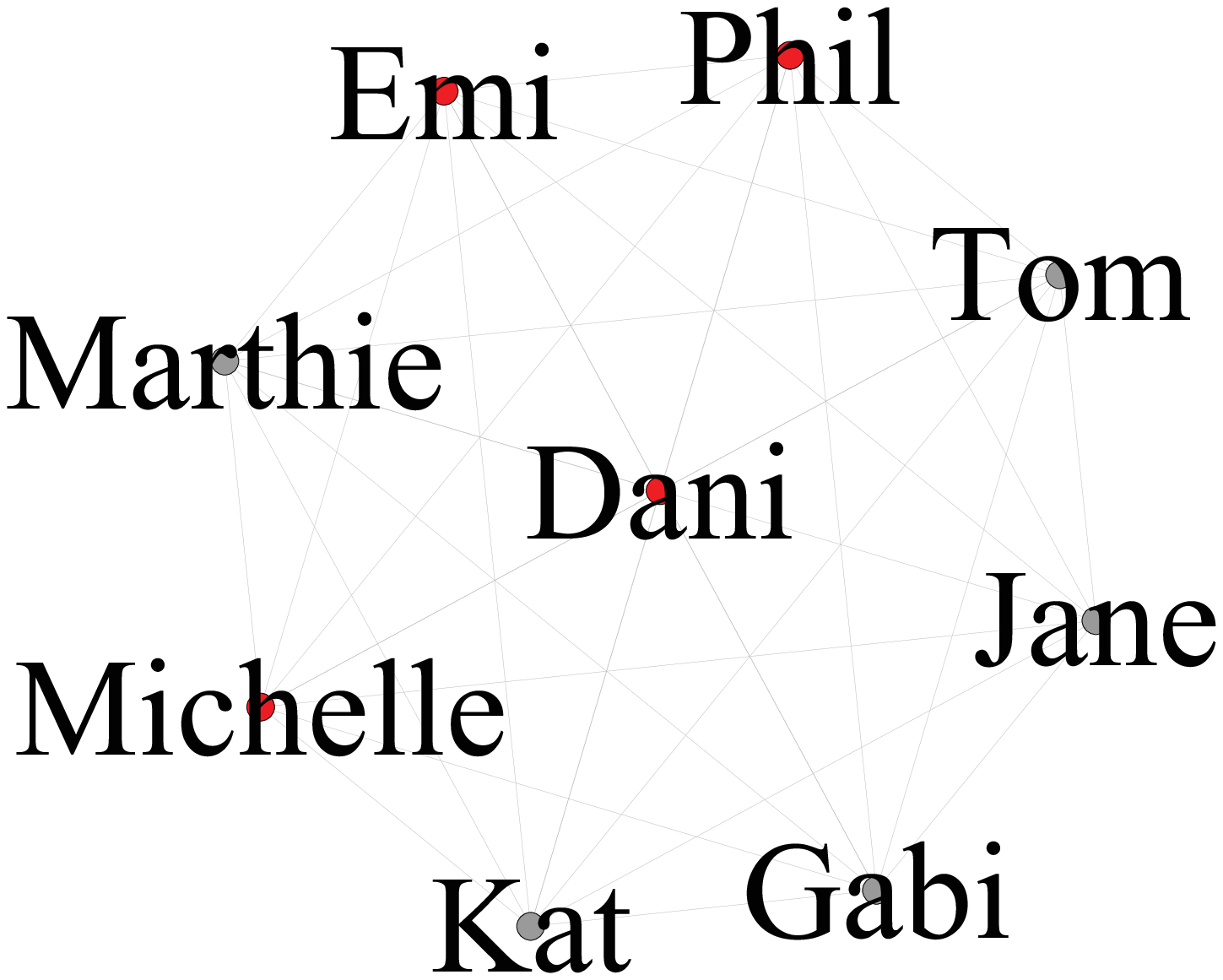}
}
\subfigure[top-$3$ MAC] { \label{fig:yelp-top3-MAC}
\includegraphics[width=0.27\columnwidth]{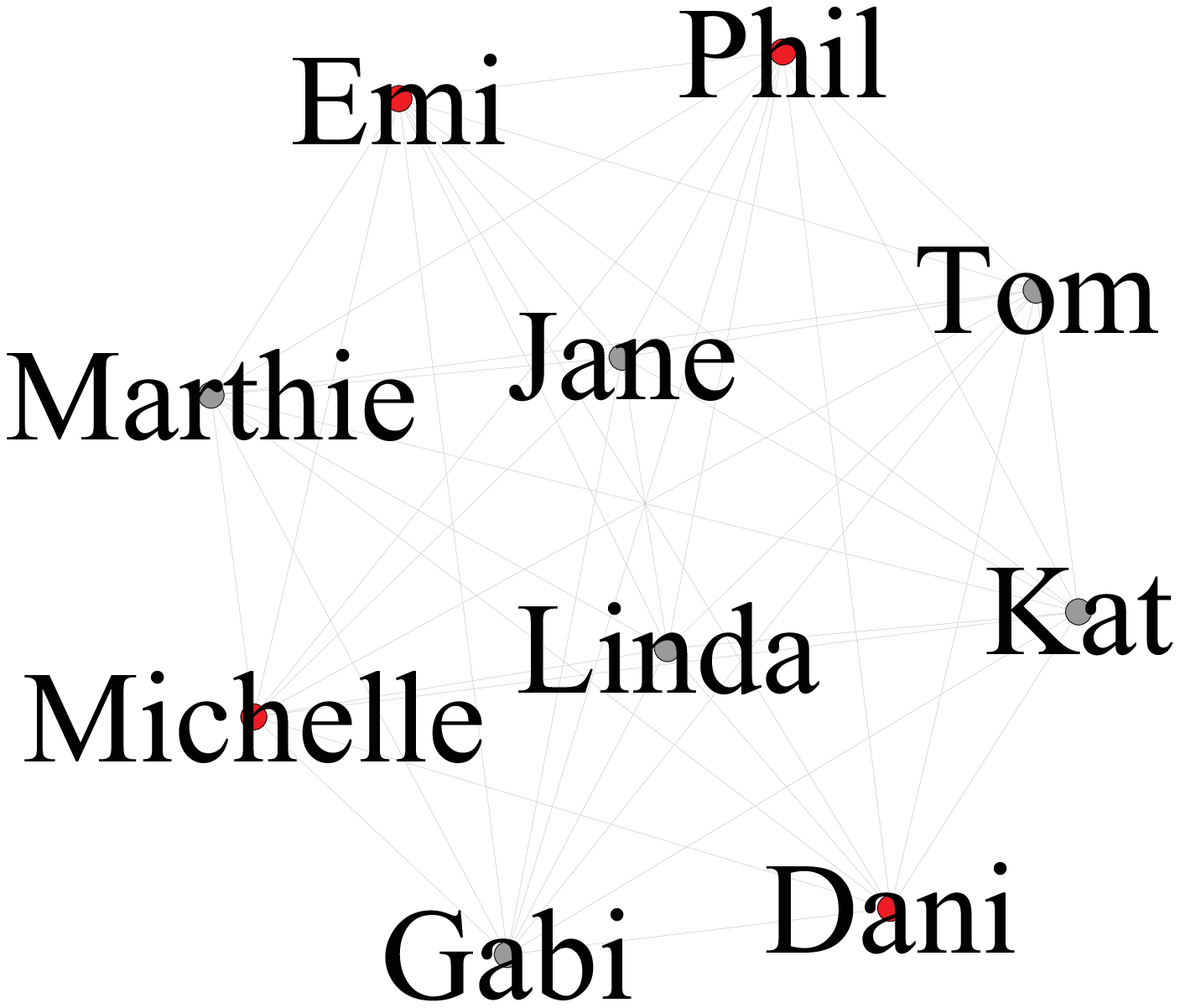}
}
\vspace{-5pt}
\caption{Case study of Yelp+SF: results for $k=6$.}
\vspace{-15pt}
\label{fig:yelp-casestudy}
\end{figure}

\noindent
\textbf{SF+Yelp.} We apply the SF and the Yelp\textsuperscript{3} for the second case study. In addition to profile information (e.g., ID, first name, etc.), users in Yelp also have real attribute data, such as average rating of all reviews, \#reviews written, \#hot compliments and so on. Note that the Yelp is more like an LBSN composed of many relatively small ego-like networks, in which ego-like users usually have more fans or followers, write more reviews and receive more compliments. That is, these ego-like users are highly active, showing high attribute values in each dimension; while ordinary users have very low attribute values in all dimensions (e.g., only browsing without posting). In fact, we find that in real world attributes are usually correlated or more, e.g., most users in Yelp have an attribute value of $0$. Thus, the number of branches in DAG will be extremely small or even unique, resulting in less half-space computation/insertion and fewer partitions in $R$. We map each user to the location in SF according to check-ins and use $Q \!=\! \{$``Emi", ``Phil", ``Dani", ``Michelle"$\}$, who are relatively active, as query vertices. To discover a group of people who are more concerned and popular, \#hot compliments, \#more compliments and \#photo compliments are used as three numerical attributes here. By setting $k\!=\!6, t\!=\!300, j\!=\!3$ and $R\!=\![0.4,0.5]\!\times\![0.1,0.2]$, Fig.~\ref{fig:yelp-casestudy} shows the top-$3$ MACs in $R$. This fully illustrates that in the real world, the diversity of (non-contained) MACs and the number of corresponding partitions w.r.t. $R$ are very small and user-friendly.

\section{Related Work}\label{section:relatedWork}
\noindent
\textbf{Community model and search.} 
A large number of community models have been proposed such as $k$-core \cite{seidman1983network,cui2014local}, $k$-truss \cite{huang2015approximate}, maximal clique \cite{cheng2011finding}, quasi-clique \cite{cui2013online}, maximal $k$-edge connected subgraph \cite{zhou2012finding,akiba2013linear,chang2013efficiently}, locally densest subgraph \cite{qin2015locally}, query-biased density \cite{wu2015robust}, etc. All these models consider only graph structural information but ignore numerical/textual attributes associated with vertices. To discover cohesive subgraphs containing the query vertices, a community search problem (CSP) was studied to find the maximal connected $k$-core in social networks \cite{sozio2010community}, for which \cite{cui2014local} proposed a more efficient local search algorithm. Recently, \cite{huang2017attribute} introduced the CSP of small-diameter $k$-truss with similar query attributes. \cite{fang2016effective} and \cite{fang2017effective} developed the CSP of $k$-core with textual attributes and smallest minimum covering circle, respectively. \cite{li2015influential} proposed an influential community with vertex's influence as one numerical attribute, based on which \cite{li2018skyline} studied a skyline community for $d$-dimensional numerical attributes. More recently, \cite{chen2018maximum} studied the CSP of $k$-truss with distance of at most $d$ for any two vertices. \cite{guo2019cohesive} and \cite{guo2020cohesive} investigated a cohesive version of CSP that brings all community members closest to the point-of-interest in road networks. \cite{chen2019contextual} and \cite{zhang2019keyword} studied two different CSPs in terms of context with only query keywords but no query vertices. In addition, \cite{liu2020vac} and \cite{luo2020efficient} made variations on the CSPs in \cite{huang2017attribute} and \cite{chen2018maximum}, respectively.

This work differs from all the prior work in the following. (1) Our multi-attributed community (MAC) model is the first one that can incorporate uncertainty of user preferences in the weight vector into $d$-dimensional numerical attributes and capture spatial cohesiveness between users in road networks. (2) The preference domain is introduced into community modeling for the first time. (3) We study the novel MAC search problems in road-social networks such that our techniques are significantly different from all previous CSP algorithms.

\noindent
\textbf{Skyline and its generalization.} The r-dominance graph used in our MAC model is relevant to the skyline \cite{borzsony2001skyline} and more to its extension, the $k$-skyband \cite{papadias2005progressive}. In traditional top-$j$ queries, if two records are inconsistent and one has no smaller value in any dimension \cite{borzsony2001skyline}, then it \emph{dominates} the other. Thus, for a dataset the skyline consists of the records which are not dominated by any other; while the $k$-skyband comprises those dominated by fewer than $k$ other records \cite{papadias2005progressive}, indicated as a superset of all records which for any weight vector may occur in the top-$j$ results. As a typical $k$-skyband computation algorithm, \emph{BBS} \cite{papadias2005progressive} adopts a spatial index in the dataset, following the branch-and-bound paradigm \cite{land1960automatic}. 


This proves an additional essential difference between \cite{li2018skyline} and our work from another perspective. Regardless of social and spatial cohesiveness, dominance in \cite{li2018skyline} between two communities comes down to a series of standard dominance tests on $d$-dimensional vectors. However, r-dominance tests adopted in our model do not suffice, 
e.g., two or more non-skyline communities may still collaboratively disqualify a skyline one if they score higher than it at different parts of $R$, collectively blocking it from being a top-$j$ result anywhere in $R$. 

\section{Conclusions}\label{section:conclusions}
In this paper, we propose a novel community model to discover normative communities suitable for multi-criteria decision making in a road-social network, in which each user is linked with location information and $d$ ($\geq\! 1$) numerical attributes. Taking a preference region of $d$-dimensional data domain as input, the resulting communities identified by our model cannot be r-dominated by other ones as long as the weight vector could fall anywhere in the region. We formalize the multi-attributed community search; distinguish two problem versions; develop solutions for corresponding processing; and using both real-world and synthetic datasets demonstrate the efficiency and scalability of our solutions and the effectiveness of our model.

\section*{Acknowledgment}
This work is supported by NSFC (No. 61932004, 61732003, 61729201 and 62072087) and Fundamental Research Funds for the Central Universities (No. N181605012). Ye Yuan is the corresponding author.

\bibliographystyle{IEEEtran}
\bibliography{IEEEabrv,sample}

\end{document}